\documentclass[11pt,a4paper]{article}
\pdfoutput=1
\usepackage{jheppub}

\usepackage{placeins}
\usepackage{latexsym}
\usepackage[bbgreekl]{mathbbol}
\usepackage{mathtools}
\usepackage{amsmath,amssymb,amsthm,amsbsy,amsfonts}
\usepackage{float}
\usepackage{graphicx,graphics}
\usepackage{psfrag}
\usepackage{subfigure}
\usepackage{nicefrac}
\usepackage[stable]{footmisc}
\usepackage{rotating}
\usepackage{afterpage}
\usepackage{bbm}

\usepackage{etoolbox}

\makeatletter
\newcounter{qrr@oldeq}
\newcounter{qrr@oldsubeq}
\newcounter{qrr@realeq}
\renewenvironment{subequations}{%
  \refstepcounter{equation}%
  \protected@edef\theparentequation{\theequation}%
  \setcounter{parentequation}{\value{equation}}%
  \setcounter{equation}{0}%
  \def\theequation{\theparentequation\alph{equation}}%
  \ignorespaces
}{%
  \setcounter{qrr@oldeq}{\value{parentequation}}%
  \setcounter{qrr@oldsubeq}{\value{equation}}%
  \setcounter{equation}{\value{parentequation}}%
  \ignorespacesafterend
}
\newenvironment{subequations*}{%
  \setcounter{qrr@realeq}{\value{equation}}%
  \let\theparentequation\theequation%
  \patchcmd{\theparentequation}{equation}{parentequation}{}{}%
  \setcounter{parentequation}{\numexpr\value{qrr@oldeq}-1}%
  \setcounter{equation}{\value{qrr@oldsubeq}}%
  \def\theequation{\theparentequation\alph{equation}}%
  \refstepcounter{parentequation}%
  \ignorespaces
}{%
  \setcounter{qrr@oldeq}{\value{parentequation}}%
  \setcounter{qrr@oldsubeq}{\value{equation}}%
  \setcounter{equation}{\value{qrr@realeq}}%
  \ignorespacesafterend
}
\makeatother

\def\NO{\nonumber}

\newcommand{\be}{\begin{equation}}
\newcommand{\ee}{\end{equation}}

\def\bea{\begin{eqnarray}}
\def\eea{\end{eqnarray}}

\def\bal#1\eal{\begin{align}#1\end{align}}

\def\bald{\begin{aligned}}
\def\eald{\end{aligned}}

\def\beqx{\begin{displaymath}}
\def\eeqx{\end{displaymath}}

\newcommand{\bmat}{\left(\begin{array}}
\newcommand{\emat}{\end{array}\right)}

\def\a{\alpha}

\def\c{\chi}
\def\d{\delta}
\def\e{\epsilon}
\def\f{\phi}
\def\g{\gamma}

\def\k{\kappa}
\def\l{\lambda}
\def\m{\mu}
\def\n{\nu}
\def\o{\omega}
    \def\om{\omega}
\def\p{\pi}

    \def\th{\theta}
\def\r{\rho}
\def\s{\sigma}
\def\t{\tau}

\def\D{\Delta}

\def\G{\Gamma}

\def\O{\Omega}
    \def\Om{\Omega}
\def\P{\Pi}

    \def\Th{\Theta}
\def\S{\Sigma}

\def\vf{\varphi}

\def\ba{\bbalpha}
\def\bk{\bbkappa}

\def\bs{\bbsigma}

\def\ca{{\cal A}}

\def\ce{{\cal E}}
\def\cf{{\cal F}}

\def\ch{{\cal H}}

\def\cl{{\cal L}}
\def\cm{{\cal M}}

\def\co{{\cal O}}
\def\cp{{\cal P}}

\def\car{{\cal R}}
\def\cs{{\cal S}}

\def\bb#1{\ensuremath{\mathbb{#1}}} 

\def\bo{{\raise-.3ex\hbox{\large$\Box$}}}               
\def\pa{\partial}                                       
\def\face{{\raise.2ex\hbox{$\displaystyle \bigodot$}\mskip-2.2mu \llap {$\ddot
        \smile$}}}                                   
\def\>{\rangle}                                      
\def\<{\langle}                                      

\newcommand{\sub}[1]{\phantom{}_{(#1)}\phantom{}}    
\def\wt#1{\widetilde{#1}}                            
\def\Hat#1{\widehat{#1}}                             
\def\leftrightarrowfill{$\mathsurround=0pt \mathord\leftarrow \mkern-6mu
        \cleaders\hbox{$\mkern-2mu \mathord- \mkern-2mu$}\hfill
        \mkern-6mu \mathord\rightarrow$}        
\def\dvec#1{\vbox{\ialign{##\crcr
        \leftrightarrowfill\crcr\noalign{\kern-1pt\nointerlineskip}
        $\hfil\displaystyle{#1}\hfil$\crcr}}}           
\def\Re{{\rm Re\,}}                                     
\def\Im{{\rm Im\,}}                                     

\def\-{\hphantom{-}}

\title{Holographic Hall conductivities from dyonic backgrounds}

\author[a]{Jonathan Lindgren}
\author[b]{Ioannis Papadimitriou}
\author[a]{Anastasios Taliotis}
\author[a]{Joris Vanhoof}

\affiliation[a]{Theoretische Natuurkunde, Vrije Universiteit Brussel and\\ International Solvay Institutes, Pleinlaan 2, B-1050 Brussels, Belgium}

\affiliation[b]{SISSA and INFN - Sezione di Trieste, Via Bonomea 265, I 34136 Trieste, Italy}

\emailAdd{ejonathanlindgren@gmail.com}
\emailAdd{ioannis.papadimitriou@sissa.it}
\emailAdd{atalioti@vub.ac.be}
\emailAdd{joris.vanhoof@vub.ac.be}

\abstract{We develop a general framework for computing the holographic 2-point functions and the corresponding conductivities in asymptotically locally AdS backgrounds with an electric charge density, a constant magentic field, and possibly non-trivial scalar profiles, for a broad class of Einstein-Maxwell-Axion-Dilaton theories, including certain Chern-Simons terms. Holographic renormalization is carried out for any theory in this class and the computation of the renormalized AC conductivities at zero spatial momentum is reduced to solving a single decoupled first order Riccati equation. Moreover, we develop a first order fake supergravity formulalism for dyonic renormalization group flows in four dimensions, allowing us to construct analytically infinite families of such backgrounds by specifying a superpotential at will. These RG flows interpolate between AdS$_4$ in the UV and a hyperscaling violating Lifshitz geometry in the IR with exponents $1<z<3$ and $\th=z+1$. For $1<z<2$ the spectrum of fluctuations is gapped and discrete. Our hope and intention is that this analysis can serve as a manual for computing the holographic 1- and 2-point functions and the corresponding transport coefficients in any dyonic background, both in the context of AdS/CMT and AdS/QCD.       
}

\keywords{AdS/CFT, gauge/gravity correspondence, D-branes, AdS/CMT, hyperscaling violation}

\preprint{SISSA 15/2015/FISI}

\begin{document}
	\maketitle

\section{Introduction and summary of results}
\label{intro}

Holographic techniques have been applied in recent years to a wide range of quantum systems exhibiting strong dynamics, both in high energy and condensed matter physics. Even though the application of holography to most real-world systems is entirely phenomenological, holographic models have been very successful in capturing certain qualitative aspects of strongly coupled systems. Indeed, such models provide an {\em effective theory} description for certain strongly coupled systems, completely on par with an effective field theory description, which, as is the case for non-BCS superconductors \cite{Gubser:2008px,Hartnoll:2008vx}, is often not known.     

A class of phenomena that have attracted particular attention lately are phenomena occurring in the presence of a strong magnetic field, and considerable effort has been devoted to understanding such phenomena both within the field theory \cite{Bernevig:2002eq,Fujita:2009kw,Hikida:2009tp, Belhaj:2010iw,Son:2013rqa,Stephanov:2014dma,Gursoy:2014aka,Chen:2014cla,Geracie:2014nka,Kharzeev:2015kna} and the holographic \cite{Erdmenger:2007bn,Goldstein:2010aw,Bergman:2010gm,Gynther:2010ed,Landsteiner:2011iq,Jokela:2011eb,Kristjansen:2012ny,Bu:2012mq,Wu:2014dha,Lippert:2014jma,Mamo:2015aia,Amoretti:2015gna,Blake:2015ina,Kim:2015wba,Fuini:2015hba} frameworks. In particular, the focus in \cite{Son:2013rqa,Chen:2014cla,Wu:2014dha,Geracie:2014nka} was on condensed matter phenomena, such as the Quantum Hall Effect \cite{Bernevig:2002eq,KeskiVakkuri:2008eb, Davis:2008nv,Alanen:2009cn, Fujita:2009kw, Hikida:2009tp,Bergman:2010gm, Jokela:2011eb,Fujita:2012fp,Kristjansen:2012ny,Belhaj:2010iw,Lippert:2014jma,Banerjee:2014pya}, while \cite{Gursoy:2014aka,Fuini:2015hba,Mamo:2015aia} study the effect of the magnetic field on heavy ion collisions and  \cite{Erdmenger:2007bn,Gynther:2010ed,Landsteiner:2011iq,Bu:2012mq,Chen:2014cla, Stephanov:2014dma,Iatrakis:2014dka,Kim:2015wba} explore the Stark, chiral magnetic, chiral vortical, and Nernst effects. In fact, many of these phenomena have topological origin, which allows for a unified understanding of both condensed matter and heavy ion physics in strong magnetic fields \cite{Kharzeev:2015kna}. 

Thermoelectric conductivities have been computed holographically from dyonic backgrounds at finite temperature in the pioneering works \cite{Hartnoll:2007ai,2007PhRvB..76n4502H,Hartnoll:2007ip}. The magnetic field plays a crucial role in regulating the DC electric conductivity at finite charge density, removing the Drude peak that appears in any system with translation invariance. Translation invariance can also be broken in a number of other ways, leading to momentum relaxation (see e.g. \cite{Kim:2015wba} and references therein). Numerical dyonic backgrounds with running scalars at zero and finite temperature have also been considered recently in the study of the Fractional Quantum Hall Effect \cite{Lippert:2014jma}. Finally, dyonic backgrounds with a mass gap provide the interesting possibility to study thermalization in heavy ion collisions in the presence of finite charge density and magnetic field, extending previous works studying thermalization in confining backgrounds as an initial value problem \cite{Kiritsis:2011yn,Craps:2013iaa,Craps:2014eba,Ishii:2015gia} .

The purpose of the present work is twofold. Firstly, we put forward a general framework for systematically extracting holographic data from asymptotically locally AdS dyonic backgrounds in the presence of running scalars, based on a radial Hamiltonian formulation of the bulk dynamics. Although the general techniques we present here are not new, we feel that their power in simplifying the extraction of the holographic data has not been appreciated enough. In this respect we emphasize three aspects of the holographic dictionary: i) holographic renormalization and its importance to preserve the Ward identities, ii) $n$-point functions can be extracted directly from the bulk solutions, without evaluating the on-shell action, and iii) there are fundamental advantages in formulating the fluctuation equations as first order Riccati equations. All three aspects fundamentally rely on a Hamiltonian formulation of the dynamics. Indeed, there is a lot of millage one can get by exploiting the manifest symplectic structure of the Hamiltonian dynamics. This is because the local RG formulation of a quantum field theory \cite{Osborn:1991gm}, which is the natural language in the context of holography, inherently equips the space of local couplings and operators with a symplectic structure.  

As we shall see, correctly renormalizing the theory is crucial for holographically deriving Ward identities like (\ref{2pt-fns-WIs}) that relate 2- and 1-point functions, and which are required for expressing all thermoelectric conductivities in terms of the electric ones only. Moreover, the derivation of such identities becomes almost a triviality in the Hamiltonian language, since they follow directly from the first class constraints reflecting local symmetries in the bulk theory. Point ii) is important here as well. Namely, the 1-point functions are identified with the renormalized canonical momenta, in the presence of sources. This has a very concrete consequence which leads to drastic simplification of the computation of e.g. the 2-point functions: the 2-point functions can be computed by further differentiation of the 1-point functions, without evaluating the on-shell action. This is linear response theory applied directly to the bulk. Point iii) highlights the fact that the Riccati equations are precisely equations for the response functions directly, eliminating the sources from the problem at the very beginning. The advantages of doing this are that a) it becomes almost trivial to correctly renormalize individual 2-point functions, computing only the terms that contribute to a given correlator \cite{Papadimitriou:2004rz}, b) the numerical solution of the fluctuation equations is considerably simpler since only a single regularity condition must be imposed in the IR and the arbitrary sources have already been eliminated from the problem \cite{Papadimitriou:2013jca}, and c) the fact that the Riccati equations follow from the Hamiltonian formulation of the dynamics implies that the response functions are automatically compatible with the symplectic structure of the theory, thus avoiding the necessity of evaluating the symplectic form to correctly identify the modes in generic coupled systems \cite{Papadimitriou:2010as}.   

The second aim of this paper is to present exact analytic dyonic backgrounds with running scalars, some of which are confining. Most of the holographic backgrounds that have been used to study phenomena at finite charge density and magnetic field are known only numerically, or in some asymptotic limit. This is particularly the case for backgrounds that involve running scalars. Here, we present infinite classes of new analytic zero temperature solutions corresponding to dyonic renormalization group (RG) flows. In fact, utilizing the radial Hamiltonian formulation of the dynamics we derive a fake supergravity description of such backgrounds, both at zero and finite temperature. In the zero temperature case, and in the context of a bottom up model, this allows us to generate new solutions simply by specifying a superpotential at will.

\subsection*{The model}

The theory we consider is a generic bottom-up Einstein-Maxwell-Axion-Dilaton (EMAD) model described by the action 
\be\label{S}
S=\frac{1}{2\k^2}\int_\cm\text{d}^{d+1}\mathbf{x}\sqrt{-g}\left(R[g]-\pa_{\mu}\f\pa^{\mu}\f-Z(\f)\pa_{\mu}\c\pa^{\mu}\c-V(\f,\c)-\S(\f)F_{\mu\nu}F^{\mu\nu}\right)+S_{GH}+S_{CS},
\ee
where $\k^2=8\p G$ is Newton's constant and 
\be
S_{GH}=\frac{1}{2\k^2}\int_{\pa\cm} \text{d}^d{\bf x} \sqrt{-\g}\;2K,
\ee
is the Gibbons-Hawking term ensuring that this action admits a Hamiltonian description. Moreover, $S_{CS}$ stands for a Chern-Simons term whose explicit form depends on the spacetime dimension. For $d=3$, which we will mostly focus on, the Chern-Simons term takes the form\footnote{We write $\e^{\mu\nu\rho\s}=e^{\mu}_{a}e^{\nu}_{b}e^{\rho}_{c}e^{\s}_{d}\wt{\e}^{abcd}$ with $\wt{\e}^{\hat{n} txy}=1$, where $\hat{n}$ is the Lorentz frame index corresponding to the direction orthogonal to the boundary of $\pa\cm$ in $\cm$.   }
\be\label{ffd}
S_{CS}=-\frac{1}{2\k^2}\int_{\cm}\text{d}^{4}\mathbf{x}\sqrt{-g}\;\P(\c)\e^{\mu\nu\rho\s}F_{\mu\nu}F_{\rho\s}.
\ee
All four functions $V(\f,\c)$, $\S(\f)$, $Z(\f)$ and $\P(\c)$ specifying the action are a priori completely general, provided that the theory admits asymptotically AdS solutions. These functions can be restricted by demanding that the action (\ref{S}) can be obtained as a consistent truncation of a gauged supergravity action, or by imposing a specific global symmetry, such as $SL(2,\bb Z)$ invariance as e.g. in \cite{Lippert:2014jma}. The equations of motion following from (\ref{S}) take the form 
\begin{subequations}\label{EOMS}
\begin{align}
& R_{\mu\nu}-\frac12 R g_{\m\n}-\left(\pa_{\mu}\f\pa_{\nu}\f-\frac{1}{2}g_{\mu\nu}\pa_{\rho}\f\pa^{\rho}\f\right)+\frac{1}{2}g_{\mu\nu}V \NO\\
&-Z\left(\pa_{\mu}\c\pa_{\nu}\c-\frac{1}{2}g_{\mu\nu}\pa_{\rho}\c\pa^{\rho}\c\right)-2\S\left(F_{\mu\rho}F_{\nu}\,^{\rho}-\frac{1}{4}g_{\mu\nu}F_{\rho\s}F^{\rho\s}\right)=0, \\
& 2\square_g\f-V_\f-Z_\f\pa_{\rho}\c\pa^{\rho}\c-\S_\f F_{\rho\s}F^{\rho\s}=0, \\
& 2\nabla_\m\left(Z\pa^\m\c\right)-V_\c-\P_\c\e^{\mu\nu\rho\s}F_{\mu\nu}F_{\rho\s}=0, \\
&\nabla_{\mu}\left(\S F^{\mu\nu}+\P\e^{\mu\nu\rho\s}F_{\rho\s}\right)=0.
\end{align}
\end{subequations}

\subsection*{Summary of results}	

In this paper we focus on homogeneous background solutions of these equations of motion with a finite electric charge density, a constant magnetic field, and  running scalars. The presence of the electric charge density and constant magnetic field imply that these background solutions break Lorentz invariance, even at zero temperature. One of our main results is a completely general description of such backgrounds in terms of a fake superpotential (\ref{superpotential}) and first order BPS-like flow equations (\ref{RGeqs}). Since Lorentz invariance is broken even at zero temperature, the fake supergravity description we develop here applies equally to both zero temperature RG flows and finite temperature black hole solutions. For solutions with zero charge density and magnetic field it reduces to the standard fake supergravity for Poincar\'e domain walls at zero temperature, but it provides a non-trivial generalization to black hole solutions. It should be emphasized that the fake supergravity description we develop here follows from Hamilton-Jacobi (HJ) theory and it is quite different from the first order formulation in e.g. \cite{Kiritsis:2012ma}. In particular, the superpotential (\ref{superpotential}) depends in general on both the scalars and the warp factor. Moreover, once a solution of the superpotential equation is given, the solutions of the equations of motion follow immediately from the first order equations (\ref{RGeqs}). This is quite distinct from the formulation of \cite{Kiritsis:2012ma}, where the superpotential effectively facilitates a change of variables in order to write the second order equations as a system of coupled first order equations. 

In the context of a bottom up model like the one we study here, where the functions specifying the action are a priori arbitrary, the HJ  origin of our fake supergravity description is particularly useful. In particular, one can very easily construct solutions of the superpotential equation (\ref{superpotential}) which immediately lead to exact solutions of the equations of motion. We present such an infinite class of exact RG flows in four dimensions specified by an arbitrary function $W_o(\f)$ in the superpotential (\ref{exactRG}) in Section \ref{bgnd}. These RG flows interpolate between AdS$_4$ in the UV and a hyperscaling violating geometry with exponents $1<z<3$ and $\th=z+1$ in the IR. By computing the Schr\"odinger potential of the fluctuations we show that the subclass of flows with $1<z<2$ has a discrete spectrum of fluctuations, while the spectrum is continuous for $2\leq z <3$. A complete analysis of these exact backgrounds will presented elsewhere \cite{rgflows}.  
	
Besides the description of general homogeneous backgrounds with finite charge density and constant magnetic field, our aim in this paper is to provide a general framework for computing the renormalized 2-point functions of the dual stress tensor, current and scalar operators in any such background and for a generic model of the form (\ref{S}). In this respect we describe the holographic renormalization of any such theory in terms of a counterterm potential $U(\f,\c)$ that is obtained by solving a superpotential equation in a Taylor series in $\f$ and $\c$, and we explicitly write the renormalized 1-point functions in the presence of sources in terms of the renormalized radial canonical momenta in (\ref{1pt-fns}) \cite{Papadimitriou:2004ap}. This identification not only provides general expressions for the VEVs in any dyonic background in (\ref{VEVs}), but also allows one to extract the 2-point functions directly from the canonical momenta, without the need to evaluate the on-shell action. Moreover, using this identification of the renormalized 1-point functions in terms of the canonical momenta enables us to immediately identify the first class constraints in the Hamiltonian formalism, which reflect local symmetries in the bulk, with the Ward identities (\ref{diff-WI}), (\ref{gauge-WI}), and (\ref{trace-WI}), reflecting global symmetries in the dual field theory. Since these Ward identities hold for the 1-point functions in the presence of arbitrary sources, differentiating with respect to the stress tensor and current sources one obtains two Ward identities for the 2-point functions, which are given in (\ref{2pt-fns-WIs}). As we show in Section \ref{Kubo}, these Ward identities together with the Kubo formulas allow one to express all thermoelectric conductivities in terms of the electric conductivities, generalizing the corresponding result of \cite{Hartnoll:2007ip} for the dyonic Reissner-Nordstr\"om black hole to more general backgrounds including running scalars.  

Another important aspect of our general framework for computing the renormalized 2-point functions is that we formulate the fluctuation equations in terms of first order Riccati equations instead of second order linear equations. The advantage of the Riccati equations is that they are differential equations for the response functions directly, with only one integration constant per equation, which can be fixed by imposing an IR condition. This drastically simplifies both the holographic renormalization of the 2-point functions \cite{Papadimitriou:2004rz} and especially the numerical solution of the fluctuation equations, since the arbitrary source is factored out of the problem from the onset \cite{Papadimitriou:2013jca}. For the finite frequency and zero spatial momentum fluctuations we consider here the fluctuation equations reduce to a single Riccati equation (\ref{Ric}) (or equivalently (\ref{smooth-riccati})), from which all conductivities can be determined. This Riccati equation generically is not integrable, but some progress can be made by considering various limiting cases. Particularly interesting is the small frequency expansion which we determine up to (and including) $\co(\o^3)$ in Section \ref{Sw}. The corresponding result we obtain for the conductivities is given in (\ref{sw-conductivities}) and agrees completely with that obtained in \cite{Hartnoll:2007ai} for the dyonic Reissner-Nordstr\"om black hole. Moreover, we find that using two different Pad\'e approximants based on the small frequency expansion of the response functions (\ref{response-small-w}) captures completely both the hydrodynamic and small magnetic field approximations discussed in \cite{Hartnoll:2007ip}, including the location of the poles closest to the origin of the complex frequency plane. 

\subsection*{Organization of the paper}

The rest of the paper is organized as follows. In Section 2 we present the radial Hamiltonian formulation of the dynamics described by the action (\ref{S}) and we set up the algorithm for holographically renormalizing any such theory. This leads to a holographic derivation of the Ward identities, which we later use to relate various 2-point functions and conductivities. In Section 3 we discuss a general class of homogeneous backgrounds with finite charge density, constant magnetic field and running scalars and we develop a fake supergravity formalism that allows us to construct an infinite family of RG flows in four dimensions.
Moreover, general expressions for the renormalized vacuum expectation values (VEVs) corresponding to these backgrounds are derived. Section 4 contains the analysis of the fluctuation equations around the backgrounds described in Section 3 for certain class of time dependent fluctuations at zero spatial momentum. In particular, after decoupling the fluctuation equations following \cite{Hartnoll:2007ai,Hartnoll:2007ip,Lippert:2014jma}, we reduce the system to a single decoupled first order Riccati equation \cite{Papadimitriou:2004rz,Papadimitriou:2013jca}. In Section 5 we show how the renormalized 2-point functions can be extracted directly from the radial canonical momenta and the solution of the Riccati equation, showing that evaluating the on-shell action is a completely redundant step. Moreover, we discuss how the various conductivities are related to the 2-point functions of the current and stress tensor through the Kubo formulas, and derive the relations implied by the Ward identities. Section 6 addresses the solution of the Riccati equation in various limiting cases. In particular, we derive the UV and IR asymptotic expansions and obtain general perturbative solutions for small and large frequency. All the machinery developed in the previous sections is applied to the dyonic Reissner-Nordstr\"om black hole in four dimensions in Section 7, where we solve numerically the Riccati equation and compare with the various approximate solutions discussed in Section 6. We end with some concluding remarks in Section 8, and three appendices where we provide a few more examples of exact RG flows in various dimensions, as well as the Gauss-Codazzi and general fluctuation equations following from the action (\ref{S}).

\section{Radial Hamiltonian formalism and the holographic dictionary} 
\label{hamiltonian-formalism}

The on-shell action is usually the starting point for any holographic computation since this is identified with the generating function of the dual field theory \cite{Witten:1998qj, Gubser:1998bc}. However, unless one wants to evaluate the free energy of a particular state, corresponding to the value of the (renormalized) on-shell action, evaluating the on-shell action is a redundant step since the radial canonical momenta are holographically identified with the one-point functions of the dual operators \cite{Papadimitriou:2004ap}, which contain the same information. In particular, the vacuum expectation values (VEVs) of the dual operators correspond to the values of the (renormalized) radial momenta on a particular solution, while higher-point functions are obtained by successively differentiating the canonical momenta with respect to the sources. All $n$-point functions can therefore be extracted directly from the bulk solutions, without the need to evaluate the on-shell action. This is especially important in Lorentzian signature where it is often much harder to evaluate the on-shell action as a function of the sources than the canonical momenta. Moreover, the radial Hamiltonian formulation of the bulk dynamics allows one to most straightforwardly carry out the procedure of holographic renormalization, either fully non-linearly \cite{deBoer:1999xf,Papadimitriou:2004ap, Martelli:2002sp} or perturbatively to the desired order that contributes to the $n$-point functions one wants to renormalize \cite{Papadimitriou:2004rz}. Finally, the holographic Ward identities follow immediately from the radial Hamiltonian dynamics since they correspond to first class constraints.      	
	
In order to formulate the bulk dynamics in a Hamiltonian language we begin by decomposing the bulk variables in components along and transversely to the radial coordinate, as in the standard ADM treatment of gravity \cite{ADM}, except that the Hamiltonian time now is the radial coordinate instead of real time. In particular the bulk metric is written in the form
\be
ds^2=(N^2+N_iN^i)dr^2+2N_idrdx^i+\g_{ij}dx^idx^j,
\ee
in terms of the lapse function $N$, the shift function $N_i$ and induced metric $\g_{ij}$, while the $U(1)$ gauge field is decomposed as 
\be
A=adr+A_idx^i.
\ee
In terms of these variables the bulk Ricci scalar becomes 
\be
R[g]=R[\g]+K^2-K_{ij}K^{ij}+\nabla_{\mu}\left(-2Kn^{\mu}+2n^{\nu}\nabla_{\nu}n^{\mu}\right),
\ee
where $K_{ij}$ is the extrinsic curvature given by 
\be
K_{ij}=\frac{1}{2N}\left(\dot{\g}_{ij}-D_iN_j-D_jN_i\right),
\ee
and $n^{\mu}=\left(\frac{1}{N},-\frac{N^i}{N}\right)$ is the unit normal vector to the constant $r$ hypersurfaces. Here and in the following a $\dot{}$ denotes a derivative with respect to the radial coordinate $r$ and $D_i$ represents the covariant derivative with respect to induced metric $\g_{ij}$. Inserting these expressions in the action (\ref{S}) and using the identities 
\be
\sqrt{-g}=N\sqrt{-\g},\quad g^{\mu\nu}=\begin{pmatrix}\frac{1}{N^2}&-\frac{N^i}{N^2}\\-\frac{N^i}{N^2}&\g^{ij}+\frac{N^iN^j}{N^2}\end{pmatrix},
\ee
we find that the action can be written in the form $S=\int\text{d}r L$ and $S_{CS}=\int\text{d}r L_{CS}$ with
\begin{align}\label{radial-lagrangian}
L=\frac{1}{2\k^2}\int\text{d}^{d}\mathbf{x}N\sqrt{-\g}&\left\{R[\g]+K^2-K_{ij}K^{ij}-\frac{1}{N^2}\left(\dot{\f}-N^i\pa_i\f\right)^2-\frac{Z(\f)}{N^2}(\dot{\c}-N^i\pa_i\c)^2\right. \NO\\
&\left.-\frac{2}{N^2}\S(\f)\g^{ij}(\dot{A}_i-\pa_ia-N^kF_{ki})(\dot{A}_j-\pa_ja-N^lF_{lj})\right. \NO\\
&\left.-\g^{ij}\pa_i\f\pa_j\f-Z(\f)\g^{ij}\pa_i\c\pa_j\c-\S(\f)F_{ij}F^{ij}-V(\f,\c)\rule{0.cm}{0.5cm}\right\},
\end{align}
and provided $d=3$, 
\be
L_{CS}=-\frac{1}{2\k^2}\int\text{d}^{3}\mathbf{x}\sqrt{-\g}4\P(\c)\e^{ijk}(\dot{A}_i-\pa_ia)F_{jk}.
\ee
This Lagrangian is the basis of the radial Hamiltonian analysis.  

The canonical momenta following from this Lagrangian are 
\begin{subequations}\label{momenta}
\begin{align}
\pi^{ij}=&\frac{\d L}{\d\dot{\g}_{ij}}=\frac{1}{2\k^2}\sqrt{-\g}\left(K\g^{ij}-K^{ij}\right), \\
\pi_{\f}=&\frac{\d L}{\d\dot{\f}}=-\frac{1}{\k^2}\frac{\sqrt{-\g}}{N}\left(\dot{\f}-N^i\pa_i\f\right), \\
\pi_{\c}=&\frac{\d L}{\d\dot{\c}}=-\frac{1}{\k^2}\frac{\sqrt{-\g}}{N}Z(\f)\left(\dot{\c}-N^i\pa_i\c\right), \\
\pi^i=&\frac{\d L}{\d\dot{A}_i}=-\frac{2}{\k^2}\frac{\sqrt{-\g}}{N}\S(\f)\left(\g^{ij}(\dot{A}_j-\pa_ja)-N_jF^{ji}\right)-\frac{2}{\k^2}\sqrt{-\g}\left(\P(\c)\e^{ijk}F_{jk}\right),
\end{align}
\end{subequations}
while the canonical momenta conjugate to the fields $N$, $N_i$, and $a$ vanish identically since the corresponding velocities do not appear in the Lagrangian. 
Note that the only contribution of the Chern-Simons term is the last term in the canonical momentum conjugate to the gauge field $A_i$. Clearly this term is present only for $d=3$. Using these expressions for the canonical momenta we can evaluate the Legendre transform of the Lagrangian to obtain the radial Hamiltonian, namely
\be
H=\int\text{d}^{d}\mathbf{x}\left(\pi^{ij}\dot{\g}_{ij}+\pi_{\f}\dot{\f}+\pi_{\c}\dot{\c}+\pi^i\dot{A}_i\right)-L=\int\text{d}^{d}\mathbf{x}\left(N\mathcal{H}+N_i\mathcal{H}^i+a\mathcal{F}\right),
\ee
where
\begin{subequations}\label{constraints}
\begin{align}
\mathcal{H}=&-\frac{\k^2}{\sqrt{-\g}}\left(2\left(\g_{ik}\g_{jl}-\frac{1}{d-1}\g_{ij}\g_{kl}\right)\pi^{ij}\pi^{kl}\right.\NO\\
&\left.+\frac{1}{4}\S^{-1}(\f)\left(\pi_i+\frac{2}{\k^2}\sqrt{-\g}\P(\c)\e_i{}^{kl}F_{kl}\right)\left(\pi^i+\frac{2}{\k^2}\sqrt{-\g}\P(\c)\e^{ipq}F_{pq}\right)+\frac{1}{2}Z^{-1}(\f)\pi_{\c}^2+\frac{1}{2}\pi_{\f}^2\right) \NO \\
&+\frac{\sqrt{-\g}}{2\k^2}\left(-R[\g]+Z(\f)\pa_i\c\pa^i\c+\S(\f)F_{ij}F^{ij}+V(\f,\c)+\pa_i\f\pa^i\f\right), \\
\mathcal{H}^i=&-2D_j\pi^{ij}+F^{ij}\left(\pi_j+\frac{2}{\k^2}\sqrt{-\g}\P(\c)\e_j{}^{kl}F_{kl}\right)+\pi_{\c}\pa^i\c+\pi_{\f}\pa^i\f, \\
\mathcal{F}=&-D_i\pi^i.
\end{align}
\end{subequations}
Since the canonical momenta for the fields $N$, $N_i$ and $a$ vanish identically, the corresponding Hamilton's equations imply the constraints 
\be\label{constraints=0}
\ch=\ch^i=\cf=0.
\ee
These are first class constraints, reflecting the diffeomorphism and gauge invariance of the bulk theory, and imply that the Hamiltonian is identically zero on-shell. These first class constraints are the starting point for the analysis of the bulk dynamics and for the construction of the holographic dictionary.   

As is well known from HJ theory, the first class constraints (\ref{constraints=0}) together with the condition that the canonical momenta be expressible as gradients of a functional $\cs[\g,A,\c,\f]$, i.e.  
\be\label{HJ-momenta}
\pi^{ij}=\frac{\d \cs}{\d\g_{ij}},\quad \pi^i=\frac{\d \cs}{\d A_i},\quad \pi_{\c}=\frac{\d \cs}{\d\c},\quad \pi_{\f}=\frac{\d \cs}{\d\f},
\ee
are completely equivalent to the full set of second order equations of motion. In particular, inserting these expressions for the canonical momenta in the constraints (\ref{constraints=0}) leads to a set of functional differential equations for the function $\cs[\g,A,\c,\f]$. These are the HJ equations for Hamilton's principal function $\cs[\g,A,\c,\f]$, which is also identified with the on-shell action evaluated with a radial cut-off. The most general solution of the second order equations of motion is described by a {\em complete integral} of the HJ equations, i.e. a solution that contains as many integration `constants' (in this context functions of the transverse coordinates) as generalized coordinates. These integration constants can be thought of as the `initial momenta' -- the renormalized momenta denoted with a $\text{ }\Hat{ }\text{ }$ in (\ref{Sren})-- and as we shall see are identified holographically with the renormalized 1-point functions.     

Given a complete integral of the HJ equations one can immediately write down first order BPS-like flow equations by identifying the expressions (\ref{momenta}) and (\ref{HJ-momenta}) for the canonical momenta. With the gauge choice $N=1$, $N_i=0$ and $a=0$, which we will adopt from now on, these flow equations become 
\begin{subequations}\label{flow}
\begin{align}
\dot{\g}_{ij}=&-\frac{4\k^2}{\sqrt{-\g}}\left(\g_{ik}\g_{jl}-\frac{1}{d-1}\g_{ij}\g_{kl}\right)\frac{\d \cs}{\d\g_{kl}}, \\
	\dot{\f}=&-\frac{\k^2}{\sqrt{-\g}}\frac{\d \cs}{\d\f}, \\
	\dot{\c}=&-\frac{\k^2}{\sqrt{-\g}}Z^{-1}(\f)\frac{\d \cs}{\d\c} ,\\
	\dot{A}_i=&-\frac{\k^2}{2\sqrt{-\g}}\S^{-1}(\f)\g_{ij}\frac{\d \cs}{\d A_j}-\S^{-1}(\f)\P(\c)\e_i\,^{jk}F_{jk}.
\end{align}
\end{subequations}
Integrating these first order equations leads to another set of integration constants, which are identified holographically with the sources of the dual operators. The resulting space of solutions is parameterized by $2n$ integration constants, where $n$ is the number of generalized coordinates, and hence spans the full space of solutions of the second oder equations of motion. This observation is practically very useful since not only it drastically simplifies the procedure of holographic renormalization, but it also often leads to new exact background solutions.

\subsection{Holographic renormalization} 
\label{hr}

Holographic renormalization \cite{Henningson:1998gx,deHaro:2000xn,Skenderis:2002wp,deBoer:1999xf,Martelli:2002sp,Papadimitriou:2004ap} can be understood as the systematic procedure for determining the boundary terms required to render the bulk variational problem {\em at infinity} well posed. Once the variational problem is well posed the finiteness of the on-shell action is automatic \cite{Papadimitriou:2005ii,Papadimitriou:2010as}. In fact, the boundary term required is always a specific solution of the radial HJ equation, not only for asymptotically locally AdS spaces \cite{deBoer:1999xf}, but even more generally \cite{Papadimitriou:2010as}. This should not be surprising given that a complete integral $\cs$ of the HJ equations coincides with the on-shell action evaluated on the most general solutions of the equations of motion. In particular, inserting an asymptotic complete integral in the flow equations (\ref{flow}) and integrating them asymptotically is a very efficient way to derive the Fefferman-Graham asymptotic expansions and their generalizations for non asymptotically AdS spaces. The key object therefore in renormalizing the theory and constructing the holographic dictionary is the general asymptotic solution of the HJ equation. 

If the bulk theory is renormalizable, all the ultraviolet (in the QFT sense) divergent terms in this general asymptotic solution of the HJ equations must be local in transverse derivatives, since they correspond to the local counterterms required to cancel the ultraviolet divergences of the on-shell action. The constraints $\ch^i=0$ and $\cf=0$ respectively impose invariance under transverse diffeomorphisms and $U(1)$ gauge transformations. It follows that as long as we look for a local, gauge and transverse diffeomorphism invariant functional $\cs[\g,A,\c,\f]$, the only equation that we have to solve is the Hamiltonian constraint $\ch=0$.  

In fact we only need to solve the Hamiltonian constraint asymptotically, but in a covariant way, i.e. every term should be a function of the induced fields, without any explicit radial dependence.\footnote{Keeping the boundary dimension $d$ arbitrary allows one to write the conformal anomaly in covariant form as well, without any explicit dependence on the radial cut-off. However, the cut-off dependence emerges via dimensional regularization by replacing the pole in the coefficients of the asymptotic solution of the HJ equations with the radial cut-off \cite{Papadimitriou:2004ap}.} This can be achieved by formally expanding the functional $\cs[\g,A,\c,\f]$ in a covariant expansion in eigenfunctions of a suitable functional differential operator $\d$. For asymptotically locally AdS spaces this can be the dilation operator \cite{Papadimitriou:2004ap}, but the form of this operator depends on the various functions of the scalars $\f$ and $\c$ that parameterize the Lagrangian, since these determine the leading asymptotic behavior of the induced fields. If we do not specify the explicit form of these functions, or even if we want to consider non asymptotically AdS backgrounds, then a covariant expansion still exists but in eigenfunctions of the operator \cite{Papadimitriou:2011qb}
\be\label{operator}
\d_\g=\int \text{d}^d{\bf x}2\g_{ij}\frac{\d}{\d\g_{ij}}.
\ee   
In the absence of a vector field this operator counts transverse derivatives, but this is not strictly the case when there is a Maxwell field in the theory. What (\ref{operator}) always counts is the number of inverse induced metrics, $\g^{ij}$, which coincides with the number of derivatives in the absence of a Maxwell field. When the solution of the HJ equations can be expanded both in eigenfunctions of the dilatation operator and of $\d_\g$, then the two expansions are simple rearrangements of  each other. For example, different powers of the scalars would appear at different orders in the expansion according to the dilatation operator while they would all be part of the zero order solution of the expansion in eigenfunctions of $\d_\g$.    

Before we proceed with the recursive solution of the HJ equation $\ch=0$ for the function $\cs$, a few comments are in order regarding the role of the Chern-Simons term. Note that using (\ref{HJ-momenta}) we can write  
\be
\pi^i+\frac{2}{\k^2}\sqrt{-\g}\P(\c)\e^{ikl}F_{kl}=\frac{1}{\k^2}\sqrt{-\g}\P'(\c)\e^{ijk}A_j\pa_k\c+\frac{\d }{\d A_i}\left(\cs+\frac{1}{\k^2}\int\text{d}^{3}\mathbf{x}\sqrt{-\g}\P(\c)\e^{ijk}A_iF_{jk}\right).
\ee
It follows that if and only if $\P(\c)$ is a constant then the effect of the Chern-Simons term is simply a shift of the on-shell action $\cs$ according to 
\be
\cs\to \wt \cs=\cs+\frac{1}{\k^2}\int\text{d}^{3}\mathbf{x}\sqrt{-\g}\P(\c)\e^{ijk}A_iF_{jk},
\ee
where $\wt\cs$ satisfies the HJ equation derived from a Lagrangian without a Chern-Simons term. Crucially, this boundary term is gauge-invariant if and only if  $\P(\c)$ is a constant. In that and only that case therefore we can simply determine $\cs$ by solving the Hamiltonian constraint in the absence of a Chern-Simons term and then replace $\cs$ with $\wt\cs$. For non-constant  $\P(\c)$, however, the Hamiltonian constraint in the presence of the Chern-Simons term cannot be reduced to that without such a term. Hence, to determine $\cs$ one must solve directly the Hamiltonian constraint in (\ref{constraints}). Nevertheless, we only need to determine the divergent part of the on-shell action $\cs$ and it is straightforward to show that provided the equations of motion admit asymptotically locally AdS solutions the Chern-Simons term is asymptotically finite.\footnote{This would not necessarily be the case for a massive gauge field though. See e.g. \cite{Jimenez-Alba:2014iia}.} For the purpose of determining the boundary counterterms therefore, even for non-constant $\P(\c)$, we can always solve the Hamiltonian constraint corresponding to the theory without a Chern-Simons term. As we shall see though, the CS term leads to a gravitational anomaly in the holographic Ward identities.    

In order to construct the asymptotic solution of the HJ equation for general potentials in the Lagrangian we therefore proceed by formally expanding $\cs$ as 
\be\label{formal-exp}
\cs=\cs\sub{0}+\cs\sub{2}+\cdots,
\ee 
where each term $\cs\sub{2k}$ in this expansion is an eigenfunction of $\d_\g$ with eigenvalue $d-2k$, i.e. 
\be
\d_\g \cs\sub{2k}=(d-2k)\cs\sub{2k}.
\ee 
The zero order term $\cs\sub{0}$ must contain no transverse derivatives and be gauge invariant and so it must take the form
\be\label{U}
\cs\sub{0}=\frac{1}{\k^2}\int \text{d}^d{\bf x} \sqrt{-\g}U(\f,\c),
\ee
for some function of the scalars $U(\f,\c)$. Inserting this in the Hamiltonian constraint $\ch=0$ leads to the `fake superpotential' equation
\be\label{sup-eq}
U_\f^2+Z^{-1}(\f)U_\c^2-\frac{d}{d-1}U^2=V(\f,\c),
\ee
where the subscripts denote partial derivatives with respect to the corresponding field. In asymptotically locally AdS spacetimes this PDE can be solved by seeking a solution $U(\f,\c)$ in the form of a Taylor expansion, with the quadratic term determining the scaling dimension of the dual operator. In particular, for asymptotically locally AdS spacetimes the potential takes the form
\be
V(\f,\c)=-\frac{d(d-1)}{L^2}+ m_\f^2\f^2+m_\c^2\c^2+\cdots,
\ee  
where the scalar masses must satisfy the Breitenlohner-Freedman bound \cite{Breitenlohner:1982bm}
\be
m_\f^2L^2 ,\,m_\c^2L^2\geq -\left(\frac{d}{2}\right)^2,
\ee
and are related to the dimensions $\D_\f$ and $\D_\c$ of the dual operators as 
\be
m^2_\f L^2=-\D_\f(d-\D_\f),\quad m^2_\c L^2=-\D_\c(d-\D_\c).
\ee
The possible solutions of (\ref{sup-eq}) are of the form
\be
U(\f,\c)=\frac{d-1}{L}+\frac{1}{2L} \m_\f\f^2+\frac{1}{2L}\m_\c\c^2+\cdots,
\ee
where $\m_\f$ and $\m_\c$ can each take two values, respectively $\D_\f$ or $d-\D_\f$ and $\D_\c$ or $d-\D_\c$. However, only a solution of the form 
\be\label{U-exp}
U(\f,\c)=\frac{d-1}{L}+\frac{1}{2L} (d-\D_\f)\f^2+\frac{1}{2L}(d-\D_\c)\c^2+\cdots,
\ee
can be used as counterterms \cite{Papadimitriou:2004rz} and therefore one must seek a solution of this form.  

Once the relevant solution $U(\f,\c)$ is determined by solving (\ref{sup-eq}), the higher order terms $\cs\sub{2k}$ are determined through the {\em linear} equations obtained by substituting the formal expansion (\ref{formal-exp}) in the Hamiltonian constraint, namely\footnote{As discussed earlier, we use the Hamiltonian constraint without a Chern-Simons term since this term only contributes at finite order in the on-shell action.}
\be
\frac{2\k^2}{\sqrt{-\g}}\left(2\p\sub{0}^i_j\p\sub{2k}^j_i-\frac{2}{d-1}\p\sub{0}\p\sub{2k}+\frac12\p_\f\sub{0}\p_\f\sub{2k}+\frac12Z^{-1}\p_\c\sub{0}\p_\c\sub{2k}\right)=\car\sub{2k},
\ee
where  
\be
\pi\sub{2k}^{ij}=\frac{\d \cs\sub{2k}}{\d\g_{ij}},\quad \pi\sub{2k}^i=\frac{\d \cs\sub{2k}}{\d A_i},\quad \pi\sub{2k}_{\c}=\frac{\d \cs\sub{2k}}{\d\c},\quad \pi\sub{2k}_{\f}=\frac{\d \cs\sub{2k}}{\d\f},
\ee
and the inhomogeneous source of this linear equation is given by
\bal
\car\sub{2}=&\frac{\sqrt{-\g}}{2\k^2}\left(-R[\g]+Z(\f)\pa_i\c\pa^i\c+\pa_i\f\pa^i\f\right),\NO\\
\car\sub{4}=&\frac{\sqrt{-\g}}{2\k^2}\S F_{ij}F^{ij}-\frac{\k^2}{\sqrt{-\g}}\left(2\p\sub{2}^i_j\p\sub{2}^j_i-\frac{2}{d-1}\p\sub{2}^2+\frac14\S^{-1}\p\sub{2}^i\p\sub{2}_i+\frac12\p_\f\sub{2}^2+\frac12Z^{-1}\p_\c\sub{2}^2\right),\NO\\
\car\sub{2k}=&-\frac{\k^2}{\sqrt{-\g}}\sum_{\ell=1}^{k-1}\left(2\p\sub{2\ell}^i_j\p\sub{2k-2\ell}^j_i-\frac{2}{d-1}\p\sub{2\ell}\p\sub{2k-2\ell}+\frac14\S^{-1}\p\sub{2\ell}^i\p\sub{2k-2\ell}_i\right.\NO\\
&\left.\hskip0.7in+\frac12\p_\f\sub{2\ell}\p_\f\sub{2k-2\ell}+\frac12Z^{-1}\p_\c\sub{2\ell}\p_\c\sub{2k-2\ell}\right),\quad k>2.
\eal
Using the explicit form (\ref{U}) of the leading order solution $\cs\sub{0}$ these recursive equations become
\be
-\frac{2}{d-1}U\p\sub{2k}+U_\f\p_\f\sub{2k}+ Z^{-1}U_\c\p_\c\sub{2k}=\car\sub{2k}.
\ee
This can be simplified further by writing 
\be
\cs\sub{2k}=\int \text{d}^d{\bf x}\cl\sub{2k},
\ee
and using the identity
\be
\d\g_{ij}\p\sub{2k}^{ij}+\d A_i\p\sub{2k}^i+\d\f\p_\f\sub{2k}+\d\c\p_\c\sub{2k}=\d\cl\sub{2k}+\pa_iv\sub{2k}^i,
\ee
for some vector field $v\sub{2k}^i$. Applying this identity to the variations under $\d_\g$ and using the freedom to define $\cl\sub{2k}$ up to a total derivative we arrive at the relation 
\be
2\p\sub{2k}=(d-2k)\cl\sub{2k},
\ee
which allows us to simplify the recursion relations to 
\be\label{recursion}
\left(U_\f\frac{\d}{\d\f}+ Z^{-1}U_\c\frac{\d}{\d\c}\right)\int \text{d}^d{\bf x}\cl\sub{2k}-\left(\frac{d-2k}{d-1}\right)U\cl\sub{2k}=\car\sub{2k}.
\ee
These linear functional PDEs can be solved systematically following the procedure developed in \cite{Papadimitriou:2011qb}. Importantly, only the inhomogeneous solution, which is unique, contributes to the divergences. The homogeneous solutions are ultraviolet finite. The counterterms are therefore defined as 
\be\label{counterterms}
S_{ct}:=-\sum_{k=0}^{[d/2]}\cs\sub{2k}.
\ee   
Carrying out this calculation keeping the dimension $d$ as a parameter one finds that certain terms contain a pole at particular dimensions, e.g. $1/(d-4)$. Such terms lead via dimensional regularization to cut-off dependence according to the replacement rule
\be
\frac{1}{d-4}\to r_o,
\ee 
where $r_o$ is the radial cut-off in the canonical radial coordinate $r$ \cite{Papadimitriou:2004ap,Papadimitriou:2011qb}. The sum of all such terms is then identified with the holographic conformal anomaly \cite{Henningson:1998gx}.

Given the local counterterms $\cs_{ct}$ the renormalized action (evaluated at a radial cut-off) is given by
\be\label{Sren}
S_{ren}:=\cs+\cs_{ct}=\int \text{d}^d{\bf x}\left(\g_{ij}\Hat\p^{ij}+A_i\Hat\p^i+\f\Hat\p_\f+\c\Hat\p_\c\right),
\ee
where the quantities $\Hat\p^{ij}$, $\Hat\p^{i}$, $\Hat\p_\f$ and $\Hat\p_\c$ are arbitrary functions that correspond to integration `constants' of the HJ equation. They correspond to the renormalized canonical momenta and can be identified with the renormalized 1-point functions of the dual operators through the relations\footnote{To avoid cluttering the notation we do not differentiate between the renormalized 1-point functions evaluated at the cut-off, as e.g. in (\ref{1pt-fns}), and the limit obtained by multiplying these 1-point functions with the appropriate factor of the cut-off and sending the cut-off to infinity, as in e.g. (\ref{VEVs}).}
\bal\label{1pt-fns}\boxed{
\bald
&\langle T^{ij} \rangle =-\frac{2}{\sqrt{-\g}}\frac{\d\cs_{ren}}{\d\g_{ij}}=-\frac{2}{\sqrt{-\g}}\Hat\p^{ij},
&&\langle J^i \rangle =\frac{1}{\sqrt{-\g}}\frac{\d\cs_{ren}}{\d A_{i}}=\frac{1}{\sqrt{-\g}}\Hat\p^{i},\\
&\langle \co_\f \rangle =\frac{1}{\sqrt{-\g}}\frac{\d\cs_{ren}}{\d\f}=\frac{1}{\sqrt{-\g}}\Hat\p_\f, &&
\langle \co_\c \rangle =\frac{1}{\sqrt{-\g}}\frac{\d\cs_{ren}}{\d\c}=\frac{1}{\sqrt{-\g}}\Hat\p_\c.
\eald}
\eal
It should be emphasized that these are the 1-point functions in the presence of arbitrary sources and so any higher-point function can simply be obtained from these by further functional differentiation. These expressions are central to our subsequent analysis, and more generally to the holographic dictionary, since they allow one to extract the $n$-point functions of the dual operators directly from the radial canonical momenta, i.e. from the bulk solution of the equations of motion, without having to evaluate the on-shell action in terms of the sources.

\subsection{Holographic Ward identities} 
\label{WIs}

The identification of the renormalized 1-point functions with the renormalized canonical momenta in (\ref{1pt-fns}) allows us to translate the constraints $\ch^i=0$ and $\cf=0$ into Ward identities for the 1-point functions in the presence of arbitrary sources. Since these constraints are linear in the canonical momenta, they hold at each order of the expansion (\ref{formal-exp}) in eigenfunctions of the operator $\d_\g$. In particular, they hold for the renormalized momenta leading respectively to the diffeomorphism, 
\be	\label{diff-WI}
D_j\langle T^{j}_i \rangle+F_{ij}\langle J^j \rangle+\langle \co_\c \rangle\pa_i\c+\langle \co_\f \rangle\pa_i\f=-\frac{2}{\k^2}\P(\c)\e^{jkl}F_{ij}F_{kl},	
\ee
and $U(1)$, 
\be\label{gauge-WI}
D_i\langle J^i \rangle=0,
\ee
Ward identities, which reflect symmetries that are always present in the bulk and so they hold irrespectively of the asymptotic form of the background. Notice that the Chern-Simons term introduces a gravitational anomaly in the dual theory, given by the RHS of (\ref{diff-WI}). For asymptotically locally AdS backgrounds we also have the trace Ward identity (for a derivation see e.g. \cite{Papadimitriou:2004ap})
\be\label{trace-WI}
\langle T^i_i \rangle+(d-\D_\c)\langle \co_\c \rangle\pa^i\c+(d-\D_\f)\langle \co_\f \rangle\pa^i\f=\ca,
\ee
where $\ca$ is the conformal anomaly, given by the coefficient of the cut-off dependent terms in $\cs_{ct}$. 

Since these Ward identities hold in the presence of arbitrary sources, we can differentiate them with respect to the sources to obtain constraints on higher-point functions. In particular, differentiating (\ref{diff-WI}) with respect to the sources of the stress tensor and the current we get respectively 
\begin{subequations}\label{2pt-fns-WIs}
\begin{align}
&D_j\langle T^{j}_i(\mathbf{x})T^{kl}(\mathbf{x}') \rangle-2D_j\left(\d_i^{(k}\langle T^{l)j}(\mathbf{x})\rangle\d^{(d)}(\mathbf{x},\mathbf{x}')\right)-\g^{kl} \langle T^j_i(\mathbf{x}) \rangle D_j\d^{(d)}(\mathbf{x},\mathbf{x}')+\langle T^{kl}(\mathbf{x}) \rangle D_i\d^{(d)}(\mathbf{x},\mathbf{x}')\NO\\
&+F_{ij}(\mathbf{x})\langle T^{kl}(\mathbf{x}') J^j(\mathbf{x})\rangle+\langle T^{kl}(\mathbf{x}')\co_\c(\mathbf{x}) \rangle\pa_i\c+\langle T^{kl}(\mathbf{x}') \co_\f(\mathbf{x})\rangle\pa_i\f=\NO\\
&\g^{kl}\d^{(d)}(\mathbf{x},\mathbf{x}')D_j\langle T^{j}_i \rangle-\frac{4}{\k^2}\P(\c)\e^{jpq}F_{ij}F_{pq}\g^{kl}\d^{(d)}(\mathbf{x},\mathbf{x}')=0,\\\NO\\
& D_j\langle T^j_i(\mathbf{x})J^k(\mathbf{x}')\rangle+\langle J^j(\mathbf{x}) \rangle\left(\d^k_j\pa_i-\d^k_i\pa_j\right)\d^{(d)}(\mathbf{x},\mathbf{x}')+F_{ij}(\mathbf{x})\langle J^j(\mathbf{x}) J^k(\mathbf{x'})\rangle+\langle \co_\c(\mathbf{x})J^k(\mathbf{x}') \rangle\pa_i\c\NO\\
&+\langle \co_\f(\mathbf{x})J^k(\mathbf{x}') \rangle\pa_i\f=-\frac{2}{\k^2}\P(\c)\e^{jpq}\left(F_{pq}(\mathbf{x})\left(\d^k_j\pa_i-\d^k_i\pa_j\right)+2F_{ij}(\mathbf{x})\d^k_q\pa_p\right)\d^{(d)}(\mathbf{x},\mathbf{x}'),
\end{align}
\end{subequations}
where all derivatives are with respect to $\mathbf{x}$, the covariant delta function is defined through
\be
\d^{(d)}(\mathbf{x},\mathbf{x}')\equiv \frac{1}{\sqrt{-\g}}\d^{(d)}(\mathbf{x}-\mathbf{x}'),
\ee
and we have used that 
\be
\langle J^j(\mathbf{x})T^{kl}(\mathbf{x}') \rangle= \langle T^{kl}(\mathbf{x}') J^j(\mathbf{x}) \rangle+\g^{kl}\d^{(d)}(\mathbf{x},\mathbf{x}')\langle J^j(\mathbf{x}) \rangle,
\ee
and similarly for the 2-point functions of the stress tensor and the scalar operators. We will revisit these two Ward identities for the 2-point functions in Section \ref{2pnt-fns}.

\section{Dyonic backgrounds}
\label{bgnd}

Having established the holographic dictionary for the model (\ref{S}), we are now interested in computing the holographic 2-point functions in backgrounds of the form 
\begin{subequations}\label{Bans}
\begin{align}
&ds^2_B=dr^2+e^{2A(r)}\left(-f(r)dt^2+dx^2+dy^2\right), \\
&A_B=\a(r)dt+\frac{H}{2}(xdy-ydx), \\
&\f_B=\f_B(r),
\qquad
\c_B=\c_B(r),
\end{align}
\end{subequations}
where $H$ is a constant background magnetic field, $f(r)$ is the blackening factor and $A(r)$ is the warp factor. The gauge field strength on such backgrounds is given by
\be\label{FB}
F_B=dA_B=\dot{\a}dr\wedge dt+H dx\wedge dy.
\ee
Inserting the ansatz (\ref{Bans}) in the Gauss-Codazzi equations (\ref{GC}) we obtain the following set of equations for backgrounds of this form:
\begin{subequations}\label{BEOM}
\begin{align}
&\ddot{A}+\dot{A}\left(d\dot{A}+\frac12f^{-1}\dot{f}\right)+\frac{V}{(d-1)}+\frac{2\S}{(d-1)}e^{-2A}f^{-1}\left(fe^{-2A}H^2+\dot{\a}^2\right)=0, \\
&(d-1)\left(\ddot{A}-\frac12f^{-1}\dot f\dot A\right)+\dot{\f}^2+Z\dot{\c}^2=0, \\
&\ddot{f}+\dot{f}\left(d\dot{A}-\frac12f^{-1}\dot f\right)-4\S e^{-2A}\left(fe^{-2A}H^2+\dot{\a}^2\right)=0, \\
& 2\ddot{\f}_B+2\left(d\dot{A}+\frac12f^{-1}\dot f\right)\dot{\f}_B-V_{\f}-Z_{\f}\dot{\c}_B^2-2\S_{\f}\frac{e^{-2A}}{f}\left(fe^{-2A}H^2-\dot{\a}^2\right)=0, \\
& 2Z\ddot{\c}_B+2Z\left(d\dot{A}+\frac12f^{-1}\dot f\right)\dot{\c}_B-V_{\c}+2Z_{\f}\dot{\f}_B\dot{\c}_B-8\P_{\c}\frac{e^{-dA}}{\sqrt{f}}\dot{\a}H=0, \\
&\pa_r\left(e^{(d-2)A}\frac{\S}{\sqrt{f}}\dot{\a}-2\P H\right)=0.
\end{align}
\end{subequations}
Note that due to the presence of the Chern-Simons term (\ref{ffd}) these equations make sense either in any $d$ provided $\P=0$,  or for any $\P$ in $d=3$. In what follows we keep $\P$ arbitrary and work in $d=3$, but our results can be adapted to general $d$ provided $\P$ is set to zero.

In the following it will often be more convenient to work with an alternative radial coordinate defined through 
\be\label{zr}
\pa_r=-\sqrt{f}e^{-A}\pa_{u},
\ee
so that the background metric in (\ref{Bans}) takes the form
\begin{align}\label{eqn:ConfCoord}
ds^2_B=e^{2A(u)}\left(\frac{du^2}{f(u)}-f(u)dt^2+dx^2+dy^2\right).
\end{align}
Denoting with a prime differentiation with respect to the radial coordinate $u$, the background equations for $d=3$ become
\begin{subequations}
\label{eom2} 
\begin{align}
& A''+2A'\left(A'+\frac12f^{-1}f'\right)+\frac{1}{2f}e^{2A}V+\S f^{-1}e^{-2A}\left(H^2+\a'^2\right)=0, \\
& 2\left(A''-A'^2\right)+\f'^2_B+Z\c'^2_B=0, \\
& f''+2A'f'-4\S e^{-2A}\left(H^2+\a'^2\right)=0, \\
& \f''_B+2\left(A'+\frac12f^{-1}f'\right)\f'_B-\frac12f^{-1}V_{\f}e^{2A}-\frac12Z_{\f}\c'^2_B-\S_{\f}e^{-2A}f^{-1}\left(H^2+\a'^2\right)=0, \\
& Z\c''_B+2Z\left(A'+\frac12f^{-1}f'\right)\c'_B-\frac12f^{-1}V_{\c}e^{2A}+Z_{\f}\f'_B\c'_B+4\P_{\c}e^{-2A}f^{-1}\a'H=0, \\
& \left(\S\a'+2\P H\right)'=0.
\end{align}
\end{subequations}
Notice that the last equation can be integrated in general leading to 
\be\label{ap}
\a'=\frac{1}{\S}(\wt Q-2\P H),
\ee
where the integration constant $\wt Q$, as we shall see momentarily, determines the time component of the vacuum expectation value of the conserved $U(1)$ current and so corresponds to the background electric charge density.

\subsection{First order flow equations and the fake superpotential}

The radial Hamiltonian formulation of the dynamics we developed in Section \ref{hamiltonian-formalism} allows us to describe any solution of the second order equations (\ref{BEOM}) in terms of first order equations and a `fake superpotential'. In order to derive these first order equations we observe that for backgrounds of the form (\ref{Bans}) the canonical momenta (\ref{momenta}) (in the gauge $N=1$, $N_i=a=0$) become 
\begin{subequations}\label{momenta-background}
\begin{align}
\pi_{ij}=&\frac{1}{2\k^2}e^{(d+2)A}f^{1/2}\left((d-1)\dot A\left(\d_{ij}-(1+f)\d_{i0}\d_{j0}\right)+\frac{\dot f}{2f}\left(\d_{ij}-\d_{i0}\d_{j0}\right)\right), \\
\pi_{\f}=&-\frac{1}{\k^2}e^{dA}f^{1/2}\dot{\f}, \\
\pi_{\c}=&-\frac{1}{\k^2}e^{dA}f^{1/2}Z(\f)\dot{\c}, \\
\pi^i=&\frac{2}{\k^2}\left(\S e^{(d-2)A}f^{-1/2}\dot{\a}-2H\P\right)\d^{i0}=-\frac{2}{\k^2}\wt Q\d^{i0}.
\end{align}
\end{subequations}	
Moreover, a variation of the on-shell action with respect to a metric of the form (\ref{Bans}) takes the form
\be
\d\cs=\p^{tt}\d\g_{tt}+\p^{ab}\d\g_{ab}=2\d A\p+f^{-1}\d f\g_{tt}\p^{tt},
\ee
and hence
\begin{subequations}
\begin{align}
&\frac{\d\cs}{\pa A}=2\p=\frac{1}{2\k^2}e^{dA}f^{1/2}2(d-1)\left(d\dot A+\frac{\dot f}{2f}\right),\\
&\frac{\d\cs}{\pa f}=f^{-1}\g_{tt}\p^{tt}=\frac{1}{2\k^2}e^{dA}f^{1/2}(d-1)f^{-1}\dot A.
\end{align}
\end{subequations}	
Eliminating $\ddot A$ from the first two equations in (\ref{BEOM}) and replacing the velocities with partial derivatives of $\cs(A,f,\f,\c,\a)$ using these expressions leads to the HJ equation for backgrounds of the form (\ref{Bans}), namely
\be
\left(\frac{\k^2 e^{dA}}{f^{1/2}}\right)^2\left(\frac{4}{d-1}f\cs_f\left(\cs_A-df\cs_f\right)-\cs_\f^2-Z^{-1}\cs_\c^2\right)+V_{eff}(A,\f,\c)=0,
\ee
where 
\be\label{Veff}
V_{eff}(A,\f,
\c)\equiv V(\f,\c)+2\S(\f) e^{-4A}H^2+2\S^{-1}(\f) e^{-2(d-1)A}\left(\wt Q-2H\P(\c)\right)^2.
\ee
Notice that the background magnetic field and electric charge makes the effective potential dependent on the warp factor $A$ in addition to the scalars $\f$ and $\c$. The dependence of $\cs$ on $f$ and $\a$ can be eliminated by the separable ansatz
\be\label{sup-ansatz}
\cs=-\frac{1}{\k^2}\int\text{d}^d\mathbf{x}\left(e^{dA}f^{1/2}W(A,\f,\c)+2\wt Q\a\right),
\ee
where the term proportional to $\wt Q$ accounts for the canonical momentum conjugate to the time component $\a$ of the vector field and the fake superpotential $W$ satisfies the equation
\be\label{superpotential}\boxed{
W_\f^2+Z^{-1}(\f)W_\c^2-\frac{1}{d-1}\left(d+\pa_A\right)W^2=V_{eff}(A,\f,\c).}
\ee
Note that this ansatz is capable of providing an {\em almost} complete integral of the HJ equation, with ``almost'' referring to the fact that this ansatz does not contain an integration constant for the generalized coordinate $f$. 
Inverting the momenta (\ref{momenta-background}) and using the ansatz (\ref{sup-ansatz}) for the on-shell action $\cs$ leads to the advertised first order equations
\begin{subequations}\label{RGeqs}
\begin{align}
&\dot A=-\frac{1}{d-1}W,\\
&\frac{\dot f}{f}=-\frac{2}{(d-1)}W_A,\\
&\dot \f=W_\f,\\
&\dot \c=Z^{-1}W_\c,\\
&\dot \a=-\S^{-1}e^{-(d-2)A}f^{1/2}\left(\wt Q-2H\P(\c)\right).
\end{align}
\end{subequations}	
As is guaranteed by the HJ construction (and can be checked explicitly using the background equations (\ref{BEOM})), given a fake superpotential that satisfies (\ref{superpotential}) any solution of the first order equations (\ref{RGeqs}) leads to a solution of the second order equations of motion (\ref{BEOM}).

\subsection{Exact families of dyonic backgrounds}
\label{exact}

In fact, the superpotential equation (\ref{superpotential}) together with the first order equations (\ref{RGeqs}) amounts to a (fake supergravity \cite{Freedman:2003ax}) solution generating technique for dyonic backgrounds with running scalars, which becomes particularly powerful in the context of a bottom up model where the potentials defining the action (\ref{S}) are a priori unspecified. To support this claim we provide two explicit examples here: a superpotential that gives the general dyonic Reissner-Nordstr\"om black hole in arbitrary dimension, and an infinite class of RG flows between AdS$_4$ in the UV and a hyperscaling violating Lifshitz geometry in the IR. Additional solutions of the superpotential equation (\ref{superpotential}), including in other dimensions, are presented in Appendix \ref{W-solutions}.   

\subsection*{Generalized dyonic Reissner-Nordstr\"om black hole}

The dyonic Reissner-Nordstr\"om black hole in arbitrary dimension can be obtained immediately from the flow equations (\ref{RGeqs}), which imply that a universal consistent solution, independently of the form of the potentials as long as they admit asymptotically AdS solutions, is given by $\f=\c=0$. In that case the superpotential equation (\ref{superpotential}) can be integrated to obtain  
\be\label{RN-sup}
W_{RN}(A)=-\frac{d-1}{L}\sqrt{1+\frac{2L^2\S^{-1}(\wt Q-2\P H)^2e^{-2(d-1)A}}{(d-1)(d-2)}-\frac{2L^2\S H^2e^{-4A}}{(d-1)(d-4)}-C e^{-dA}},
\ee
where $C$ is an integration constant and $\P$ is understood to be zero unless $d=3$. Introducing the new radial coordinate $u$ through\footnote{This is a priori a different radial coordinate than the one introduced in (\ref{zr}), but for the Reissner-Nordstr\"om black hole the two coincide.} $e^{-A}=u/L$ one can integrate the first order equations (\ref{RGeqs}) to obtain the general form of the backgrounds corresponding to the superpotential (\ref{RN-sup}), namely
\begin{subequations}\label{RN_d}
\begin{align}
& ds^2_{d+1}=\frac{L^2}{u^2}\left(f^{-1}(u)du^2-f(u)dt^2+d\mathbf{x}^2\right),\\
& f(u)=1+\frac{2L^{-2(d-2)}\S^{-1}(0)(\wt Q-2\P(0) H)^2}{(d-1)(d-2)}u^{2(d-1)}-\frac{2L^{-2}\S(0) H^2}{(d-1)(d-4)}u^4-C L^{-d}u^d,\\
&\a=\a_0+\frac{L^{-(d-3)}}{d-2}\S^{-1}(0)(\wt Q-2\P(0) H)u^{d-2}.
\end{align}
\end{subequations}
From these expressions we identify the integration constant $C$ in (\ref{RN-sup}) with the black hole mass, namely $M=L^{-d}C$, 
while the integration constant $\a_0$ in the gauge field is identified with the chemical potential $\m$.  

\subsection*{Exact dyonic RG flows}

For $d=3$ (AdS$_4$), the superpotential equation (\ref{superpotential}) admits a solution of the form
\be\label{exactRG}\boxed{
W(A,\f)=W_o(\f)\sqrt{1+q^2e^{-4A}},}
\ee
where the parameter $q$ is defined through
\be
H^2\S_0+(\wt Q-2\P_0 H)^2\S^{-1}_0=q^2L^{-2},
\ee
and the scalar potentials $V(\f)$ and $\S(\f)$ are determined in terms of a single arbitrary function $W_o(\f)$ as
\bal
&V(\f)=W_o'^2-\frac32W_o^2,\NO\\
&H^2\S(\f)+(\wt Q-2\P_0 H)^2\S^{-1}(\f)=\frac{q^2}{2}\left(W_o'^2+\frac12W_o^2\right).
\eal
In order to solve explicitly the second equation for $\S(\f)$ we need to distinguish two cases, depending on whether the ratio $|\wt Q-2\P_0H|/|H|$
is smaller or greater than 1. To do this it is convenient to introduce two parameters, $\t_e$ and $\t_m$ through the identifications
\be
\frac{\wt Q-2\P_0H}{\S_0 H}=:\left\{\begin{matrix}
\coth(\t_e/2), & |\wt Q-2\P_0H|>|H|,\\
\tanh(\t_m/2), & |\wt Q-2\P_0H|<|H|,
\end{matrix}\right.
\ee
so that the corresponding expressions for $\S(\f)$ are 
\bal
&|\wt Q-2\P_0H|>|H|:\NO\\
&\S^{-1}(\f)=\frac12L^2\S_0^{-1}(1+\tanh^2(\t_e/2))\left(\frac12\left(W_o'^2+\frac12W_o^2\right)+\sqrt{\frac14\left(W_o'^2+\frac12W_o^2\right)^2- L^{-4}\tanh^2\t_e}\right),\NO\\\NO\\
&|\wt Q-2\P_0H|<|H|:\NO\\
&\S(\f)=\frac12L^2\S_0(1+\tanh^2(\t_m/2))\left(\frac12\left(W_o'^2+\frac12W_o^2\right)+\sqrt{\frac14\left(W_o'^2+\frac12W_o^2\right)^2- L^{-4}\tanh^2\t_m}\right).
\eal
Note that since $W_o\sim -2/L+\co(\f^2)$ in the UV, the quantities under the square root are always non-negative. Moreover, since $\S(\f)$ depends explicitly on the parameters $\t_e$ or $\t_m$, these parameters (i.e. the ratio of the electric charge density to the magnetic field) must be considered as a specification of the theory -- not of the solutions. 

For either of the two cases, inserting the superpotential (\ref{exactRG}) in the first order equations (\ref{RGeqs}) allows us to obtain the explicit form of the metric in the form
\be
ds^2=\frac{d\f^2}{W_o'^2\left(1+q^2e^{-4A}\right)}+e^{2A}\left(-\left(1+q^2e^{-4A}\right)dt^2+dx^2+dy^2\right),
\ee
where the warp factor is expressed in terms of the scalar via  
\be
A=-\frac12\int^\f d\bar\f\frac{W_o(\bar\f)}{W_o'(\bar\f)}.
\ee
This geometry is asymptotically AdS$_4$ in the UV. Assuming that in the IR\footnote{The case $\f\to-\infty$ is completely analogous, with $\l<0$.} $\f\to\infty$ and the function $W_o(\f)$ behaves as 
\be
W_o(\f)\sim w_o e^{\l\f},\quad \l>0,
\ee
the flow equations imply
\be
A\sim-\frac{1}{2\l}\f.
\ee
Introducing the radial coordinate 
\be
v=e^{-(\l+1/2\l)\f},
\ee
in the IR the metric asymptotes to the hyperscaling violating Lifshitz geometry \cite{Charmousis:2010zz,Huijse:2011ef,Iizuka:2011hg,Ogawa:2011bz,Shaghoulian:2011aa,Dong:2012se,Iizuka:2012iv,Gouteraux:2012yr}
\be
ds^2_{IR}= v^{\th-2}\left(\frac{dv^2}{q^2\l^2 w_o^2 (\l+1/2\l)^2}-q^2v^{-2(z-1)}dt^2+dx^2+dy^2\right),
\ee
with Lifshitz and hyperscaling violating exponents  
\be\boxed{
z=\frac{\l^2+3/2}{\l^2+1/2},\quad \th=z+1,\quad 1<z<3.}
\ee
These are type IIIb hyperscaling violating backgrounds for $1<z\leq 2$ and type IVb for $2<z<3$ according to the classification of solutions of the null energy conditions in \cite{Chemissany:2014xsa}. It is interesting to note that for the purely electric solution ($H=0$), these IR geometries correspond to solutions of an Einstein-Maxwell-Dilaton theory with a vanishing scalar potential \cite{Charmousis:2010zz,Gouteraux:2012yr,Gath:2012pg,Chemissany:2014xsa}. In the present case, however, the scalar potential is not zero! The reason for this apparent puzzle is that the scalar potential is asymptotically subleading in the IR in the purely electric case, and there are necessarily subleading terms, contrary to the case of pure exponential potentials. Another point we must emphasize is that, contrary to what is often claimed in the literature, these geometries are not singular, since the concept of a curvature singularity in the presence of a diverging scalar is inherently ambiguous. Indeed, as it was shown in \cite{Chemissany:2014xsa}, these geometries are perfectly well behaved in the `dual frame' where they become Lifshitz and the singularity is completely absorbed in the diverging scalar. It follows that there is no need for applying Gubser's criterion for this kind or IR geometries, in the same way that this criterion is not applicable to RG flows between two AdS fixed points.     

In the far IR, the radial coordinate $v$ is related to the conformal coordinate $u$ introduced in (\ref{zr}) as 
\be
v\sim \left\{\begin{matrix}
u^{\frac{2\l^2+1}{2\l^2-1}}, & u\to+\infty & \l^2<\frac12, \\
e^{-c_0^2 u}, & u\to+\infty, &\l^2=\frac12,\\
(u_*-u)^{\frac{2\l^2+1}{2\l^2-1}}, & u\to u_*^-, & \l^2>\frac12,
\end{matrix}\right.
\ee
where $c_0$ is a non vanishing constant that depends on the charges. Computing the Schr\"odinger potential for the fluctuation equations (\ref{Eqn:Fluctuation})
one finds that for both $H\neq 0$ and $H=0$ cases it behaves in the IR as 
\be
V_{Sch}=\frac12 h_1'+\frac14 h_1^2-h_0\sim k^2 v^{-\frac{2(2\l^2-1)}{2\l^2+1}},
\ee
where $k^2\sim q^2$ is non-zero. We therefore expect that the spectrum of fluctuations is discrete and gapped for $\l^2>1/2$ ($1<z<2$), continuous and gapped for $\l^2=1/2$ ($z=2$), and continuous and ungapped for $\l^2<1/2$ ($2<z<3$). A full analysis of these RG flows, including the possibility of embedding them in gauge supergravity will appear elsewhere \cite{rgflows}.

\subsection{Vacuum expectation values}

Using the identification (\ref{1pt-fns}) of the 1-point functions with the renormalized canonical momenta we can now evaluate the vacuum expectation values (VEVs) of the dual operators in these backgrounds. Combining (\ref{1pt-fns}) and (\ref{momenta-background}) we find that the general form of the renormalized 1-point functions evaluated in backgrounds of the form (\ref{Bans})	take the from
\begin{subequations}\label{VEVs}
\begin{align}
\langle T_{ij} \rangle &=-\lim_{r\to\infty}e^{(d-2) r/L}\left(\frac{e^{2A}}{\k^2}\left((d-1)\dot A\left(\d_{ij}-(1+f)\d_{i0}\d_{j0}\right)+\frac{\dot f}{2f}\left(\d_{ij}-\d_{i0}\d_{j0}\right)\right)+\frac{2e^{-3A}}{\sqrt{f}}\p_{ij}^{ct}\right),\\
\langle \co_\f \rangle &=\lim_{r\to\infty}e^{\D_\f r/L}\left(-\frac{1}{\k^2}\dot{\f}+e^{-3A}f^{-1/2}\p_\f^{ct}\right), \\
\langle \co_\c \rangle &=\lim_{r\to\infty}e^{\D_\c r/L}\left(-\frac{1}{\k^2}Z(\f)\dot{\c}+e^{-3A}f^{-1/2}\p_\c^{ct}\right),\\
\langle J^i \rangle &=-\frac{2}{\k^2}\wt Q\d^{i0}.
\end{align}
\end{subequations}	

To simplify these expressions we need the form of the counterterms $\cs_{ct}$ and the corresponding canonical momenta $\p_{ij}^{ct}$, $\p_\f^{ct}$ and $\p_\c^{ct}$ evaluated in backgrounds of the form (\ref{Bans}). These counterterms can be computed in full generality using the algorithm presented in Section \ref{hr}, but this analysis can be simplified by focusing on background solutions only. In particular, in Section \ref{hr} terms involving the field strength $F_{ij}$ were counted as derivative terms and do not appear in the leading order solution (\ref{U}) parameterized by the function $U(\f,\c)$. Since the field strength is constant in backgrounds of the form (\ref{Bans}) however, we can take its effect into account by a modified function $U(A,\f,\c)$ that satisfies the same equation (\ref{superpotential}) as the fake superpotential $W(A,\f,\c)$, instead of the simpler equation (\ref{sup-eq}). In fact, since we need to determine only the divergent part of $U(A,\f,\c)$ we can drop the term proportional to $\wt Q-2H\P$ in the effective potential (\ref{Veff}) since this term would only affect $U$ at order $e^{-2(d-1)A}$, which is always subleading relative to $e^{-dA}$. The counterterm function $U(A,\f,\c)$ for general backgrounds of the form (\ref{Bans}) therefore can always be obtained as a Taylor expansion solution of the following PDE:     
\be\label{U-eq}
U_\f^2+Z^{-1}(\f)U_\c^2-\frac{1}{d-1}\left(d+\pa_A\right)U^2=V(\f,\c)+2\S(\f) e^{-4A}H^2.
\ee
As we emphasized in Section \ref{hr}, the relevant solution must have a Taylor expansion of the form (\ref{U-exp}). From the form of (\ref{U-eq}) we immediately deduce that the counterterm function $U(\f,\c)$ will be a function only of the scalars $\f$ and $\c$ as long $d<4$ since in that case the term involving the background magnetic field contributes at subleading order relative to $e^{-dA}$. It follows that for $d<4$ the function $U(\f,\c)$ can be obtained by solving the simpler equation (\ref{sup-eq}). For the marginal case $d=4$ this term will contribute to $U$ with a coefficient that has a pole of the form $1/(d-4)$ (see (\ref{RN-sup})) which via the dimensional regularization described in Section \ref{hr} leads to a cut-off dependent divergence and a related conformal anomaly. For $d>4$ the magnetic field in (\ref{U-eq}) will contribute to the ultraviolet divergences in the standard way.   

Finally, having determined the counterterm function $U(\f,\c)$ for backgrounds of the form (\ref{Bans}) through (\ref{U-eq}), we obtain the renormalized 1-point functions   
\begin{subequations}\label{VEVs-U}
\begin{align}
\langle T_{ij} \rangle &=-\frac{1}{\k^2}\lim_{r\to\infty}e^{ dr/L}\left(\left((d-1)\dot A-U\right)\left(\d_{ij}-(1+f)\d_{i0}\d_{j0}\right)+\left(\frac{\dot f}{2f}-\frac{U_A}{d-1}\right)\left(\d_{ij}-\d_{i0}\d_{j0}\right)\right)\NO\\
&=\frac{1}{\k^2}\lim_{r\to\infty}e^{ dr/L}\left(\left(W+U\right)\left(\d_{ij}-(1+f)\d_{i0}\d_{j0}\right)+\frac{1}{d-1}\left(W_A+U_A\right)\left(\d_{ij}-\d_{i0}\d_{j0}\right)\right),\\
\langle \co_\f \rangle &=-\frac{1}{\k^2}\lim_{r\to\infty}e^{\D_\f r/L}\left(\dot{\f}+U_\f\right)=-\frac{1}{\k^2}\lim_{r\to\infty}e^{\D_\f r/L}\left(W_\f+U_\f\right), \\
\langle \co_\c \rangle &=-\frac{1}{\k^2}\lim_{r\to\infty}e^{\D_\c r/L}\left(Z\dot{\c}+U_\c\right)=-\frac{1}{\k^2}\lim_{r\to\infty}e^{\D_\c r/L}\left(W_\c+U_\c\right),\\
\langle J^i \rangle &=-\frac{2}{\k^2}\wt Q\d^{i0}.
\end{align}
\end{subequations}		
As was mentioned earlier, we see from these expressions that the integration constant $\wt Q$ corresponds to the conserved electric charge density of the background. Moreover, the energy density $\e$ is given by 
\be\label{energy-density}
\e:=\langle T_{t}^t\rangle=-\langle T_{tt}\rangle =-\frac{1}{\k^2}\lim_{r\to\infty} e^{dr/L}\left((d-1)\dot A-U\right), 
\ee
while the pressure density corresponds to (no index summation)
\be\label{pressure-density}
\cp:=\langle T_{ii}\rangle =-\frac{1}{\k^2}\lim_{r\to\infty} e^{dr/L}\left((d-1)\dot A-U+\frac{\dot f}{2f}-\frac{U_A}{d-1}\right).
\ee
	
\section{Fluctuation equations at zero spatial momentum}
\label{fluct}

We next consider linear fluctuations around general backgrounds of the form (\ref{Bans}), and for arbitrary potentials $V(\f,\c)$, $\S(\f)$, $Z(\f)$ and $\P(\c)$. In order to be able to decouple the fluctuation equations (which we present in full generality for $d=3$ in Appendix \ref{gen-fluct}) we follow \cite{Hartnoll:2007ai,Hartnoll:2007ip,Lippert:2014jma} and consider only fluctuations that are independent of the spatial transverse coordinates and we set certain components of the fluctuations consistently to zero.  
	
In particular, denoting the most general fluctuations that preserve the gauge $N=1$, $N_i=a=0$ by
\be
\g_{ij}=\g_{Bij}+h_{ij},\quad 
A_i=A_{Bi}+a_i, \quad
\f=\f_B+\vf,\quad
\c=\c_B+\t,
\ee
with $S_i^j\equiv\g_B^{jk}h_{ki}$,  we switch off the fluctuations $S_t^{t}=S_x^x=S_y^y=S_x^y=\vf=\t=a_t=0$ and only keep the components $a_x=a_x(r,t)$, $a_y=a_y(r,t)$, $S_t^x=S_t^x(r,t)$ and $S_t^y=S_t^y(r,t)$. For such fluctuations the only non-trivial equations are 
\begin{subequations} \label{eqn:Fluctuations}
	\begin{align}
&\mbox{\bf Einstein $xt$:}\NO\\
&\left(\pa_r^2+\left(3\dot{A}-\frac12f^{-1}\dot{f}\right)\pa_r-4\S e^{-4A}H^2\right)S_t^x=-4\S e^{-2A}\left(\dot{\a}\dot{a}_x+H e^{-2A}\pa_ta_y\right),\label{eqn:Stx} \\
&\left(\pa_r^2+\left(3\dot{A}+\frac32f^{-1}\dot{f}\right)\pa_r+4\S e^{-2A}f^{-1}\dot\a^2\right)S_x^t=4\S e^{-2A}f^{-1}\left(\dot{\a}\dot{a}_x+H e^{-2A}\pa_ta_y\right),\label{eqn:Stx-reverse} \\
&\mbox{\bf Einstein $yt$:}\NO\\
&\left(\pa_r^2+\left(3\dot{A}-\frac12f^{-1}\dot{f}\right)\pa_r-4\S e^{-4A}H^2\right)S_t^y=-4\S e^{-2A}\left(\dot{\a}\dot{a}_y- He^{-2A}\pa_ta_x\right), \label{eqn:Sty} \\
&\left(\pa_r^2+\left(3\dot{A}+\frac32f^{-1}\dot{f}\right)\pa_r+4\S e^{-2A}f^{-1}\dot\a^2\right)S_y^t=4\S e^{-2A}f^{-1}\left(\dot{\a}\dot{a}_y-H e^{-2A}\pa_ta_x\right),\label{eqn:Sty-reverse} \\
&\mbox{\bf Einstein $rx$:}\NO\\
&\pa_t\dot S^x_t=-4\S e^{-2A}\left(H\dot\a S^y_t+Hf\dot a_y+\dot\a\pa_t a_x\right),\label{eqn:Stxdot} \\
&\mbox{\bf Einstein $ry$:}\NO\\
&\pa_t\dot S^y_t=-4\S e^{-2A}\left(-H\dot\a S^x_t-Hf\dot a_x+\dot\a\pa_t a_y\right),\label{eqn:Stydot} \\
&\mbox{\bf Maxwell $x$:}\NO\\
&\pa_r\left(\S f^{-1/2}e^A\left(\dot\a S^x_t+f\dot a_x\right)\right)=\S f^{-1/2}e^{-A}(\pa_t^2a_x+H\pa_tS^y_t)+2\P_\c\dot\c_B\pa_ta_y,\label{eqn:Maxx}\\
&\mbox{\bf Maxwell $y$:}\NO\\
&\pa_r\left(\S f^{-1/2}e^A\left(\dot\a S^y_t+f\dot a_y\right)\right)=\S f^{-1/2}e^{-A}(\pa_t^2a_y-H\pa_tS^x_t)-2\P_\c\dot\c_B\pa_ta_x.\label{eqn:Maxy}
\end{align}
\end{subequations}
Note that (\ref{eqn:Stx}) and (\ref{eqn:Stx-reverse}) as well as (\ref{eqn:Sty}) and (\ref{eqn:Sty-reverse}) are trivially related since 
\be
S^t_x=-f^{-1}S^x_t,\quad S^t_y=-f^{-1}S^y_t.
\ee
The remaining equations can be decoupled by introducing the complexified variables
\be
S_t^\pm\equiv S^x_t\pm iS^y_t,\quad h_{t\pm}\equiv h_{tx}\pm ih_{ty},\quad a_\pm\equiv a_x\pm ia_y,
\ee
so that we can write 
\begin{subequations} \label{eqn:Fluctuations-complex}
	\begin{align}
		&\left(\pa_r^2+\left(3\dot{A}-\frac12f^{-1}\dot{f}\right)\pa_r\right)S_t^\pm=-4\S e^{-2A}\left(\dot{\a}\dot{a}_\pm \mp iH e^{-2A}\pa_ta_\pm-H^2 e^{-2A}S^\pm_t\right),\label{eqn:Stz} \\
		&\pa_t\dot S^\pm_t=4\S e^{-2A}\left(\pm iH(\dot\a S^\pm_t+f\dot a_\pm)-\dot\a\pa_t a_\pm\right),\label{eqn:Stzdot} \\
		&\pa_r\left(\S f^{-1/2}e^A\left(\dot\a S^\pm_t+f\dot a_\pm\right)\right)=\S f^{-1/2}e^{-A}(\pa_t^2 a_\pm\mp iH\pa_tS^\pm_t)\mp2i\P_\c\dot\c_B\pa_ta_\pm,\label{eqn:Maxz}
	\end{align}
\end{subequations}
or after Fourier transforming in time ($\pa_t\to i\o$)
\begin{subequations} \label{eqn:Fluctuations-fourier}
	\begin{align}
		&\pa_r\left(e^{3A}f^{-1/2}\dot S^\pm_t \right)=-4\S e^{A}f^{-1/2}\left(\dot{\a}\dot{a}_\pm\pm\o H e^{-2A}a_\pm-H^2 e^{-2A}S^\pm_t\right),\label{eqn:Stz-fourier} \\
		&\o\dot S^\pm_t=4\S e^{-2A}\left(\pm H(\dot\a S^\pm_t+f\dot a_\pm)-\o\dot\a a_\pm\right),\label{eqn:Stzdot-fourier} \\
		&\pa_r\left(\S f^{-1/2}e^A\left(\dot\a S^\pm_t+f\dot a_\pm\right)\right)= \S f^{-1/2}e^{-A}(-\o^2 a_\pm\pm\o HS^\pm_t)\pm2\o\P_\c\dot\c_Ba_\pm.\label{eqn:Maxz-fourier}
	\end{align}
\end{subequations}
Note that multiplying (\ref{eqn:Stzdot-fourier}) with $e^{3A}f^{-1/2}$, taking the radial derivative and substituting (\ref{eqn:Maxz-fourier}) in the resulting expression gives back (\ref{eqn:Stz-fourier}), which is therefore not independent, unless $\om=0$. 

In order to decouple these equations for $\om\neq 0$ we replace $a_\pm$ with the linear combinations
\be
\ce_\pm\equiv \om a_\pm\mp H S^\pm_t,
\ee
in terms of which (\ref{eqn:Fluctuations-fourier}) become 
\begin{subequations} \label{eqn:Fluctuations-fourier-s}
	\begin{align}
		&\o\pa_r\left(e^{3A}f^{-1/2}\dot S^\pm_t \right)=-4\S e^{A}f^{-1/2}\left(\dot{\a}\dot\ce_\pm\pm\dot\a H\dot S^\pm_t\pm\o H e^{-2A}\ce_\pm\right),\label{eqn:Stz-fourier-s} \\
		&\left(\o^2-4\S e^{-2A}fH^2\right)\dot S^\pm_t=4\S e^{-2A}\left(\pm Hf\dot \ce_\pm-\o\dot\a\ce_\pm \right),\label{eqn:Stzdot-fourier-s} \\
		&\o\S f^{-1/2}e^A\dot\a \dot S^\pm_t+\pa_r\left(\S f^{1/2}e^A(\dot \ce_\pm\pm H\dot S^\pm_t)\right)= -\o^2\S f^{-1/2}e^{-A}\ce_\pm \pm2\o\P_\c\dot\c_B\ce_\pm,\label{eqn:Maxz-fourier-s}
	\end{align}
\end{subequations}
where we have used the last equation in (\ref{BEOM}) to obtain (\ref{eqn:Maxz-fourier-s}). Substituting now the expression for $\dot S^\pm_t$ from (\ref{eqn:Stzdot-fourier-s}) into (\ref{eqn:Stz-fourier-s}) leads to the two decoupled equations for $\ce_\pm$
\be\label{eqn:decoupled}\boxed{
\ddot\ce_\pm+g_1(\o,H)\dot\ce_\pm+g_0(\o,\pm H)\ce_\pm=0,}
\ee 
where
\begin{subequations}\label{eqn:gs}
	\begin{align}
	g_1(\o,H)= & \pa_r\log\left|\S e^Af^{1/2}\Om^{-1}\right|,\\
	g_0(\o,\pm H)= & f^{-1}e^{-2A}(\Om-4\S\dot\a^2)\mp\frac{\o}{H}f^{-1}\dot\a\pa_r\log\left|\S e^A f^{-1/2}\dot\a\Om^{-1}\right|,
 	\end{align}
\end{subequations}
with
\be
\label{Omega}
 \Om\equiv\o^2-4\S H^2fe^{-2A}.
\ee
In terms of the $u$ coordinate these equations read
\be\label{Eqn:Fluctuation}\boxed{
\ce_\pm''+h_1(\o,H)\ce_\pm'+h_0(\o,\pm H)\ce_\pm=0,}
\ee
where
\begin{subequations} \label{eqn:hs}
 	\begin{align}
 		h_1(\o,H)=&\pa_u\log\left|\S f\Om^{-1}\right|,\label{h0}\\
 		h_0(\o,\pm H)=&f^{-2}(\Om-4\S f e^{-2A}\a'^2)\mp\frac{\o}{H}f^{-1}\a'\pa_u\log\left|\S \a'\Om^{-1}\right|.\label{h1}
 	\end{align}
\end{subequations}
 
\subsection*{Riccati form of the fluctuation equations}

Finally, these linear second order fluctuation equations can be expressed in first order Riccati form by introducing the response functions $\car_{\pm}$ \cite{Papadimitriou:2004rz,Papadimitriou:2013jca}, namely
\be\label{response}\boxed{
\dot\ce_\pm=\car_\pm\ce_\pm,}
\ee
so that (\ref{eqn:decoupled}) take the Riccati form
\be\label{Ric}\boxed{
\dot\car_\pm+\car^2_\pm+g_{1}(\o,H)\car_\pm+g_0(\o,\pm H)=0.}
\ee
Since these are first order differential equations there is only one integration constant for each, which is fixed by imposing suitable boundary conditions in the interior of the geometry. This is one of the advantages of the Riccati formulation of the fluctuation equations, since they compute directly the response functions, without any dependence on the arbitrary sources.

\section{Renormalized Green's functions, Kubo formulas, and transport coefficients} 
\label{2pnt-fns}

Another advantage of the Riccati form (\ref{Ric}) of the fluctuation equations is that, as we now show, the 2-point functions can be determined in general in terms of the response functions $\car_{\pm}$. Therefore, solving the Riccati equations directly determines the 2-point functions, without the need to evaluate the on-shell action and then take functional derivatives. The Riccati  formulation of the fluctuation equations implements linear response theory in the bulk.

\subsection{Holographic Green's functions and Ward identities}

In order to unravel the relation between the response functions $\car_{\pm}$ and the 2-point functions we need to determine the canonical momenta (\ref{momenta}), which as we have seen in Section \ref{hamiltonian-formalism} are identified with the 1-point functions, to linear order in the fluctuations. For the fluctuations we considered in Section \ref{fluct} the only non-trivial components of the canonical momenta to linear order in the fluctuations are 
\begin{subequations}
\begin{align}
&\overset{(1)}{\p^{ij}}=-\frac{1}{2\k^2}\sqrt{-\g_B}\left(\frac12\dot S^i_k\g_B^{kj}+2\dot AS^{ij}+\frac{\dot f}{2f}\left(S^{ij}-S^i_t\g_B^{tj}\right)\right),\\
&\overset{(1)}{\p^i}=\frac{2}{\k^2}\sqrt{-\g_B}\S\left(\dot\a S^{it}-\g_B^{ij}\dot a_j\right)+\frac{4i\om}{\k^2}\P\left(\d^i_xa_y-\d^i_ya_x\right),
\end{align}
\end{subequations}
whose only non-zero components are 
\begin{subequations}
\begin{align}
\overset{(1)}{\pi^\pm_h}=&-\frac{1}{4\k^2}\sqrt{-\g_B}\g_B^{tt}\left(\dot{S}^\pm_t+4\dot{A}S^\pm_t\right),\\
\overset{(1)}{\pi^\pm_a}=&\frac{2}{\k^2}\sqrt{-\g_B}\g_B^{tt}\left(\S\left(\dot{\a}S^\pm_t+f\dot{a}_\pm\right)\mp2\o\P e^{-A}f^{1/2}a_\pm\right),
\end{align}
\end{subequations}
where $\p^{\pm}_h=\p^{xt}\pm i\p^{yt}$,  $\p^\pm_a=\p^{x}\pm i\p^{y}$. Using the defining relations (\ref{response}) for the response functions $\car_{\pm}$ we can express the velocities $\dot S_t^\pm$ and $\dot a_\pm$ in terms of the fluctuations $\ce_\pm$ as
\begin{subequations}\label{dots}
	\begin{align}
		&\dot S^\pm_t=4\S e^{-2A}\Om^{-1}\left(\pm Hf\car_\pm-\o\dot\a\right)\ce_\pm,\label{Sdot}\\
		&\dot a_\pm=\Om^{-1}\left(\o \car_\pm\mp4H\S e^{-2A}\dot\a\right)\ce_\pm.\label{adot}
	\end{align}
\end{subequations}
Hence, the only non-zero components of the momenta to linear order in the fluctuations take the form
\be\label{linear-momenta}
\overset{(1)}{\pi^\pm_h}=\frac12\sqrt{-\g_B}\left(C_{hh}^\pm h_{t\pm}+C_{ha}^\pm a_\pm\right),\quad
\overset{(1)}{\pi^\pm_a}=\sqrt{-
	\g_B}\left(C_{ha}^\pm h_{t\pm}+C_{aa}^\pm a_\pm\right),
\ee
where 
\begin{subequations}\label{Cs-new}
	\begin{align}
		C_{hh}^\pm(r,\o)&=\frac{2}{\k^2}\g_B^{tt}e^{-2A}\left(\pm H\S e^{-2A} \Om^{-1}\left(\pm Hf\car_\pm-\om\dot\a\right)-\dot A\right), \\
		C_{ha}^\pm(r,\o)&=-\frac{2}{\k^2}\g_B^{tt}\S \Om^{-1}e^{-2A}\o\left(\pm Hf\car_\pm-\om\dot\a\right), \\
		C_{aa}^\pm(r,\o)&=\pm\frac{2}{\k^2}\g_B^{tt}\frac{\o}{H}\left(\S\Om^{-1}\o\left(\pm H f\car_\pm-\o\dot\a\right)-\wt Qe^{-A}f^{1/2}\right),
	\end{align}
\end{subequations}
and we have used the fact that $e^{A}f^{-1/2}\S\dot\a-2\P H=-\wt Q$ is a constant (see (\ref{ap}) and (\ref{zr})). 

In order to renormalize these expressions for the canonical momenta linear in the fluctuations we must take into account the contribution of the boundary counterterms, which for $d=3$ take the form   
\begin{align}
	\mathcal{S}_{ct}&=-\frac{1}{\k^2}\int\text{d}^{3}\mathbf{x}\,\sqrt{-\g}\left(U(\f,\c)+\Th(\f,\c)R[\g]\right),
\end{align}
where $U$ is determined through (\ref{sup-eq}), while $\Th$ is determined by the algorithm described in Section \ref{hr}. However, for backgrounds of the form (\ref{Bans}), where the Ricci scalar of the induced metric vanishes, and for fluctuations of the form discussed in Section \ref{fluct} it is straighforward to check that the counterterm involving the Ricci scalar does not contribute (see last equation in (\ref{fluct-exp})). In particular, when expanded to second order in the fluctuations the counterterms for $d=3$ take the form
\be
\overset{(2)}{S_{ct}}=\frac{1}{2\k^2}\int\text{d}^2\mathbf{x}\int\text{d}\o\,\sqrt{-\g_B}\g_B^{tt}Ue^{-2A}h_{t+}(\o)h_{t-}(-\o).
\ee
These counterterms lead to the renormalized response functions 
\begin{subequations}\label{Cs-ren}
	\begin{align}
		C_{hh}^{ren \pm}(r,\o)&=C_{hh}^{\pm}(r,\o)+\frac{1}{\k^2}e^{-2A}\g_{B}^{tt}U\NO\\
		&=\frac{2}{\k^2}\g_B^{tt}e^{-2A}\left(\pm H\S e^{-2A} \Om^{-1}\left(\pm Hf\car_\pm-\om\dot\a\right)-\dot A+\frac12U\right), \\
		C_{ha}^{ren \pm}(r,\o)&=C_{ha}^\pm(r,\o)=-\frac{2}{\k^2}\g_B^{tt}\S \Om^{-1}e^{-2A}\o\left(\pm Hf\car_\pm-\om\dot\a\right), \\
		C_{aa}^{ren \pm}(r,\o)&=C_{aa}^\pm(r,\o)=\pm\frac{2}{\k^2}\g_B^{tt}\frac{\o}{H}\left(\S\Om^{-1}\o\left(\pm H f\car_\pm-\o\dot\a\right)-\wt Qe^{-A}f^{1/2}\right).
	\end{align}
\end{subequations}
Notice that only the coefficients $C_{hh}^{\pm}$ get renormalized. 

Using (\ref{1pt-fns}) we can now express the renormalized 1-point functions in terms of the renormalized momenta, namely   
\bal\label{1pt-fns-momenta}
&\langle T^{t\pm}\rangle \equiv \langle T^{tx}\rangle\pm i\langle T^{ty}\rangle=-\frac{2}{\sqrt{-\g_B}}\Hat\p_h^\pm,\NO\\
&\langle J^{\pm}\rangle \equiv \langle J^{x}\rangle\pm i\langle J^{y}\rangle=\frac{1}{\sqrt{-\g_B}}\Hat\p_a^\pm,
\eal
or, after lowering the indices
\be
\langle T_{t\pm}\rangle=\frac{2}{\sqrt{-\g_B}}e^{4A}f\Hat\p_h^\pm= 2e^Af^{1/2}\Hat\p_h^\pm,\quad
\langle J_{\pm}\rangle =\frac{1}{\sqrt{-\g_B}}e^{2A}\Hat\p_a^\pm= e^{-A}f^{-1/2}\Hat\p_a^\pm.
\ee
The 2-point functions can now be obtained by functional differentiation of these 1-point functions with respect to the fluctuations. The precise coefficients of the functional derivatives with respect to $h_{t\pm}$ and $a_\pm$ that correspond to an insertion of respectively $T_{t\mp}$ and $J_\mp$ can be determined by computing the variation
\be\label{variation}
\d\cs_{ren}=2{\Hat\p^{xt}}\d h_{ xt}+2{\Hat\p^{yt}}\d h_{yt}+{\Hat\p^x}\d a_x+{\Hat\p^y}\d a_y=\frac12\left(2{\Hat\p^+_h}\d h_{t- }+2{\Hat\p^{-}_h}\d h_{t+ }+{\Hat\p^+_a}\d a_{-}+{\Hat\p^{-}_a}\d a_{+}\right),
\ee
where the factor of 2 in the metric momenta is due to the fact that we must sum over the two possible index combinations. It follows that 
\be\label{1pt-fns-der}
\langle T_{t\pm}(\o)\rangle=2e^Af^{1/2}\frac{\d\cs_{ren}}{\d h_{t\mp}(-\o)},\quad
\langle J_{\pm}(\o)\rangle =2e^{-A}f^{-1/2}\frac{\d\cs_{ren}}{\d a_{\mp}(-\o)}.
\ee

We now have all the ingredients in order to evaluate the renormalized 2-point functions. Combining (\ref{1pt-fns-momenta}) and (\ref{linear-momenta}) we get
\be\label{1pt-fns-linear}\boxed{
\langle T_{t\pm}\rangle=e^{4A}\left(C_{hh}^{\pm ren} h_{t\pm}+C_{ha}^{\pm ren} a_\pm\right),\quad
\langle J_{\pm}\rangle =e^{2A}\left(C_{ha}^{\pm ren} h_{t\pm}+C_{aa}^{\pm ren} a_\pm\right).}
\ee
From (\ref{1pt-fns-der}) then follows that the 2-point functions are obtained as
\bal
&\langle T_{t+}(\om)T_{t-}(-\om)\rangle= 2e^A\frac{\d}{\d h_{t+}(\o)}\langle T_{t+}(\o)\rangle= 2e^{5A}C_{hh}^{+ren}(\o),\NO\\
&\langle T_{t+}(\om)J_-(-\om)\rangle=2e^{-A}\frac{\d}{\d a_{+}(\o)}\langle T_{t+}(\o)\rangle=2e^{3A}C_{ha}^{+ren}(\o),\NO\\
&\langle J_+(\om)J_-(-\om)\rangle =2e^{-A}\frac{\d}{\d a_{+}(\o)}\langle J_{+}(\o)\rangle=2e^AC_{aa}^{+ren}(\o),
\eal
and similarly for the fluctuations $h_{t-}$ and $a_-$. In particular, the full set of renormalized 2-point functions that can be computed with the fluctuations we considered are 
\begin{subequations}\label{2-pt-fns-+/-}
	\begin{align}
		\langle T_{t+}(\om)T_{t-}(-\om)\rangle  & = \,  
		2\lim_{r\to\infty}\left(e^{7r/L}C_{hh}^{+ren}(r,\o)\right), \\
		\langle T_{t-}(\om)T_{t+}(-\om)\rangle &  =\,  
		2\lim_{r\to\infty}\left(e^{7r/L}C_{hh}^{-ren}(r,\o)\right), \\
		\langle T_{t+}(\om)J_-(-\om)\rangle  &  =\,  
		2\lim_{r\to\infty}\left(e^{5r/L}C_{ha}^{+ren}(r,\o)\right), \\
		\langle T_{t-}(\om)J_+(-\om)\rangle & =\,  
		2\lim_{r\to\infty}\left(e^{5r/L}C_{ha}^{-ren}(r,\o)\right), \\
		\langle J_+(\om)J_-(-\om)\rangle  & =\, 
		2\lim_{r\to\infty}\left(e^{3r/L}C_{aa}^{+ren}(r,\o)\right),\\
		\langle J_-(\om)J_+(-\om)\rangle   &=\,
		2\lim_{r\to\infty}\left(e^{3r/L}C_{aa}^{-ren}(r,\o)\right),
	\end{align}
\end{subequations}
with all other 2-point functions vanishing identically. In the $x,y$ basis these become 
\begin{subequations}\label{2-pt-fns-lim}
	\begin{align}
		&\langle T_{tx}(\om)T_{tx}(-\om)\rangle \hskip-0.5cm & = \,  
		&\hskip0.5cm\langle T_{ty}(\om)T_{ty}(-\om)\rangle \hskip-0.5cm & =\,  
		&\frac12\lim_{r\to\infty}\left(e^{7r/L}\left(C_{hh}^{+ren}(r,\o)+C_{hh}^{-ren}(r,\o)\right)\right), \\
		&\langle T_{tx}(\om)T_{ty}(-\om)\rangle \hskip-0.5cm & = \,  
		&-\langle T_{ty}(\om)T_{tx}(-\om)\rangle \hskip-0.5cm & =\,  
		&\frac i2\lim_{r\to\infty}\left(e^{7r/L}\left(C_{hh}^{+ren}(r,\o)-C_{hh}^{-ren}(r,\o)\right)\right), \\
		&\langle T_{tx}(\om)J_x(-\om)\rangle \hskip-0.5cm & =\, 
		&\hskip0.5cm\langle T_{ty}(\om)J_y(-\om)\rangle \hskip-0.5cm & =\,  
		&\frac12\lim_{r\to\infty}\left(e^{5r/L}\left(C_{ha}^{+ren}(r,\o)+C_{ha}^{-ren}(r,\o)\right)\right), \\
		&\langle T_{tx}(\om)J_y(-\om)\rangle \hskip-0.5cm & =\, 
		&-\langle T_{ty}(\om)J_x(-\om)\rangle \hskip-0.5cm & =\,  
		&\frac i2\lim_{r\to\infty}\left(e^{5r/L}\left(C_{ha}^{+ren}(r,\o)-C_{ha}^{-ren}(r,\o)\right)\right), \\
		&\langle J_x(\om)J_x(-\om)\rangle \hskip-0.5cm & =\,
		&\hskip0.5cm\langle J_y(\om)J_y(-\om)\rangle \hskip-0.5cm & =\, 
		&\frac12\lim_{r\to\infty}\left(e^{3r/L}\left(C_{aa}^{+ren}(r,\o)+C_{aa}^{-ren}(r,\o)\right)\right),\\
		&\langle J_x(\om)J_y(\om)\rangle \hskip-0.5cm & =\,
		&-\langle J_y(\om)J_x(\om)\rangle \hskip-0.5cm & =\, 
		&\frac i2\lim_{r\to\infty}\left(e^{3r/L}\left(C_{aa}^{+ren}(r,\o)-C_{aa}^{-ren}(r,\o)\right)\right).
	\end{align}
\end{subequations}

The limits on the RHS can be evaluated explicitly by considering asymptotically AdS backgrounds of the form (\ref{Bans}) so that 
\begin{subequations}\label{bgnd-UV}
	\begin{align}
		&\a=\a_0+\S_0^{-1}(\wt Q-2H\P_0)Le^{-r/L}+\co\left(e^{-2r/L}\right),\\
		& A=\frac rL+\co\left(e^{-r/L}\right),\\
		& f=1-ML^{-3}e^{-3r/L}+\co\left(e^{-4/L}\right),
	\end{align}
\end{subequations}
and asymptotically $\S\sim\S_0$, $\P\sim\P_0$ for some constants $\S_0$ and $\P_0$. Note that this asymptotic form of $f$ follows from the third equation in (\ref{eom2}), even for backgrounds with running scalars. As we shall see, these asymptotic conditions for the background imply that the response functions $\car_\pm$ asymptotically behave as 
\be
\car_\pm(r,\o)=\car_{\pm(1)}(\o) e^{-r/L}+\co(e^{-2r/L}),
\ee
where the functions $\car_{\pm(1)}(\o)$ are determined by the boundary conditions in the interior of the bulk spacetime. Evaluating the limits (\ref{2-pt-fns-lim}) using these asymptotics we finally obtain 
\begin{subequations}\label{2pt-fns-WIs-omega}
	\begin{align}
		&\langle T_{tx}(\om)T_{tx}(-\om)\rangle \hskip-1.2cm & = \,  
		&\hskip0.5cm \langle T_{ty}(\om)T_{ty}(-\om)\rangle \hskip-1.2cm & =\,  
		&\hskip0.5cmi\frac{H}{\om}\langle T_{tx}(\om)J_y(-\om)\rangle+\langle T_{tt}\rangle, \\
		&\langle T_{tx}(\om)T_{ty}(-\om)\rangle \hskip-1.0cm & = \,  
		&-\langle T_{ty}(\om)T_{tx}(-\om)\rangle \hskip-1.0cm & =\,  
		&-i\frac{H}{\o}\langle T_{tx}(\om)J_x(-\om)\rangle, \\
		&\langle T_{tx}(\om)J_x(-\om)\rangle \hskip-1.2cm & =\, 
		&\hskip0.5cm \langle T_{ty}(\om)J_y(-\om)\rangle \hskip-1.2cm & =\,  
		&\hskip0.5cm i\frac{H}{\o}\langle J_x(\om)J_y(-\om)\rangle-\langle J^t\rangle, \\
		&\langle T_{tx}(\om)J_y(-\om)\rangle \hskip-1.2cm & =\, 
		&-\langle T_{ty}(\om)J_x(-\om)\rangle \hskip-1.2cm & =\,  
		&-i\frac{H}{\o}\langle J_x(\om)J_x(-\om)\rangle,
	\end{align}
\end{subequations}
where
\begin{subequations}\label{2-pt-fns}
	\begin{align}
		&\langle J_x(\om)J_x(-\om)\rangle \hskip-0.9cm & =\,
		&\hskip0.5cm\langle J_y(\om)J_y(-\om)\rangle \hskip-0.9cm & =\, 
		&-\frac{2}{\k^2}\S_0\car^+_{(1)}(\o)\hskip0.1cm,\\
		&\langle J_x(\om)J_y(-\om)\rangle \hskip-0.9cm & =\,
		&-\langle J_y(\om)J_x(-\om)\rangle \hskip-0.9cm & =\, 
		&-\frac{2i}{\k^2}\left(\S_0\car^-_{(1)}(\o)-2\P_0\o\right)\hskip-0.0cm,
	\end{align}
\end{subequations}
with
\be
\car_{(1)}^\pm(\o)\equiv\frac12\left(\car_{+(1)}(\o)\pm\car_{-(1)}(\o)\right),
\ee
and the renormalized 1-point functions $\langle T_{tt}\rangle$ and $\langle J^t\rangle$, corresponding respectively to the energy and charge densities, are given in (\ref{energy-density}) and (\ref{VEVs-U}). Notice that the only independent 2-point functions are the current-current ones, while all other non identically vanishing 2-point functions are expressed in terms of the current-current 2-point functions and the 1-point functions of the background. In fact, the relations (\ref{2pt-fns-WIs-omega}) are nothing but the Fourier transform of the 2-point function Ward identities (\ref{2pt-fns-WIs}), and so they are purely kinematic. 

A number of comments are in order here. Firstly, note that the Ward identities 
(\ref{2pt-fns-WIs-omega}) imply that certain 2-point functions can potentially diverge as $\o\to 0$ for non-zero magnetic field. However, as we shall see in Section \ref{Sw}, the leading behavior of the response functions $\car_{\pm(1)}(\o)$ for small $\o$ is (see (\ref{response-small-w})) 
\be
\car_{\pm(1)}(\o)=\pm\frac{1}{\S_0}\left(2H\P_0-\wt Q\right)\frac{\o}{H}+\co(\o^2),
\ee
independently of the regularity condition imposed in the IR.\footnote{More precisely, it is independent of the IR boundary conditions, as long as the constant $c_1$ in Section \ref{Sw} is non-zero.} This ensures that the 2-point functions do not diverge in the small frequency limit for $H\neq 0$. Another remark is that all the expressions for the 2-point functions we have discussed so far, including the Ward identities (\ref{2pt-fns-WIs-omega}), are independent of the IR boundary conditions. In order to compute the conductivities we will impose infalling boundary conditions on the horizon in order to obtain the retarded 2-point functions \cite{Son:2002sd}, but any other thermal 2-point function can be obtained by imposing the corresponding boundary condition on the horizon. As we shall show in Section \ref{IR}, imposing infalling boundary conditions on the horizon implies that the response functions $\car_\pm(\o)$ are related via $\car_+(\omega)=\car_-(-\omega)^*$, and hence 
\be\label{reflex}
\car_{+(1)}(\o)=\car_{-(1)}(-\o)^*.
\ee
Finally, it should be stressed that we have computed the renormalized 2-point functions without ever evaluating the on-shell action, contrary to the usual procedure in the literature. What allowed us to do this is the fact that the renormalized canonical momenta are identified with the 1-point functions in the presence of sources via (\ref{1pt-fns}). This is practically very important in Lorentzian holography, since the proper evaluation of the on-shell action for thermal correlators involves multiple boundaries \cite{Herzog:2002pc,Skenderis:2008dg,vanRees:2009rw}.

\subsection{Kubo formulas and transport coefficients}
\label{Kubo}

The thermoelectric conductivities are defined in linear response theory via the relations 
\be\label{JQ}
\left(
\begin{array}{c}
	\langle\vec{J}\rangle\\
	\langle\vec{Q}\rangle
\end{array}
\right)
=
\left(
\begin{array}{cc}
	\Hat\bs & T\Hat\ba \\
	T \Hat{\bar\ba} & T\Hat{\bar\bk}
\end{array}
\right)
\left(
\begin{array}{c}
	\vec E\\
	-\vec\nabla \log T
\end{array}
\right)\,,
\ee
where $\vec J$ is the electromagnetic current and $\vec Q$ is the mixed thermoelectric current defined by
\be\label{Q}
\langle Q_a\rangle =\langle T_{at}\rangle-\m \langle J_a\rangle,\quad  a,b=x,y,
\ee
and $\mu$ is the chemical potential. $\vec E$ here is the applied electric field and $\vec \nabla T$ is the spatial temperature gradient. The linear responses in the currents are encoded in the matrix of thermoelectric conductivities, where $\Hat\bs$ is electric conductivity, $\Hat\ba$ and $\Hat{\bar\ba}$ are the thermoelectric coefficients and $\Hat{\bar\bk}$ is the thermal conductivity. Each of these quantities is a $2 \times 2$ antisymmetric matrix. The hats indicate that these quantities are the conductivities in the presence of a magnetization current \cite{2007PhRvB..76n4502H}. 

Combining this definition of the conductivities with the expressions (\ref{1pt-fns-linear}) for the 1-point functions of the stress tensor and electromagnetic current and the 2-point functions (\ref{2-pt-fns-lim}) we arrive at the Kubo formulas
\begin{subequations}\label{kubos}
\begin{align}
 \Hat\s_{ab} &= i\o^{-1} \langle J_{a}(\om)J_{b}(-\om)\rangle_R,\label{s1}\\\NO\\
 T \Hat \a_{ab}&= i \o^{-1 }\langle Q_{a}(\om)J_{b}(-\om)\rangle_R=i \o^{-1}\langle T_{at}(\om)J_{b}(-\om)\rangle_R- \mu \s_{ab}, \label{Ta1}\\\NO\\
 T\Hat{\bar \k}_{ab}&=  i \o^{-1 }\langle Q_{a}(\om)Q_{b}(-\om)\rangle_R=i \o^{-1 }\langle Q_{a}(\om)T_{bt}(-\om)\rangle_R-\m T\Hat\a_{ab}\NO\\
&= i \o^{-1}\langle T_{at}(\om)T_{bt}(-\om)\rangle_R-\m i \o^{-1}\langle J_{a}(\om)T_{bt}(-\om)\rangle_R-\m T\Hat\a_{ab},\label{k}
\end{align}
\end{subequations}
where the subscript $R$ indicates that the retarded correlators must be used. 

We can now use the Ward identities (\ref{2pt-fns-WIs-omega}) --which we emphasize were derived holographically-- to express the thermoelectric coefficients and the thermal conductivity in terms of the electric conductivities only. From    (\ref{2pt-fns-WIs-omega}) we get
\begin{subequations}\label{Ta} 
	\begin{align}
		-i\o T\Hat\a_{xx}&=H\s_{xy}+i\o\m\s_{xx}-\r,\\
		\o T\Hat\a_{xy}&=-iH\s_{xx}-\o\m\s_{xy},
	\end{align}
\end{subequations}
where $ \r\equiv \langle J^t\rangle=-2\wt Q/\k^2$ is the electric charge density we found in (\ref{VEVs}). Combining these two relations as 
\be
\o T\left(\Hat\a_{xy}\pm i\Hat\a_{xx}\right)=\mp H\left(\s_{xy}\pm i\s_{xx}\right)-\o\m\left(\s_{xy}\pm i\s_{xx}\right)\pm\r,
\ee
and defining
\be \label{sah}
\s_{\pm}=\s_{xy} \pm i\s_{xx}, \quad \Hat \a_{\pm}= \Hat \a_{xy} \pm i  \Hat \a_{xx},
\ee
we arrive at the expressions 
\be\label{a=s}\boxed{
\o T  \Hat \a_{\pm} = -(\o \mu\pm H)\s_{\pm} \pm \rho,}
\ee
reproducing the result of \cite{Hartnoll:2007ip}. However, here we have arrived at this result through a holographic derivation of the Ward identities (\ref{2pt-fns-WIs}), which hold in a generic theory including scalar operators. Similarly, combining (\ref{k}) and (\ref{2pt-fns-WIs-omega}) we get
\be\label{k=s}\boxed{
\o T\Hat{\bar\k}_\pm=-(\o\m\pm H)T\Hat\a_\pm\pm(\e-\m\r),}
\ee
where $\Hat{\bar\k}_{\pm}= \Hat{\bar\k}_{xy} \pm i  \Hat{\bar\k}_{xx}$, and $\e$ is the energy density defined in (\ref{energy-density}).

It follows that the only conductivities we actually need to compute are the electric conductivities, which are holographically expressed in terms of the response functions through (\ref{2-pt-fns}) as
\begin{subequations} \label{sxy}
	\begin{align}
		&\s_{xx}=\s_{yy}=-\frac{2i}{\k^2}\S_0\om^{-1}\car^+_{(1)}(\o),\\
		&\s_{xy}=-\s_{yx}=\frac{2}{\k^2}\left(\S_0\om^{-1}\car^-_{(1)}(\o)-2\P_0\right),
	\end{align}
\end{subequations}
where the response functions are computed with ingoing boundary conditions on the horizon. Note that there is no magnetization subtraction for the electric conductivities, which is why we have dropped the \hskip0.1cm$\Hat{}$\hskip0.1cm.

\section{Response functions from the Riccati equation}
\label{R}

In the preceding sections we have shown that the renormalized 2-point functions and the corresponding conductivities for generic asymptotically AdS backgrounds of the form (\ref{Bans}) can be expressed in terms of the response functions $\car_{\pm(1)}(\o)$, which are as yet undetermined. These response functions are computed by solving the Riccati equations (\ref{Ric}) and imposing appropriate regularity conditions in the interior of the bulk spacetime. Typically these equations can only be solved numerically, but certain analytic results can be obtained by taking various limits. In this section we determine the general solution of the Riccati equations in the small and large frequency limits, as well as the ultraviolet and infrared asymptotic solutions, before presenting an algorithm for obtaining the exact solution numerically.

\subsection{UV asymptotic solutions}
\label{UV}

We begin by determining the UV behavior of the response functions $\car_\pm(r,\o)$. Inserting the UV expansions (\ref{bgnd-UV}) for the background fields in the coefficients (\ref{eqn:gs}) of the Riccati equations we obtain 
\be\label{eqn:gs-UV}
	g_1(\o,H)=\frac1L+\co(e^{-r/L}),\quad
	g_0(\o,\pm H)= \o^2e^{-2r/L}+\co(e^{-3r/L}).
\ee
It then follows trivially from the Riccati equations (\ref{Ric}) that the general UV behavior of the response functions takes the form
\be\label{RUV}
\car_\pm(r,\o)=\car_{(1)\pm}(\o)e^{-r/L}+\co(e^{-2r/L}),
\ee
where $\car_{(1)\pm}(\o)$ are the only integration constants of these first order equations. As we have seen, these integration constants, which are determined by imposing suitable regularity conditions in the IR, are the quantities encoding all the dynamical information in the 2-point functions and the corresponding transport coefficients.

\subsection{IR asymptotic solutions}
\label{IR}

The IR behavior of the response functions $\car_\pm(r,\o)$ depends crucially on the type of background considered. The ansatz (\ref{Bans}) includes both zero temperature (no horizon) and finite temperature backgrounds and the results for the renormalized Green's functions obtained in Section \ref{2pnt-fns} are applicable to both types of backgrounds. However, the conductivities involve the retarded Green's functions at finite temperature and so they are relevant observables for backgrounds with a horizon. We will therefore consider explicitly only the IR asymptotics for finite temperature backgrounds of the form (\ref{Bans}) here. Confining backgrounds of the form (\ref{Bans}) at zero temperature are discussed e.g. in \cite{Lippert:2014jma,rgflows}.  

Assuming the geometry exhibits a horizon at $u=u_h$ (corresponding ot the smallest root of the equation $f(u)=0$) giving rise to a finite temperature $T$, the function $f$ increases from zero at $u=u_h$ in the IR to the value $f=1$ at $u=0$ in the UV. In particular, $f(u)$ admits a Taylor expansion near $u_h$ of the form 
\be \label{fIR}
f(\r)=4 \pi T \r+\co\left(\r^2\right),\quad \r \equiv u_h-u,
\ee
while the warp factor satisfies $e^{A(\r)}=\co(1)$ as $\r\to 0^+$. These asymptotic conditions, together with the assumption that $\S(\f)$ and $\P(\c)$ remain finite at the horizon, allows one to determine the leading asymptotic form of the coefficients (\ref{eqn:hs}), namely
\be\label{nhh}
h_{1}=-\frac{1}{\r}+\co\left(1\right),\quad
h_0=\frac{\o^2}{(4\pi T)^2\r^2}+\co\left(\r^{-1}\right).
\ee
It follows that the near horizon behavior of the general solution of the fluctuation equations \eqref{Eqn:Fluctuation} takes the form
\be\label{eqn:IRExpansion}
\ce_\pm(\r)=c_\pm^{in}(\o)\r ^{-\frac{i\o}{4\pi T}}\left(1+\co(\r)\right)+c^{out}_\pm(\o)\r^{\frac{i\o}{4\pi T}}\left(1+\co(\r)\right),
\ee
where $c^{in}_\pm$ and $c^{out}_\pm$ are arbitrary integration constants, multiplying the two linearly independent solutions -- respectively infalling and outgoing -- of the second order equations \eqref{Eqn:Fluctuation}. The retarded Green's functions are computed by setting the outgoing mode to zero at the horizon, while the advanced Green's functions correspond to setting the infalling mode to zero \cite{Son:2002sd}. It follows that the near horizon behavior of the response functions
\be
\car_\pm=\pa_r\log|\ce_\pm|=f^{1/2}e^{-A}\pa_\r\log|\ce_\pm|,
\ee
must be of the form
\be\label{nhR}
\car_\pm(\r,\o)=\left\{\begin{tabular}{ll}
$-i\o e^{-A(u_h)}(4\pi T\r)^{-1/2}+\co(\r^{1/2})$, & Retarded,\\
$i\o e^{-A(u_h)}(4\pi T\r)^{-1/2}+\co(\r^{1/2})$, & Advanced,\\
\end{tabular}\right.
\ee
depending on whether we want to compute the retarded or advanced 2-point functions. These conditions on the horizon determine the sole integration constants $\car_{(1)\pm}(\o)$ in the solution of the Riccati equations (\ref{Ric}). Note that these boundary conditions are invariant under the combined transformation $\o\to-\o$ and complex conjugation. This leads to the relation (\ref{reflex}) we mentioned earlier between the response functions. Another important remark is that the expansion (\ref{eqn:IRExpansion}) as written here is only strictly valid for nonzero $\o$. This is because the $\co(\r)$  terms that appear in this expansion contain inverse powers of $\o$, thus rendering the limit $\o\rightarrow0$ ill defined. We need to keep this fact in mind when determining the small $\o$ behavior in Section (\ref{Sw}).

\subsection{Universal large $\o$ solution}

Besides the UV and IR asymptotic expansions, the large and small frequency solutions of the fluctuation equations can be obtained analytically. These determine respectively the large and small frequency behavior of the response functions $\car_{(1)\pm}(\o)$, and hence of the conductivities.   

In the large frequency limit the coefficients (\ref{eqn:hs}) become	
\be
h_0=\frac{\o^2}{f^2}+\co(\o),\quad
h_1=\pa_u\log|\S f|+\co\left(\o^{-2}\right).
\ee
Assuming these expansions hold uniformly in $[0,u_h]$, the fluctuation equations (\ref{Eqn:Fluctuation}) to leading order in $\o$ simplify to 
\be
f\pa_u(\S f\ce_\pm')+\o^2\S\ce_\pm=0,
\ee
whose general solution is
\be
\ce_\pm=\left(c_1+\co(\o^{-1})\right)\exp\left(i\o\int_0^u\frac{\text{d}u'}{f(u')}\right)+\left(c_2+\co(\o^{-1})\right)\exp\left(-i\o\int_0^u\frac{\text{d}u'}{f(u')}\right),
\ee
with $c_1$ and $c_2$ arbitrary integration constants. From the near horizon behavior of $f$ in (\ref{fIR}) we deduce that infalling boundary conditions on the horizon corresponds to setting $c_{2}=0$. This yields
\be
\car_\pm=\pa_r\log|\ce_\pm|=-f^{1/2}e^{-A}\pa_u\log|\ce_\pm|=-i\o f^{-1/2}e^{-A}+\co(\o^0),
\ee
and so from (\ref{RUV}) we conclude that 
\be\label{lwR}
\car_{\pm(1)}(\o)=\left\{\begin{tabular}{ll}
$-i\o +\co(\o^0)$, & Retarded,\\
$i\o +\co(\o^0)$, & Advanced.\\
\end{tabular}\right.
\ee
From (\ref{sxy}) then follows that for large frequencies the conductivities behave as
\be\label{lw-sigma}\boxed{
	\s_{xx}(\o)=\s_{yy}(\o)=-\frac{2}{\k^2}\S_0+\co(\o^{-1}), \quad
	\s_{xy}(\o)=-\s_{yx}(\o)=-\frac{4}{\k^2}\P_0+\co(\o^{-1}).
}
\ee
Notice that this result, which agrees with e.g. eq. (6) in \cite{Hartnoll:2007ip} and eq. (67) in \cite{Alanen:2009cn} in the respective conventions, is independent of the particular background considered here and is thus universally valid for asymptotically locally AdS backgrounds of the form (\ref{Bans}), generalizing previous results by adding potentially running scalars with arbitrary potentials. Moreover, as expected, it is independent of the magnetic field and charge density, which are relevant deformations of the theory and hence do not affect the UV physics.

\subsection{Universal small $\o$ solution and its Pad\'e approximant}
\label{Sw}

A universal result for the response functions $\car_{\pm(1)}(\o)$ can also be obtained in the small frequency limit. From  (\ref{h0}) and (\ref{h1}), using the background equations (\ref{eom2}), we deduce that
\begin{subequations}
\begin{align}
	h_1(\o,H)&=2A'+\frac{1}{4H^2}\left(\frac{e^{2A}}{f\S}\right)'\o^2+\co\left(\o^{4}\right),\\
	h_0(\o,\pm H)&=-\frac{e^{-2A}}{f}\left(e^{2A}f'\right)'\mp\frac{e^{-2A}}{H}\left(\frac{e^{2A}\a'}{f}\right)'\o+\frac{\o^2}{f^2}\mp\frac{1}{4H^3}\frac{\a'}{f}\left(\frac{e^{2A}}{\S f}\right)'\o^3+\co\left(\o^4\right).
\end{align}
\end{subequations}
Inserting an expansion of the form 
\be\label{small-w-exp}
	\ce_\pm=\ce_\pm^{(0)}+\o\,\ce_\pm^{(1)}+\o^2\ce_\pm^{(2)}+\o^3\ce_\pm^{(3)}+\co\left(\o^4\right),
\ee
in the fluctuation equations (\ref{Eqn:Fluctuation}) leads to the following equations up to $\co(\o^3)$: 
\begin{subequations}
\begin{align}
&\co(\o^0): && \ce''^{(0)}_\pm+2A'\ce'^{(0)}_{\pm}-\frac{e^{-2A}}{f}\left(e^{2A}f'\right)'\ce^{(0)}_{\pm}=\frac{e^{-2A}}{f}\left(e^{2A}f^2\left(\frac{\ce_\pm^{(0)}}{f}\right)'\right)'=0,\\
&\co(\o^1): && \left(e^{2A}f^2\left(\frac{\ce_\pm^{(1)}}{f}\right)'\right)'=\pm\frac{f}{H}\left(\frac{e^{2A}\a'}{f}\right)'\ce_\pm^{(0)},\\
&\co(\o^2): && \left(e^{2A}f^2\left(\frac{\ce_\pm^{(2)}}{f}\right)'\right)'=-\frac{e^{2A}}{f}\ce_\pm^{(0)}-\frac{e^{2A}f}{4H^2}\left(\frac{e^{2A}}{f\S}\right)'\ce'^{(0)}_{\pm}\pm\frac{f}{H}\left(\frac{e^{2A}\a'}{f}\right)'\ce_\pm^{(1)},\\
&\co(\o^3): && \left(e^{2A}f^2\left(\frac{\ce_\pm^{(3)}}{f}\right)'\right)'=\pm\frac{e^{2A}}{4H^3}\a'\left(\frac{e^{2A}}{\S f}\right)'\ce_\pm^{(0)}-\frac{e^{2A}}{f}\ce_\pm^{(1)}-\frac{e^{2A}f}{4H^2}\left(\frac{e^{2A}}{f\S}\right)'\ce'^{(1)}_{\pm}\pm\frac{f}{H}\left(\frac{e^{2A}\a'}{f}\right)'\ce_\pm^{(2)}.
\end{align}
\end{subequations}
Note that the $\co(\o^0)$ equation is a homogeneous second order equation and hence the corresponding general solution contains two integrations constants. Moreover, the subleading in $\o$ equations are second order inhomogeneous equations but with the same homogeneous solutions as the $\co(\o^0)$ equations. Without loss of generality we will absorb all freedom in choosing the homogeneous solutions into the $\co(\o^0)$ solution by allowing the integration constants to potentially depend on the frequency. 

At $\co(\o^0)$ the two linearly independent solutions are 
\be
	\ce^{(0)}_\pm=c_{1}f+c_{2}fp,\quad 
	p(u)\equiv\int_{0}^u\frac{\text{d}u'}{e^{2A}f^2},
\ee
where the lower limit of integration in $p(u)$ has been chosen so that the integral is well defined. In the UV these behave as
\be
	f(u)=1-Mu^3+\co\left(u^4\right),\quad 
	p(u)=\frac{u^{3}}{3L^2}+\co\left(u^{4}\right),
\ee
while near the horizon we have
\be
	f(u)=4\pi T\r+\co\left(\r^2\right),\quad
	p(\r)=\frac{e^{-2A(u_h)}}{(4\pi T)^2\r}+\co\left(\log\r\right).
\ee
Both $f$ and $fp$ are therefore regular on the horizon and so we cannot a priori exclude any of the two solutions at this order in the frequency. As we shall see, we need to determine the expansion (\ref{small-w-exp}) up to $\co(\o^2)$ in order to find suitable linear combination of the integration constants $c_1$ and $c_2$ that corresponds to the desired IR boundary conditions. The inhomogeneous solution at $\co(\o^1)$ takes the form 
\be
	\ce^{(1)}_\pm=\pm\frac{f}{H}\int_0^u\frac{\text{d}\bar u}{e^{2A}f^2}\left(c_{1}\int_{u_h}^{\bar u}f^2\left(\frac{e^{2A}\a'}{f}\right)'\text{d}\bar{\bar u}+c_{2}\int_0^{\bar u}f^2p\left(\frac{e^{2A}\a'}{f}\right)'\text{d}\bar{\bar u}\right).
\ee
where again the lower limits of integration have been chosen so that the integrals are well defined. Near the horizon these the $\co(\o^1)$ solution behaves as
\be
	\ce^{(1)}_\pm\sim\pm\frac{(\wt Q-2H\P_h)}{H\S_h}\left(c_1\r\log\r-c_2\frac{e^{-2A(u_h)}}{(4\p T)^2}\log\r\right),
\ee
where $\P_h\equiv\P(\c(u_h))$. In the UV we have 
\be
\ce^{(1)}_\pm=c_1\left(\pm\frac{1}{H\S_0}\left(\wt Q -2H\P_0\right)u+\co\left(u^2\right)\right)+c_2\co(u^4).
\ee
Combining the $\co(\o^0)$ and $\co(\o^1)$ results, near the horizon the small frequency expansion behaves as
\be
\ce_\pm\sim c_1\left(4\p T\r+\o \co(\r\log\r)+\co(\o^2)\right)+c_2\left(\frac{e^{-2A(u_h)}}{4\p T}+\o\co(\log\r)+\co(\o^2)\right). 
\ee
It follows that for any value of the integration constants $c_1$ and $c_2$ this expansion breaks down when $-|\o|\log\r \approx 1$. However, as long as $\r>> e^{-1/|\o|}$ the expansion (\ref{small-w-exp}) is well defined. Going one order higher, the $\co(\o^2)$ inhomogeneous solution takes the form
\be
	\ce^{(2)}_\pm=-f\int_0^u\frac{\text{d}\bar u}{e^{2A}f^2}\int_{u_*}^{\bar u}\left(\frac{e^{2A}}{f}\ce_\pm^{(0)}+\frac{e^{2A}f}{4H^2}\left(\frac{e^{2A}}{f\S}\right)'\ce'^{(0)}_{\pm}\mp\frac{f}{H}\left(\frac{e^{2A}\a'}{f}\right)'\ce_\pm^{(1)}\right)\text{d}\bar{\bar u},
\ee
where the lower limit of the first integration, $0<u_*<u_h$, is an arbitrary point between the horizon and the boundary since the integrand diverges both at the horizon and the boundary. Near the horizon this behaves as 
\be
\ce^{(2)}_\pm=c_1\left(-\frac{e^{2A(u_h)}}{4H^2\S_h}\log\r+\co(1)\right) + c_2\co((\log\r)^2), 
\ee
where $\S_h\equiv \S(\f(u_h))$, while in the UV
\be
\ce^{(2)}_\pm=\frac{3L^2}{4H^2\S_0}\left(M c_1 - \frac{1}{3L^2}c_2\right)u+\co(u^2).
\ee

At this point we can determine the relation between the integration constants $c_1$ and $c_2$ corresponding to ingoing boundary conditions on the horizon. Provided $\r>> e^{-1/|\o|}$, the ingoing solution in (\ref{eqn:IRExpansion}) can be expanded as  
\be
\ce_\pm=c_\pm^{in}(\o)\left(1-\frac{i}{4\pi T}\o\log\r+\co(\o\log\r)^2)\right).
\ee
On the other hand, in the same limit the small $\o$ solution we found behaves as 
\be
	\ce_\pm\sim c_2\frac{e^{-2A(u_h)}}{4\p T}\left(1\mp\o\frac{(\wt Q-2H\P_h)}{4\p TH\S_h}\log\r-\frac{c_1}{c_2}\o^2\frac{e^{4A(u_h)}4\p T}{4H^2\S_h}\log\r + \o^2\co((\log\r)^2)\right),
\ee 
and hence we conclude that 
\be\boxed{
c_1=c_2\frac{4He^{-4A(u_h)}}{(4\p T)^2\o}\left(i H\S_h \mp(\wt Q-2H\P_h)\right).
}
\ee
Applying the same argument to subleading orders in the expansion determines the point $u_*$ in $\ce^{(2)}_\pm$. 

Finally, observing that
\be
\ce'^{(1)}_\pm\mp\frac{\a'}{Hf}\ce^{(0)}_\pm=\co(u^3),
\ee
as $u\to 0$, it is straightforward to show that $\ce^{(3)}_\pm=\co(u^2)$ and hence
in the UV we get
\be
\ce_\pm= c_1\left(1\pm\frac{\o}{H\S_0}\left(\wt Q -2H\P_0\right)u+\frac{3L^2\o^2}{4H^2\S_0}\left(M - \frac{1}{3L^2}\frac{c_2}{c_1}\right)u+\co(u^2,\o^4)\right).
\ee
It follows that 
\be
\car_\pm=\pa_r\log|\ce_\pm|=-f^{1/2}e^{-A}\pa_u\log|\ce_\pm|\sim\frac{u}{L}\left(\mp\frac{\o}{H\S_0}\left(\wt Q -2H\P_0\right)-\frac{3L^2\o^2}{4H^2\S_0}\left(M - \frac{1}{3L^2}\frac{c_2}{c_1}\right)\right),
\ee
from which we read off the response functions
\be\label{response-small-w}\boxed{
\car_{(1)\pm}(\o)=\mp\frac{\o}{H\S_0}\left(\wt Q -2H\P_0\right)-\frac{3ML^2\o^2}{4H^2\S_0}+\frac{(4\p T)^2e^{4A(u_h)}\o^3}{16\S_0 H^3(iH\S_h\mp(\wt Q-2H\P_h))}+\co(\o^4).}
\ee
Notice that the leading $\co(\o)$ part of this expression is universal and independent of the IR boundary conditions as long as $c_1\neq 0$. This temperature independent part of the response functions determines the universal DC Hall conductivity and it is crucial to remove the apparent pole at $\o=0$ in the conductivities (\ref{sxy}). From (\ref{sxy}) it follows that the small frequency behavior of the conductivities is
\begin{subequations}\label{sw-conductivities}
\begin{align}
	&\s_{xx}=\s_{yy}=\frac{3ML^2}{2H^2\k^2}i\o-\frac{(4\p T)^2e^{4A(u_h)}\S_h\o^2}{8H^2\k^2(H^2\S_h^2+(\wt Q-2H\P_h)^2)}+\co\left(\o^3\right), \\
	&\s_{xy}=-\s_{yx}=\frac{\rho}{H}-\frac{(4\p T)^2e^{4A(u_h)}(\wt Q-2H\P_h)\o^2}{8H^3\k^2(H^2\S_h^2+(\wt Q-2H\P_h)^2)}+\co\left(\o^3\right).
\end{align}
\end{subequations}
Given that $M$ is related to the temperature $T$, the second order correction in $\o$ of the response function brings in a temperature dependence in the leading nontrivial correction of the transport coefficients. From the Ward identities (\ref{2pt-fns-WIs-omega}) then we obtain 
\begin{subequations}
\begin{align}
	&\langle T_{tx}(\om)J_x(-\om)\rangle=\langle T_{ty}(\om)J_y(-\om)\rangle=-\frac{(4\p T)^2e^{4A(u_h)}(\wt Q-2H\P_h)\o^2}{8H^2\k^2(H^2\S_h^2+(\wt Q-2H\P_h)^2)}+\co\left(\o^3\right), \\
	&\langle T_{tx}(\om)J_y(-\om)\rangle=-\langle T_{ty}(\om)J_x(-\om)\rangle=-\frac{3ML^2}{2H\k^2}i\o+\frac{(4\p T)^2e^{4A(u_h)}\S_h\o^2}{8H\k^2(H^2\S_h^2+(\wt Q-2H\P_h)^2)}+\co\left(\o^3\right),\\
	&\langle T_{tx}(\om)T_{tx}(-\om)\rangle=\langle T_{ty}(\om)T_{ty}(-\om)\rangle=-\e+\frac{3ML^2}{2\k^2}+\frac{(4\p T)^2e^{4A(u_h)}\S_hi\o}{8\k^2(H^2\S_h^2+(\wt Q-2H\P_h)^2)}+\co\left(\o^2\right),\\
	&\langle T_{tx}(\om)T_{ty}(-\om)\rangle=-\langle T_{ty}(\om)T_{tx}(-\om)\rangle=\frac{(4\p T)^2e^{4A(u_h)}(\wt Q-2H\P_h)i\o}{8H\k^2(H^2\S_h^2+(\wt Q-2H\P_h)^2)}+\co\left(\o^2\right).
\end{align}
\end{subequations}
These agree with equations (49)-(51) in \cite{Hartnoll:2007ai} for the  dyonic Reissner-Nordstr\"om black hole, but the present derivation holds even for backgrounds with non-trivial scalar profiles.

\subsection*{Pad\'e approximant}

The small frequency expansion (\ref{response-small-w}) of the response functions can be considerably improved by means of a Pad\'e approximant, which is capable of capturing certain poles of the conductivities. It turns out there are two different Pad\'e approximants that correctly capture the behavior of the conductivities at different limits of parameter space. The two Pad\'e approximants correspond to two different terms dominating the response function (\ref{response-small-w}). Writing (\ref{response-small-w}) as
\be
\car_{(1)\pm}(\o)=\car_{(1)\pm}^{[1]}\o+\car_{(1)\pm}^{[2]}\o^2+\car_{(1)\pm}^{[3]}\o^3+\co(\o^4),
\ee
the two relevant Pad\'e approximants are
\be\label{pades}
\car_{(1)\pm}(\o)=\left\{\begin{matrix} \frac{\o(\car_{(1)\pm}^{[1]})^2}{\car_{(1)\pm}^{[1]}-\car_{(1)\pm}^{[2]}\o}+\car_{(1)\pm}^{[3]}\o^3+\co(\o^4), & &\car_{(1)\pm}^{[1]}\o\sim \car_{(1)\pm}^{[2]}\o^2 >>\car_{(1)\pm}^{[3]}\o^3,\\
\car_{(1)\pm}^{[1]}\o+\frac{\o^2(\car_{(1)\pm}^{[2]})^2}{\car_{(1)\pm}^{[2]}-\car_{(1)\pm}^{[3]}\o}+\co(\o^4), & &\car_{(1)\pm}^{[1]}\o << \car_{(1)\pm}^{[2]}\o^2 \sim \car_{(1)\pm}^{[3]}\o^3.\\
\end{matrix}\right.
\ee
The first Pad\'e approximant is a good approximation to the response functions at low temperature, i.e. near extremality, and leads to a pole at 
\be\label{poles-1}
\o_{*\pm}\approx \pm \frac{4H(\wt Q-2H\P_0)}{3L^2M},
\ee
which agrees with (81) of \cite{Hartnoll:2007ip} in the small $H$ limit keeping $\o/H$ fixed. The second Pad\'e approximant leads to poles at 
\be\label{poles-2}
\o_{*\pm}\approx-\frac{12L^2MH}{(4\p T)^2}e^{-4A(u_h)}(iH\S_h\mp (\wt Q-2H\P_h)),
\ee 
which agrees with (74) of \cite{Hartnoll:2007ip} in the hydrodynamic limit where $H$ and $\wt Q$ are sent to zero keeping $H/\o$ and $\wt Q/\o$ fixed. As we shall see in Section \ref{RN} these Pad\'e approximants capture the poles closest to the origin of the complex $\o$ plane to very good approximation.

\subsection{Numerical solution of the Riccati equation}
\label{RSolve}

The Riccati equations (\ref{Ric}), or equivalently the second order linear equations (\ref{Eqn:Fluctuation}), are not integrable in general and so one needs to solve these equations numerically. One can solve either the Riccati equations or the second order equations, but the fact that the Riccati equations directly determine the response function by imposing only IR boundary conditions is a clear advantage compared to the second order equations where one must keep track of the arbitrary source.\footnote{It is worth pointing out that the boundary value problem for the second order equations (\ref{Eqn:Fluctuation}) is well defined, despite the apparent singularity of the the coefficients $h_0$ and $h_1$ at the zeros of $\Om = \o^2 -4 H^2 \S f e^{-2A}$. Indeed, it is straightforward to see that $h_0$ and $h_1$ have a simple pole at the zeros of $\O$ and therefore these correspond to regular singular points of the second order equations \cite{Edalati:2009bi, Lippert:2014jma}.} Moreover, the relation $\car_+(\omega)=\car_-(-\omega)^*$ implies that instead of computing $\car_+$ and $\car_-$ for positive $\omega$ we can compute only $\car_+(\omega)$ for all $\omega$. Thus, in order to compute the conductivities one only needs to solve one first order ordinary differential equation which is simpler than solving a set of coupled second order differential equations as was previously done in the literature. 

In order to solve the Riccati equations (\ref{Ric}) numerically it is convenient to introduce the new dependent variables  
\be\label{Theta}
\Th_\pm=\O^{-1}\left[\S H e^A f^{1/2}\car_\pm\pm\o\left(\wt Q-2\P H\right)\right],
\ee
in terms of which the Riccati equations take the simpler form
\be\label{smooth-riccati}\boxed{
f\left(H\Th'_\pm+4H^2e^{-2A}\Th_\pm^2\right)-\S H^2-\S^{-1}\left(\o\Th_\pm\mp\left(\wt Q-2\P H\right)\right)^2=0.}
\ee
For retarded Green's functions the horizon condition (\ref{nhR}) translates to 
\be
\Th_\pm\sim \o^{-1}\left(-i\S_h H\pm \left(\wt Q-2\P_h H\right)\right),
\ee
on the horizon, while in the UV
\be
\Th_\pm\sim \o^{-2}\left(H\S_0 \car_{(1)\pm}\pm\o\left(\wt Q-2\P_0 H\right)\right),
\ee
from which one can read off the renormalized response functions $\car_{(1)\pm}$. Equation \eqref{smooth-riccati} can be integrated using any standard solver for ordinary differential equations. However, the horizon is not a regular singular point and imposing boundary conditions on the horizon requires some care in the numerical analysis. A standard technique is to use a Taylor expansion in the vicinity of the horizon, and match the numerical solution at some small distance away from the horizon. However, we found that even a near-horizon expansion to $\co(u_h-u)^4$ was not sufficient for stabilizing the numerics for \eqref{smooth-riccati} with NDSolve in Mathematica.\footnote{We also solved \eqref{smooth-riccati} using the ODE integrator from the Python library scipy, finding the same results.} Instead, using a Pad\'e approximant based on the near-horizon expansion to $\co(u_h-u)^2$ in order to impose the IR boundary condition sufficiently away from the horizon worked very well with NDSolve.

\section{Example: Dyonic Reissner-Nordstr\"om black hole} 
\label{RN}

In this section we will apply the above general analysis to the dyonic Reissner-Nordstr\"om black hole (\ref{RN_d}), which was first studied in \cite{Hartnoll:2007ai,Hartnoll:2007ip}. The dyonic Reissner-Nordstr\"om black hole is a solution of the background equations (\ref{eom2}) either for a model with constant potentials, i.e. $V(\f,\c)=-6/L^2$, $\S(\f)=\S_0$, $Z(\f)=Z_0$ and $\P(\c)=\P_0$ as in \cite{Hartnoll:2007ai,Hartnoll:2007ip}, or in a theory with arbitrary potentials that admits asymptotically AdS solutions, provided the scalars are set to their vacuum AdS value, which can be taken without loss of generality to be zero. Setting the scalars to zero is a consistent truncation of the background equations (\ref{eom2}) provided $V(\f,\c)$, $\S(\f)$ and $\P(\c)$ don't have a linear term in their respective variables in a Taylor expansion around $\f=\c=0$. However, while in the theory with constant potentials the dyonic Reissner-Nordstr\"om black hole -- as we shall demonstrate momentarily -- is the only asymptotically AdS solution of the form (\ref{Bans}), this is not generically the case in the theory with non constant potentials. In particular, a generic theory will admit in addition to the dyonic Reissner-Nordstr\"om black hole a hairy black hole with the same charges and a phase transition between the two solutions will generically occur at some critical temperature. 

In order to show that the dyonic Reissner-Nordstr\"om black hole is the only asymptotically AdS solution of the form (\ref{Bans}) in the case of constant potentials, let us consider the background equations (\ref{eom2}), which in  this case reduce to 
\begin{subequations}
\begin{align}
		& (e^{2A}fA')'-\frac{3e^{4A}}{L^2}+\S_0\left(H^2+\a'^2\right)=0, \label{eqn:eomA1} \\
		& 2\left(A''-A'^2\right)+\f'^2_B+Z_0\c'^2_B=0, \label{eqn:eomA2} \\
		& (e^{2A}f')'-4\S_0\left(H^2+\a'^2\right)=0, \\
		& (e^{2A}f\f'_B)'=0, \\
		& (e^{2A}f\c'_B)'=0, \\
		& (\S_0\a'+2\P_0 H)'=0.
	\end{align}
\end{subequations}
The general solution of the last equation is
\be
	\a(u)=\a_0-Qu,
\ee
where the constant $Q$ is related to $\wt Q$ that we introduced before by $\S_0Q=2\P_0H-\wt Q$. Moreover, the scalar fields can be expressed in terms of $f(u)$ and $A(u)$ as,
\be
		\f_B(u)=\f_0+C_{\f}\int_0^u\frac{\text{d}u'}{e^{2A(u')}f(u')}, \quad
		\c_B(u)=\c_0+C_{\c}\int_0^u\frac{\text{d}u'}{e^{2A(u')}f(u')}.
\ee
where $\f_0$, $\c_0$, $C_\f$ and $C_\c$ are integration constants, while using the solution for $\a$, the blackening factor can be expressed in terms of $A(u)$ as
\be
	f(u)=f_0+\int_0^u\text{d}u'\left(3M+4\S_0\left(H^2+Q^2\right)u'\right)e^{-2A(u')},
\ee
where $M$ is another integration constant. Finally, eliminating $f$ from the first two equations we obtain a second order equation for the warp factor $A(u)$, which is the only remaining unknown quantity, namely 
\be
\left(C_\f^2+Z_0 C_\c^2\right)\left(A''+2A'^2\right)^2+2\left(A''-A'^2\right)\left(3MA'+(4A'u+1)\S_0(H^2+Q^2)-\frac{3e^{4A}}{L^2}\right)^2=0.
\ee
However, since we are interested in asymptotically AdS solutions we must have 
$ A\sim \log (L/u)$ as $u\to 0$. The last equation then requires that  $C_{\f}=C_{\c}=0$ for asymptotically AdS solutions and hence the scalars must be necessarily constant and so, without loss of generality, we can set them to zero. The resulting solution is the dyonic Reissner-Nordstr\"om black hole (\ref{RN_d}) in four dimensions for which
\begin{align}
	\f_B(u)&=0, & A(u)&=\log(L/u), & f(u)&=1-Mu^{3}+\frac{\S_0}{L^2}\left(H^2+Q^2\right)u^{4}, \NO \\
	\c_B(u)&=0, & \a(u)&=\a_0-Qu.
\end{align}
The horizon radius $u_h$ is the smallest positive root of the quartic equation $f(u_h)=0$, i.e. 
\be \label{M}
Mu_h^{3}=1+\frac{\S_0}{L^2}\left(H^2+Q^2\right)u_h^{4}.
\ee
This expression can be used to express the mass $M$ in terms of the horizon radius $u_h$, namely
\be
	f(u)=\left(1-\frac{u^{3}}{u_h^{3}}\right)-\frac{\S_0}{L^2}\left(H^2+Q^2\right)u_hu^{3}\left(1-\frac{u}{u_h}\right).
\ee
Finally, the Hawking temperature can be found by the usual argument demanding that the Euclidean section be free of conical defects, leading to the expression 
\be \label{T}
	T=-\frac{f'(u_h)}{4\pi}=\frac{1}{4\pi}\left(\frac{3}{u_h}- \frac{\S_0}{L^2}\left(H^2+Q^2\right)u_h^{3} \right),
\ee
while the energy and pressure densities, respectively (\ref{energy-density}) and (\ref{pressure-density}), become
\be
\e=\frac{ML^2}{\k^2},\quad \cp=-\frac{ML^2}{2\k^2}, \quad \e+2\cp=0.
\ee

\begin{figure}
\centering
\includegraphics[scale=0.46]{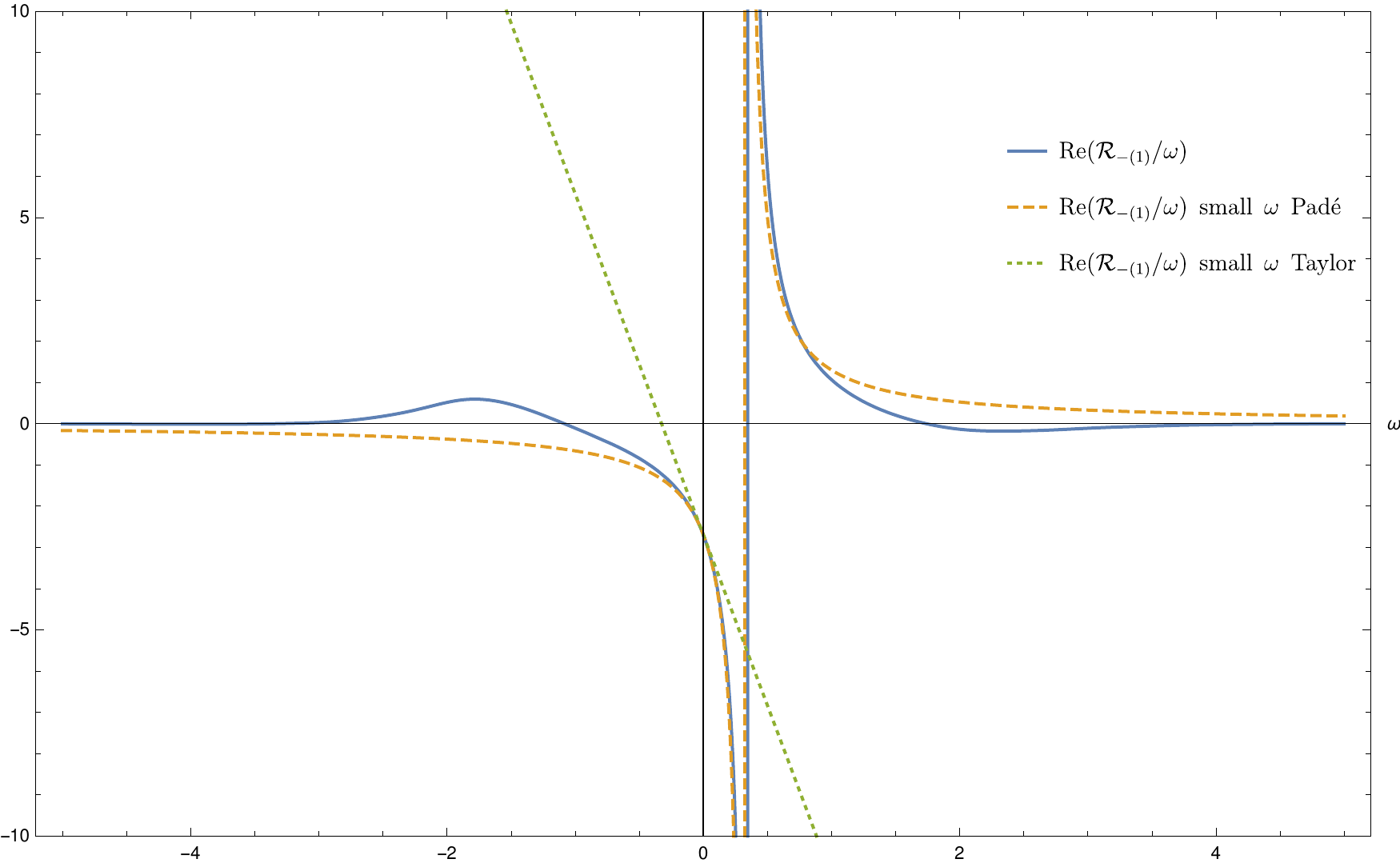}
\includegraphics[scale=0.46]{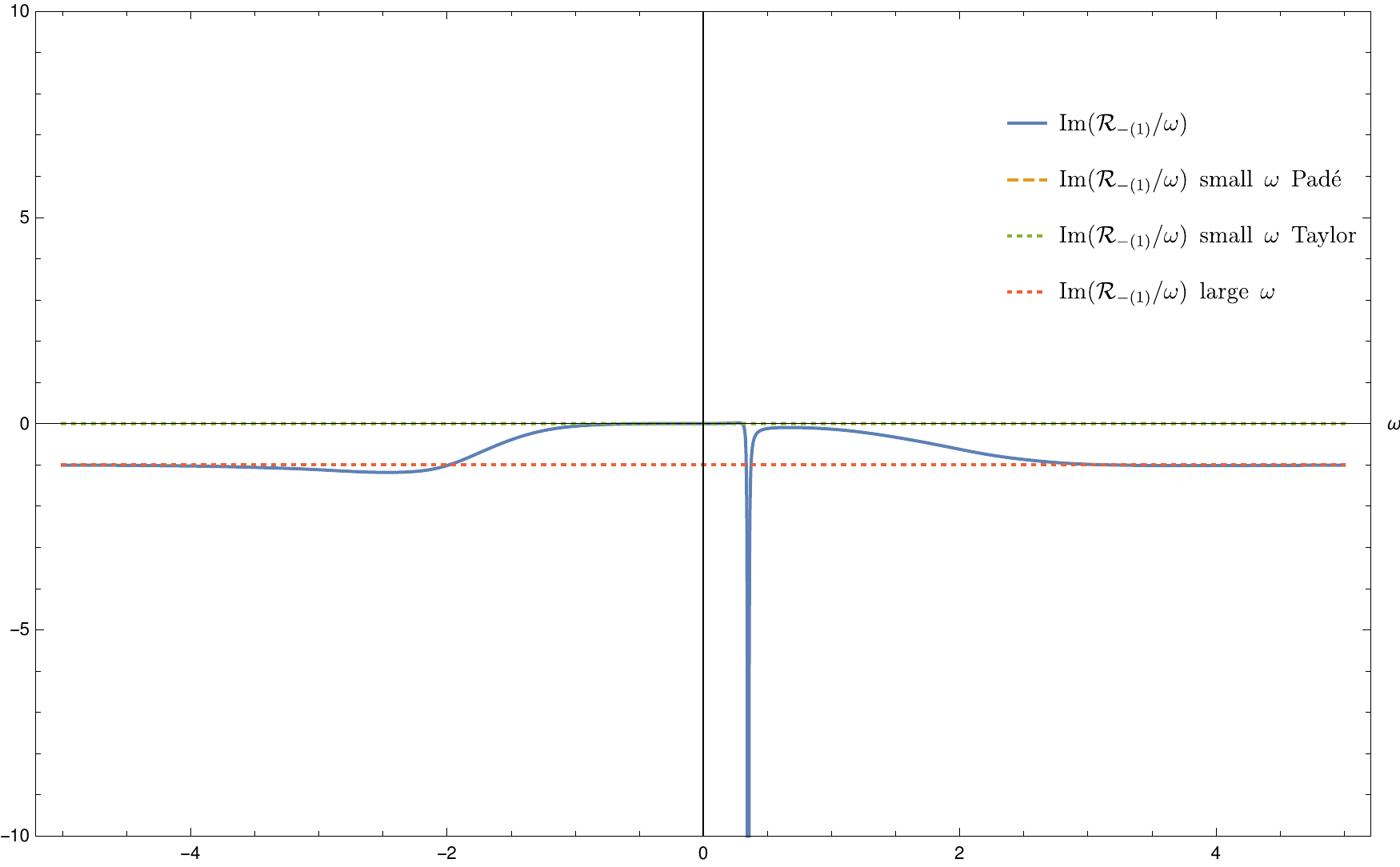}
\caption{Plots of the real and imaginary parts of the response function $\car_{-(1)}(\o)$ as a function of a real frequency $\o$ for $H=0.6$, $Q=1.6156$, $T=0.00239$, together with the small frequency expansion (\ref{response-small-w}), the first Pad\'e approximant in (\ref{pades}), as well as the asymptotic behavior (\ref{lwR}). Both the location of the horizon $u_h$ and the AdS radius $L$ are set to 1. As is evident from this plot, the Pad\'e approximant is a drastic improvement of the small $\o$ expansion in this regime of parameter space. In particular, it captures very well the pole in the Real part of $\car_{-(1)}(\o)$, but not the delta function at the same frequency (as predicted by the Kramers-Kronig relations) in the imaginary part.}
\label{fig1} 
\end{figure}

\begin{figure}
\centering
\includegraphics[scale=0.46]{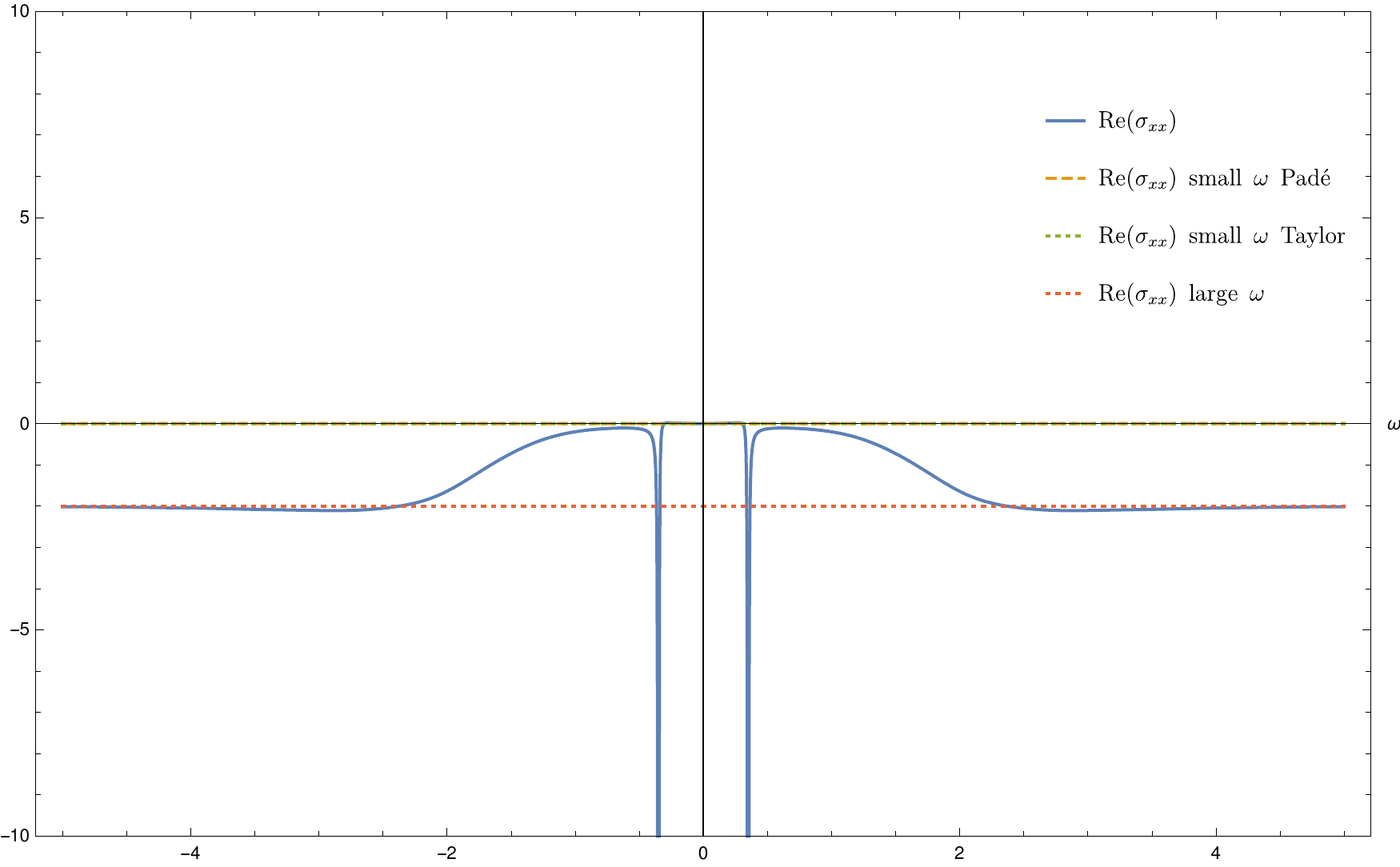}
\includegraphics[scale=0.46]{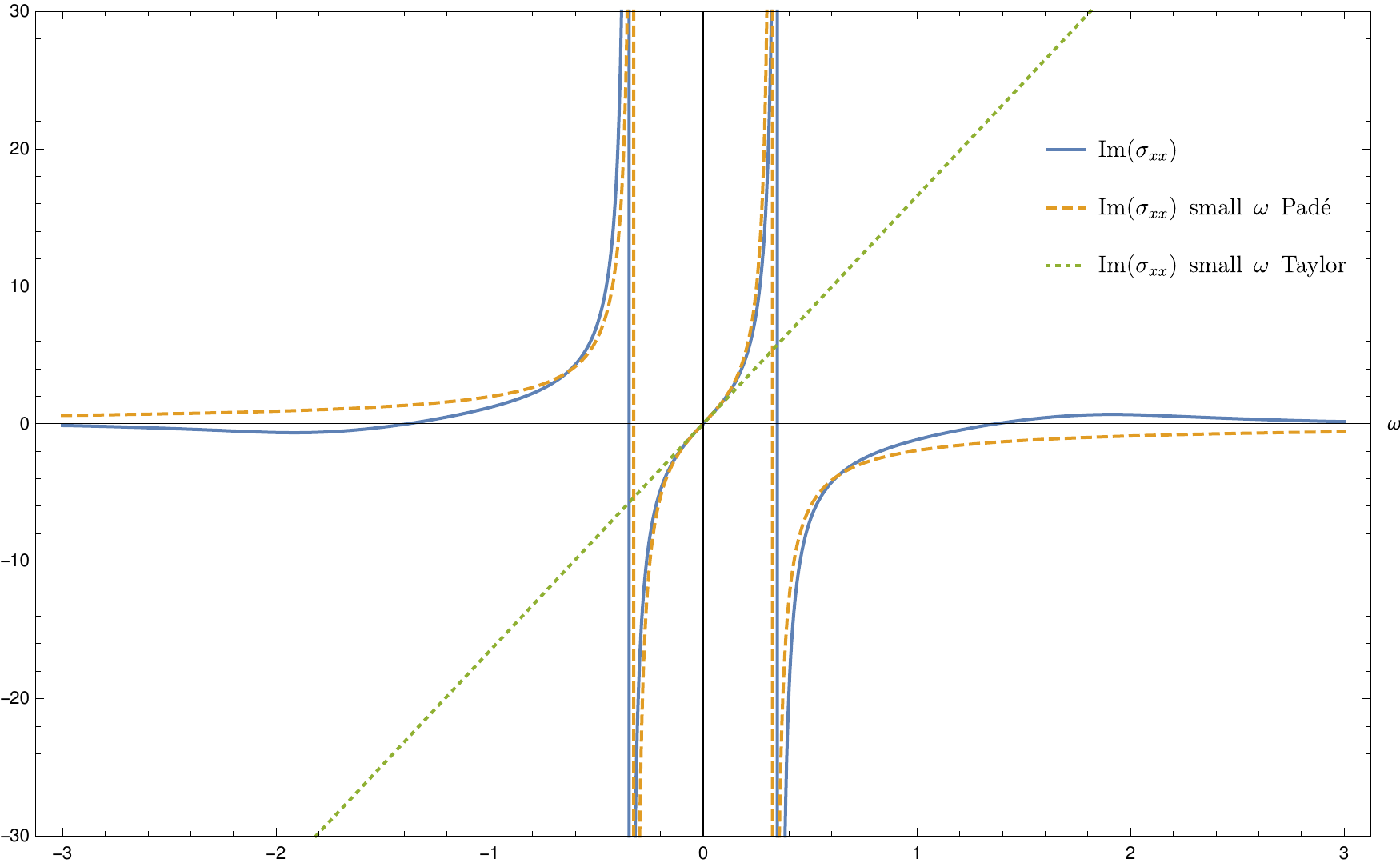}
\includegraphics[scale=0.46]{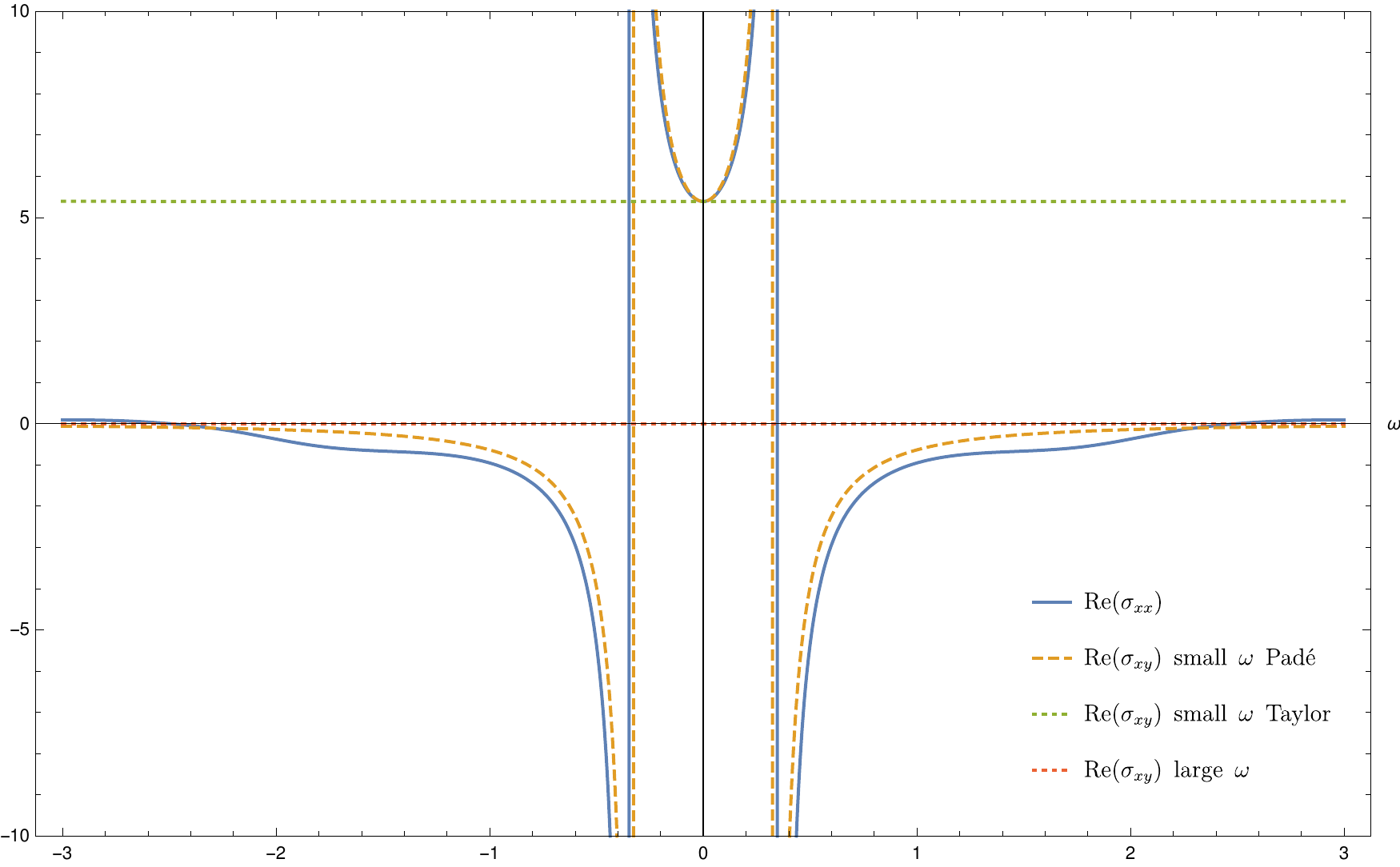}
\includegraphics[scale=0.46]{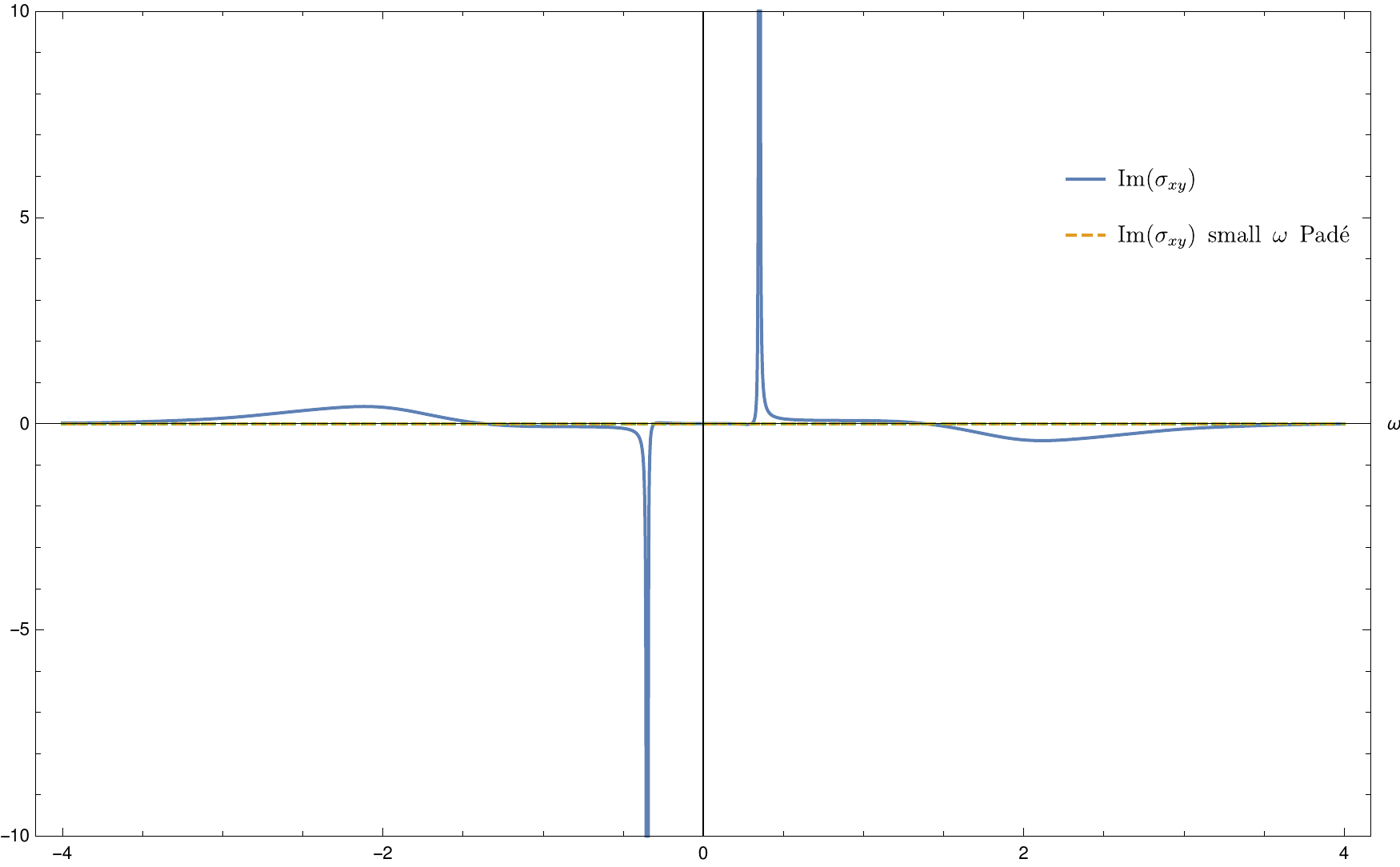}
\caption{Plots of the real and imaginary parts of the conductivities $\s_{xx}$ and $\s_{xy}$ as a function of a real frequency $\o$ for $H=0.6$, $Q=1.6156$, $T=0.00239$, together with the small frequency expansion (\ref{response-small-w}), the first Pad\'e approximant in (\ref{pades}), as well as the large $\o$ asymptotic behavior (\ref{lw-sigma}). Again the location of the horizon $u_h$ and the AdS radius $L$ are set to 1. The location of the poles in $\Im\s_{xx}$ and $\Re\s_{xy}$ are well approximated by the Pad\'e approximant and are given by (\ref{poles-1}). However, the delta functions in $\Re\s_{xx}$ and $\Im\s_{xy}$, which are related to the aforementioned poles via the Kramers-Kroning relations, are not captured by the Pad\'e approximant. As $H\to 0$ the delta functions move towards the origin of the complex $\o$ plane, giving rise to the well known Drude peak, which reflects translation invariance. The fact that the poles (and hence the delta functions) in these plots are away from the origin is a consequence of broken translation invariance due to the magnetic field \cite{Hartnoll:2007ai,Hartnoll:2007ip}. A similar effect occurs when translation invariance is broken by impurities \cite{Horowitz:2012ky}.}
\label{fig2}
\end{figure}

\begin{figure}
\centering
\includegraphics[scale=0.3]{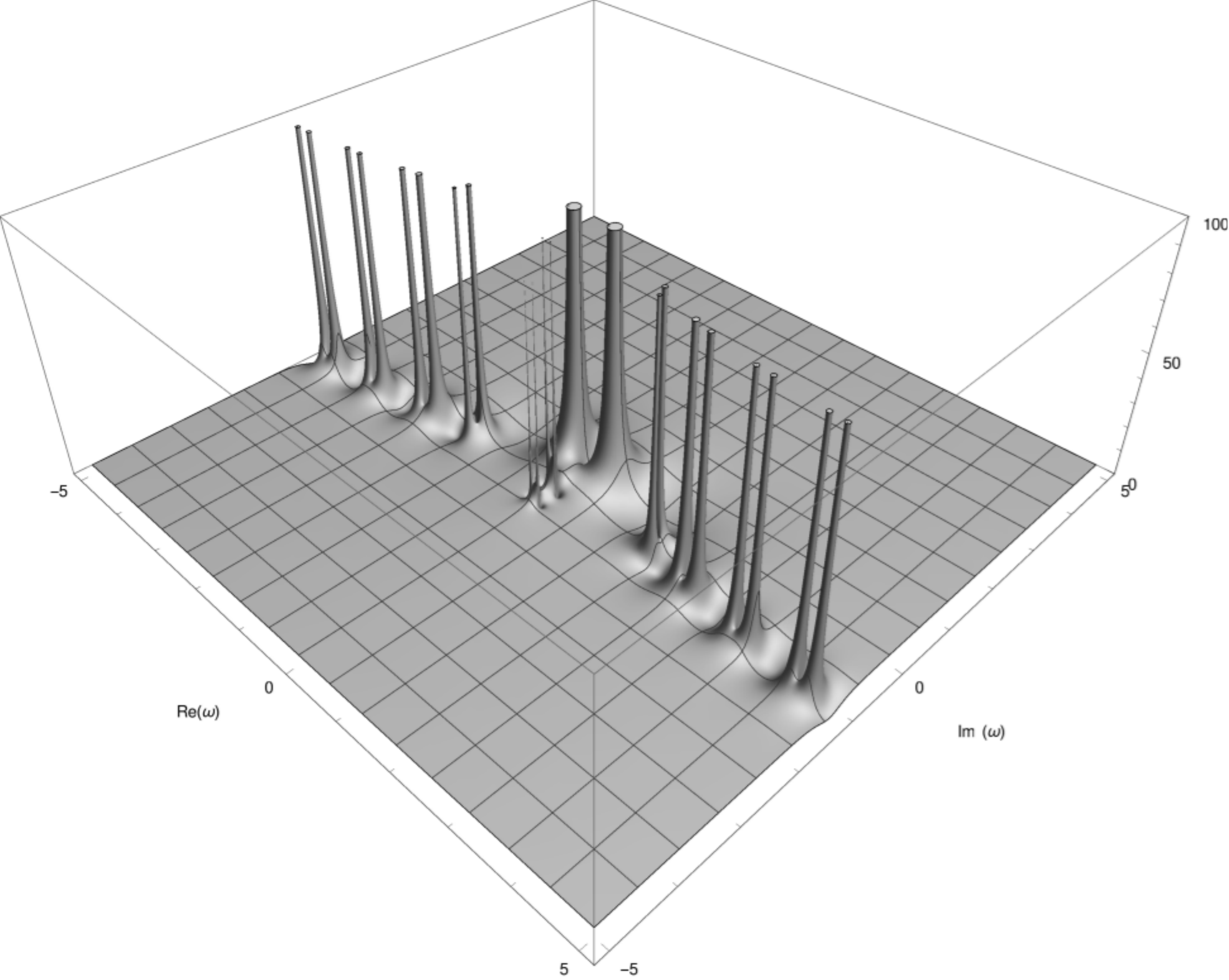}
\includegraphics[scale=0.3]{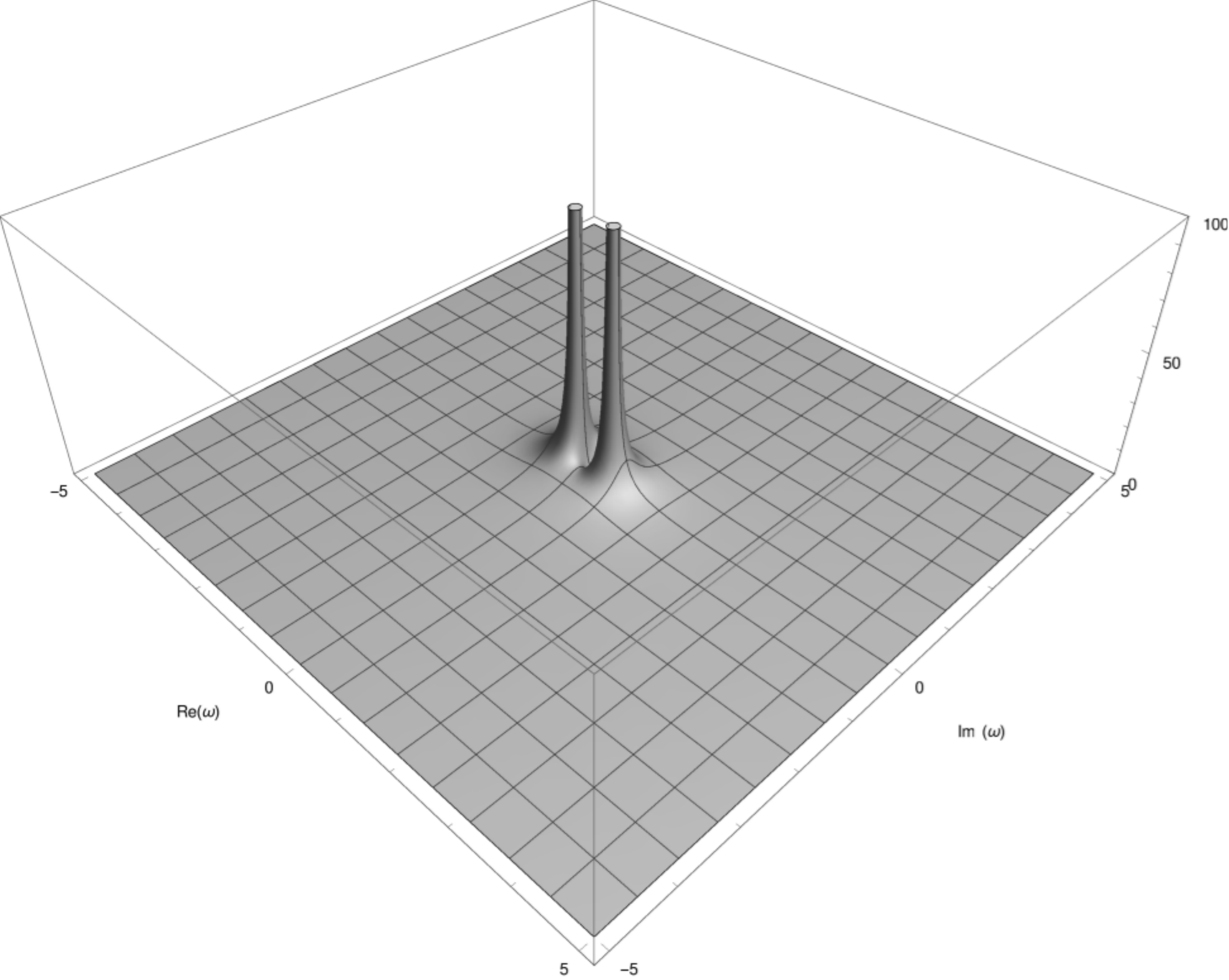}
\caption{Plot of $|\s_{xx}|^2$ as a function of a complex $\o$ for $H=0.6$, $Q=1.6156$, $T=0.00239$ (left), together with the first Pad\'e approximant in (\ref{pades}) (right). As is evident from these plots, the Pad\'e approximant captures the poles closest to the origin extremely accurately.}
\label{fig3} 
\end{figure}

\begin{figure}
\centering
\includegraphics[scale=0.3]{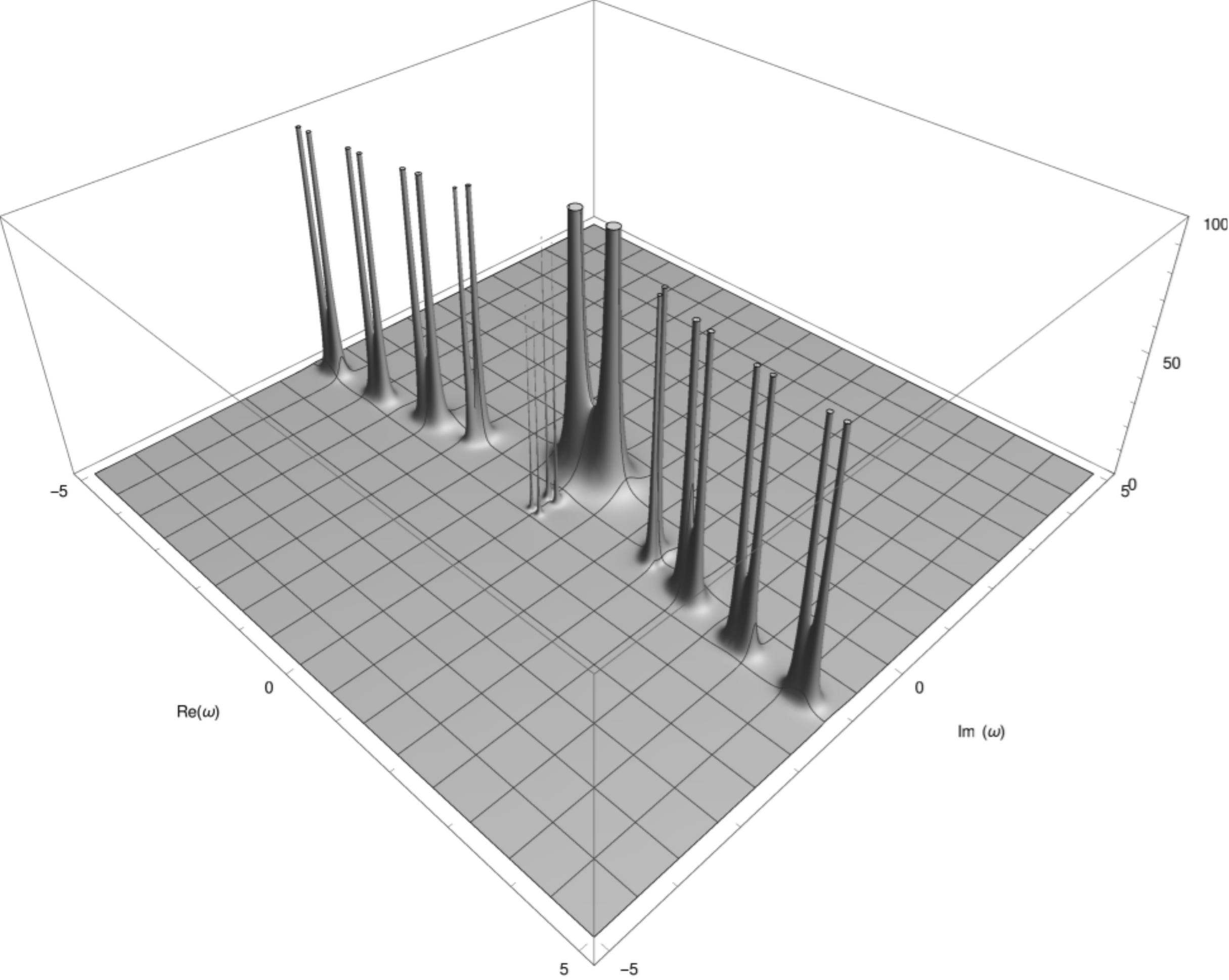}
\includegraphics[scale=0.3]{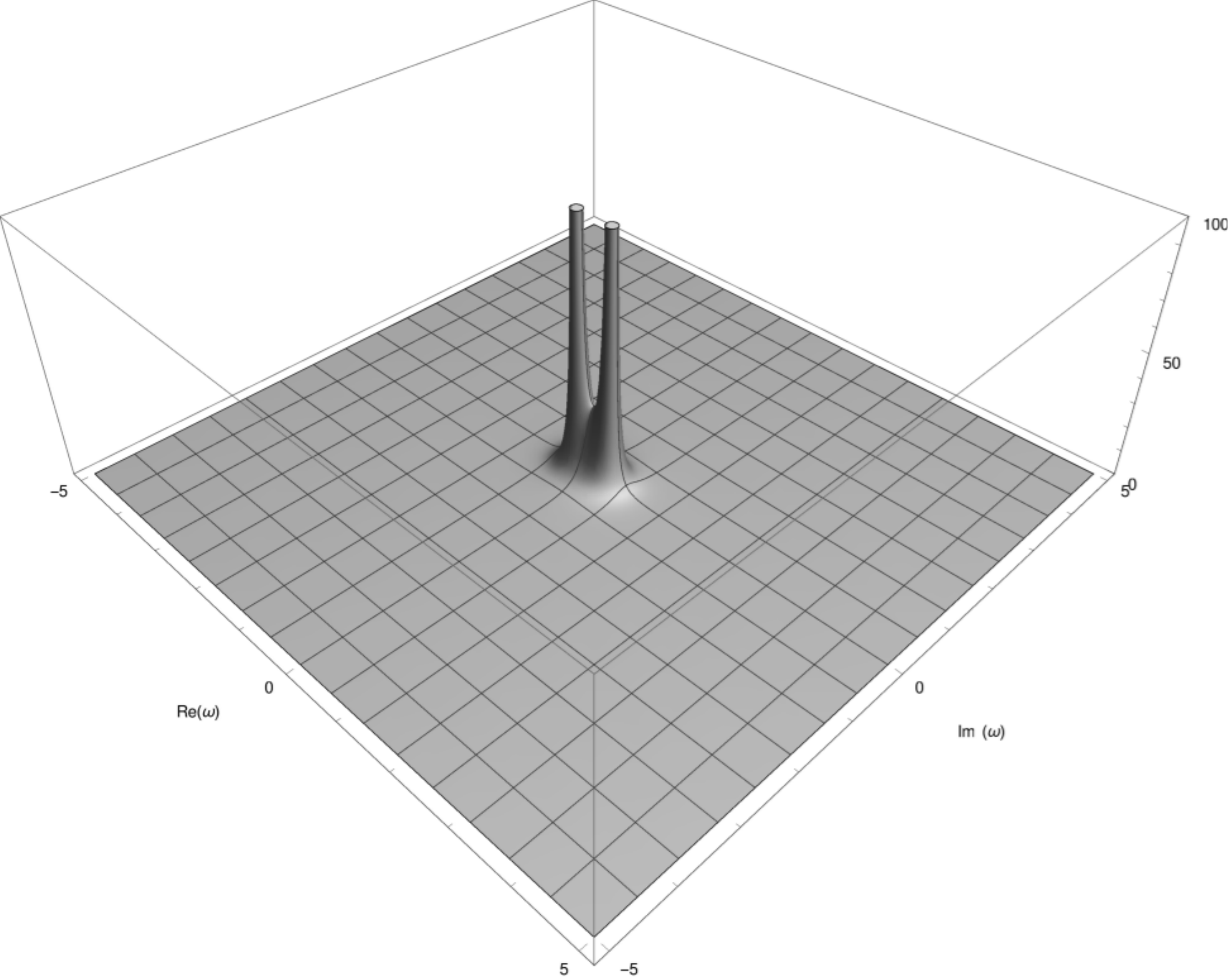}
\caption{Plot of $|\s_{xy}|^2$ as a function of a complex $\o$ for $H=0.6$, $Q=1.6156$, $T=0.00239$ (left), together with the first Pad\'e approximant in (\ref{pades}) (right).}
\label{fig4} 
\end{figure}

\begin{figure}
\centering
\includegraphics[scale=0.65]{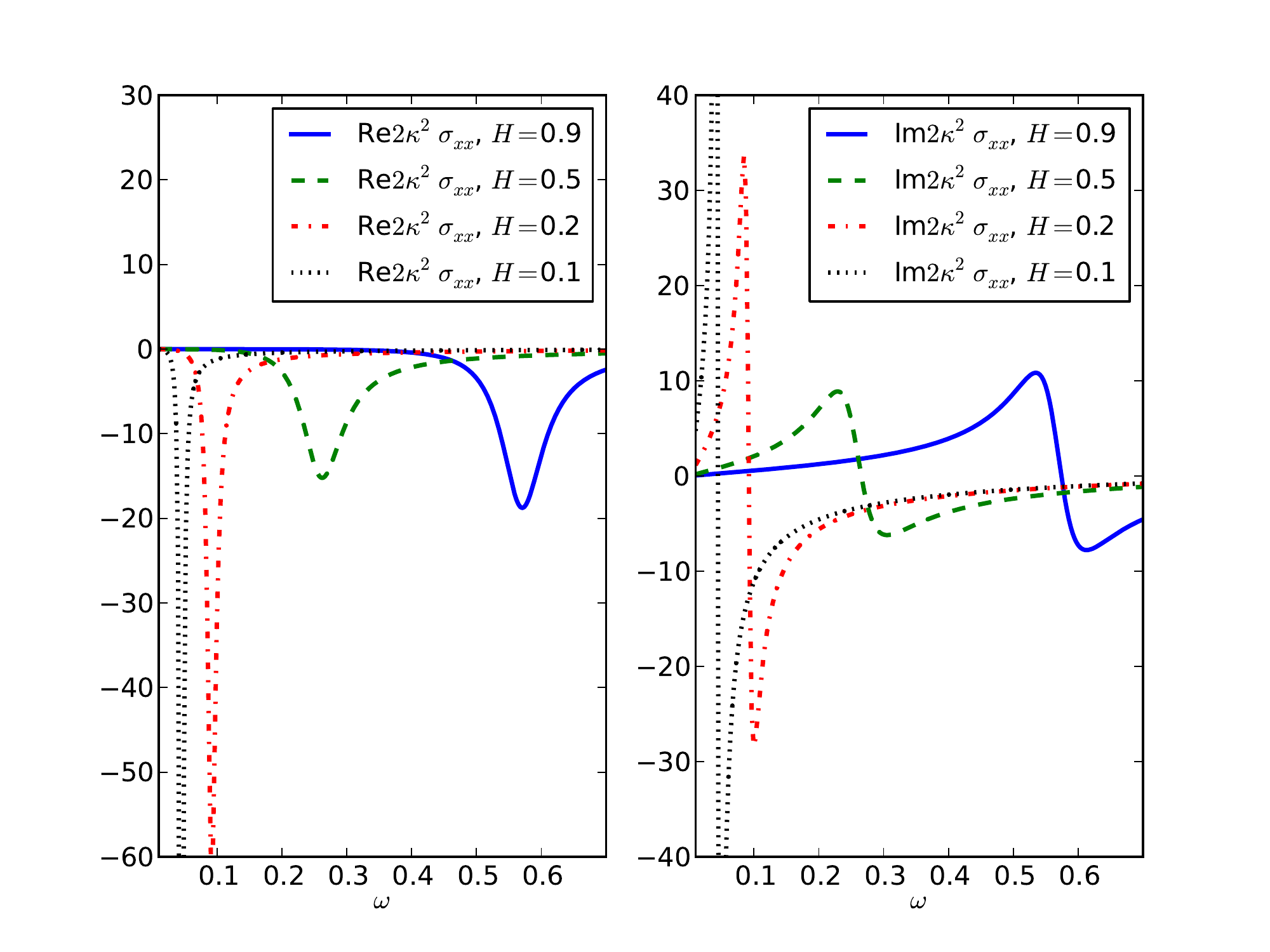}
\caption{The parameters used in these plots are $Q=1$, $M=3$, $\S_0=1$, $\P_0=0$ and $L=1$. {\bf Left panel:} $\Re \s_{xx}$ as a function of $\o$ for several values of the magnetic field $H$. As $H$ decreases, the real part of $\s_{xx}$ tends to behave as a delta function keeping the area below the curve fixed but rising sharply and becoming steeper. This behavior occurs for small but nonzero $\o$ but for even smaller $\o$ it vanishes steeply. {\bf Right panel:} $\Im \s_{xx}$ as a function of $\o$ for several values of the magnetic field $H$. As $H$ decreases, the imaginary part of $\s_{xx}$ tends to behave as $\sim \frac{1}{\o}$ for small but nonzero $\o$ but for even smaller $\o$ it vanishes steeply. We conclude that the magnetic field regulates the delta function which is present at $\o=0$ for zero magnetic field in which case there is a Drude peak involved. In particular, the Drude peak predicts $\Im \s_{xx} \sim 1/\o$ and $\Re \s_{xx} \sim \d(\o)$ which is what we find by taking a sequence of decreasing $H$ configurations.}
\label{R0vsH}
\end{figure}

\begin{figure}
\centering
\includegraphics[scale=0.46]{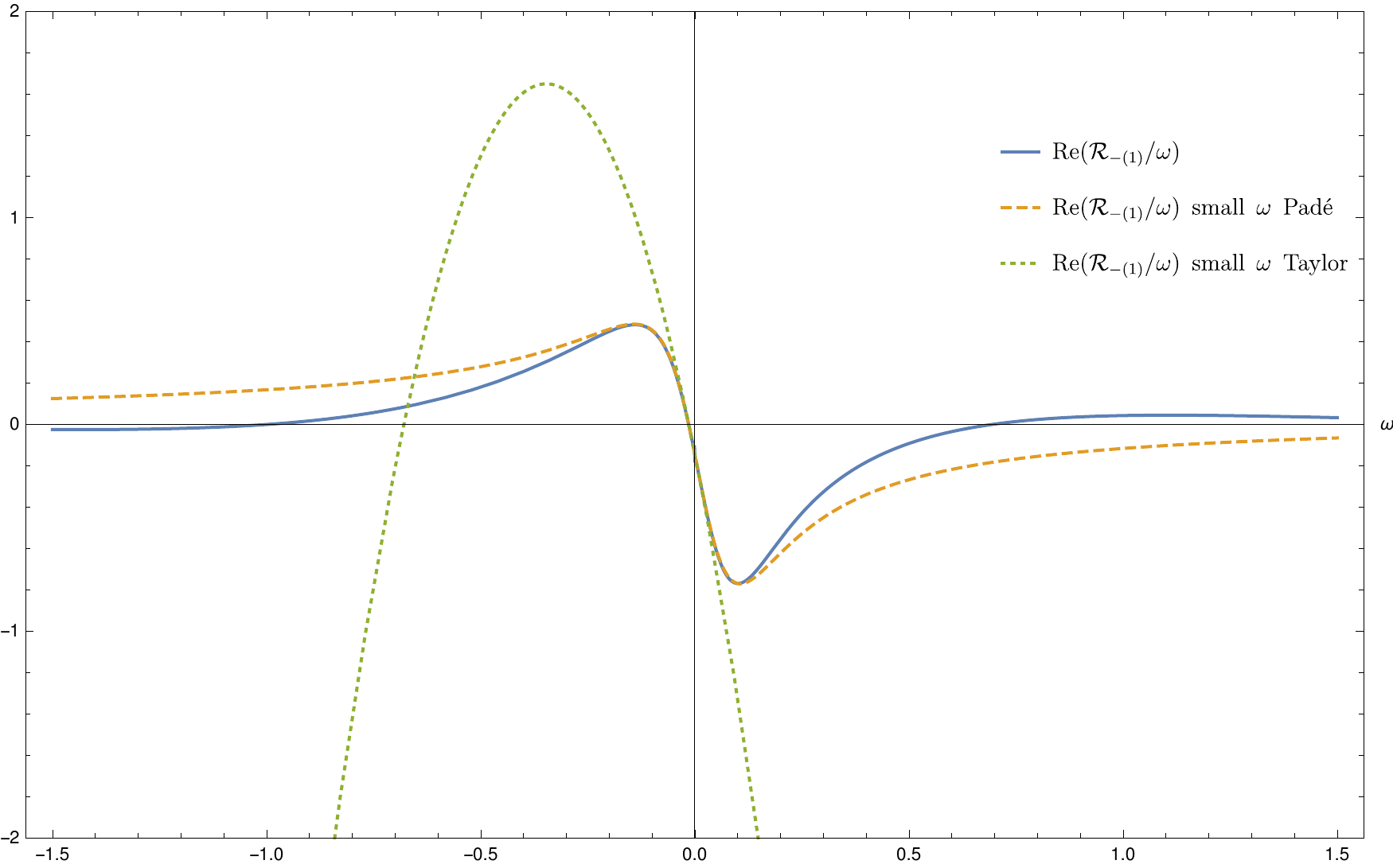}
\includegraphics[scale=0.46]{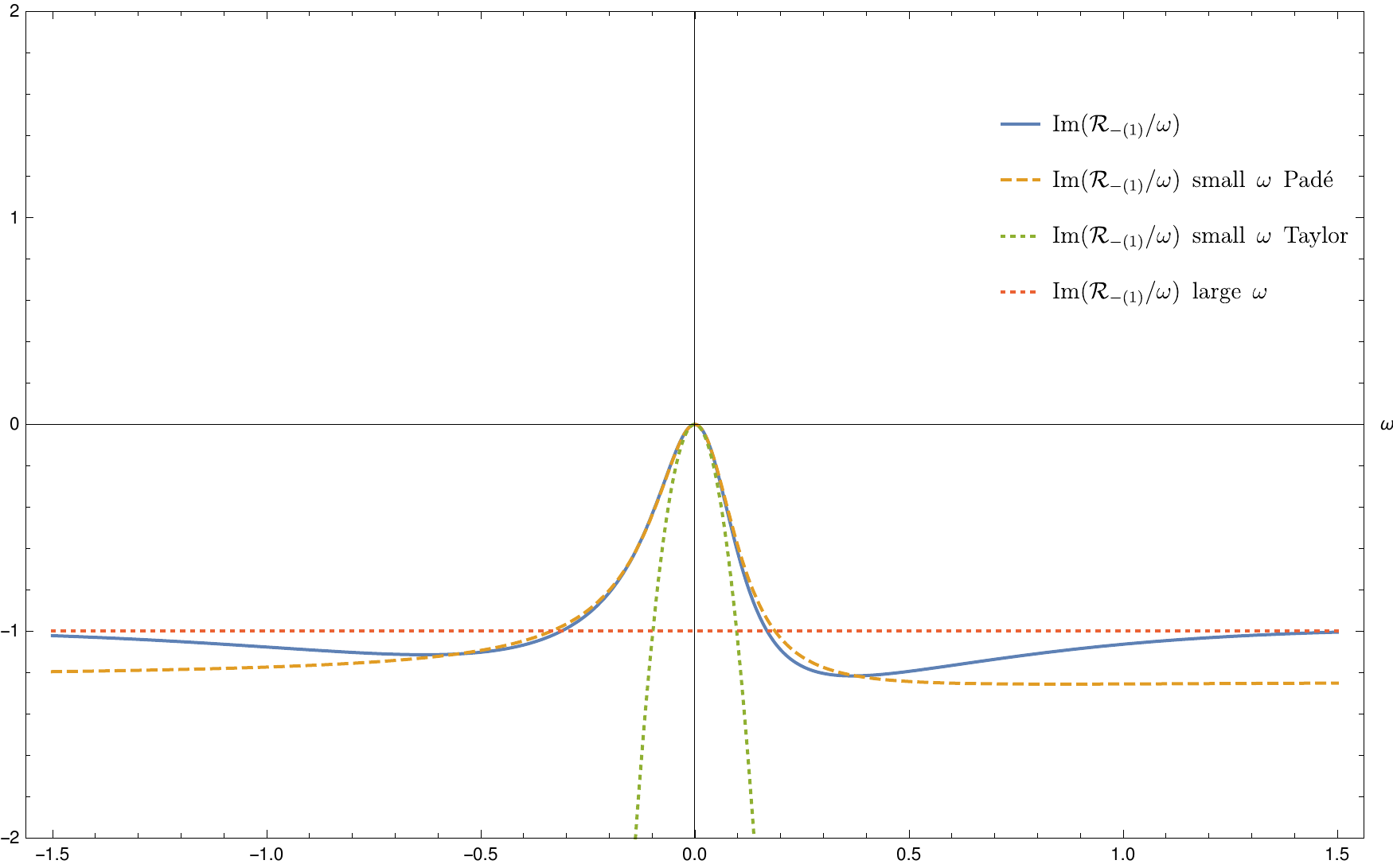}
\caption{Plots of the real and imaginary parts of the response function $\car_{-(1)}(\o)$ as a function of a real frequency $\o$ for $H=0.28$, $Q=0.04$, $T=0.2324$ and $u_h=L=1$, together with the small frequency expansion (\ref{response-small-w}), the second Pad\'e approximant in (\ref{pades}), as well as the asymptotic behavior (\ref{lwR}). Again, the Pad\'e approximant is a drastic improvement compared to the small $\o$ expansion in this regime of parameter space.}
\label{fig6} 
\end{figure}

\begin{figure}
\centering
\includegraphics[scale=0.46]{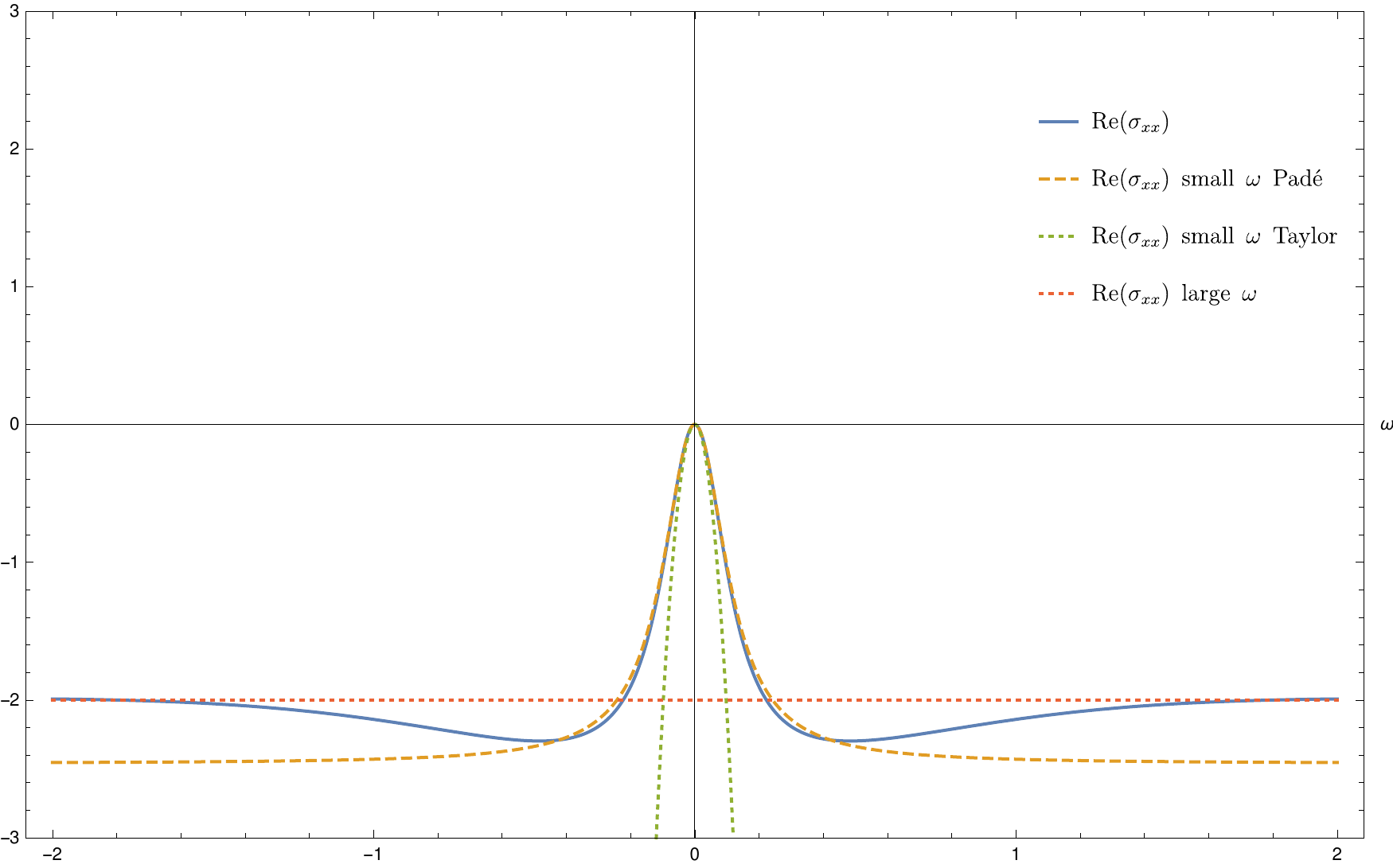}
\includegraphics[scale=0.46]{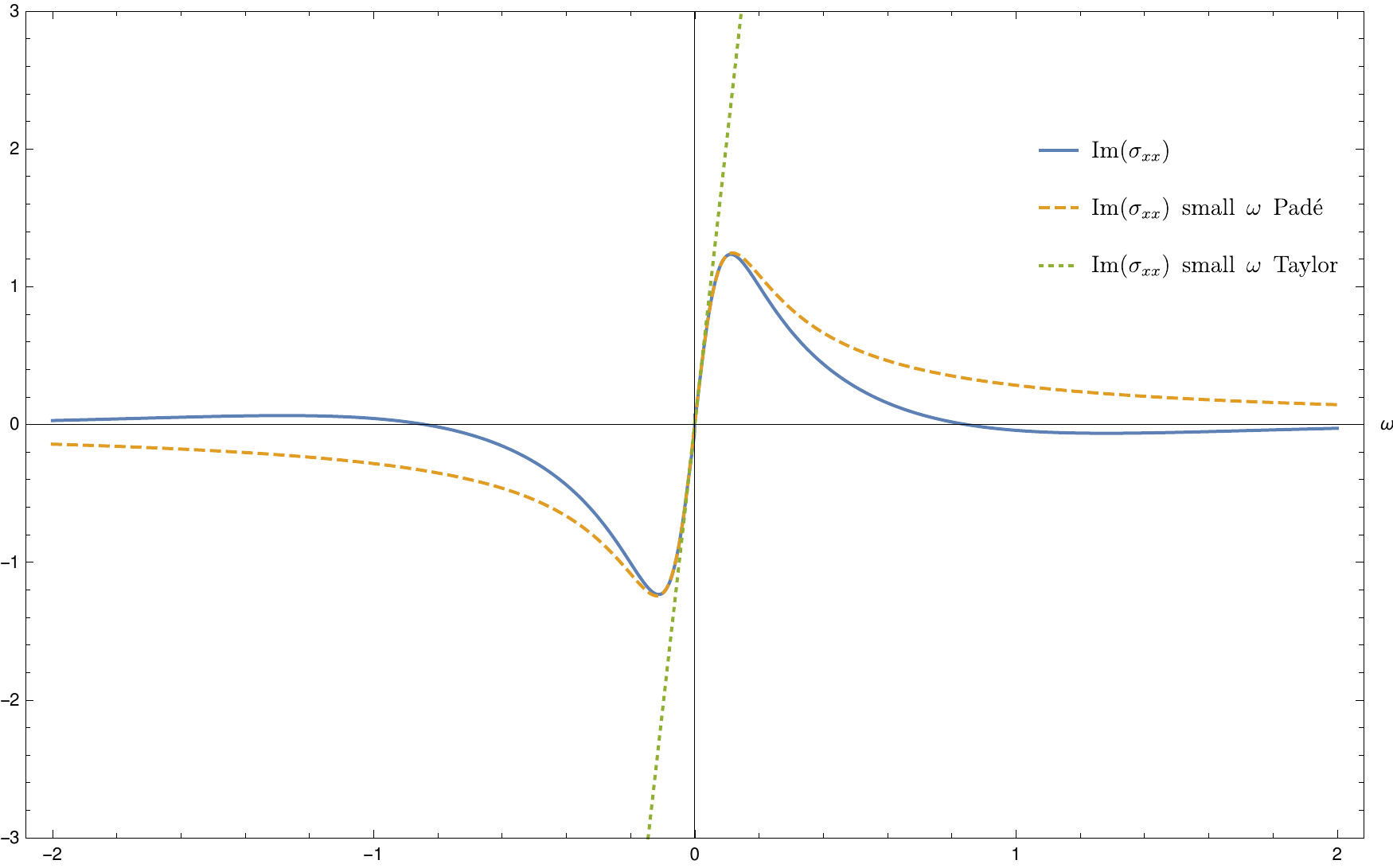}
\includegraphics[scale=0.46]{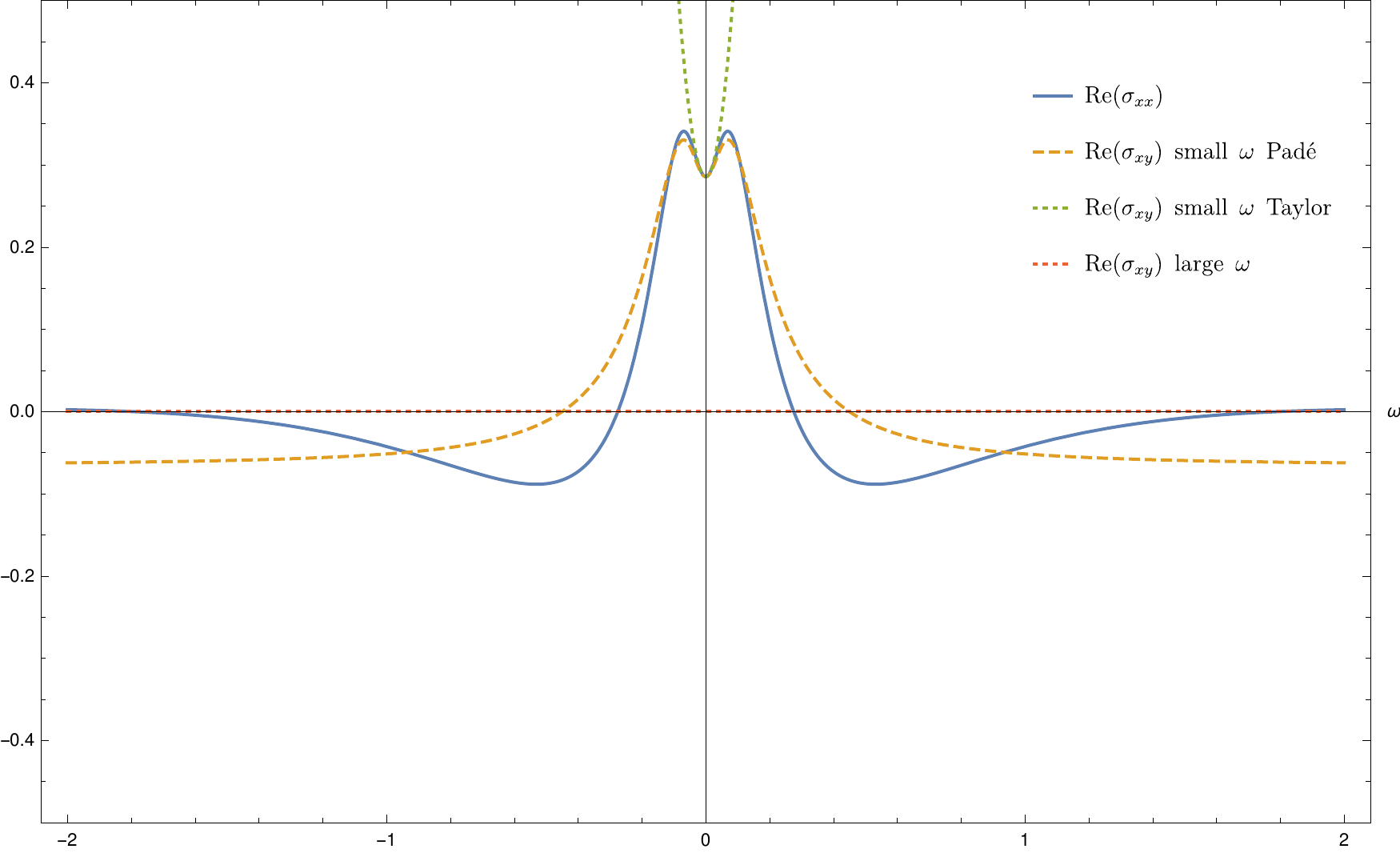}
\includegraphics[scale=0.46]{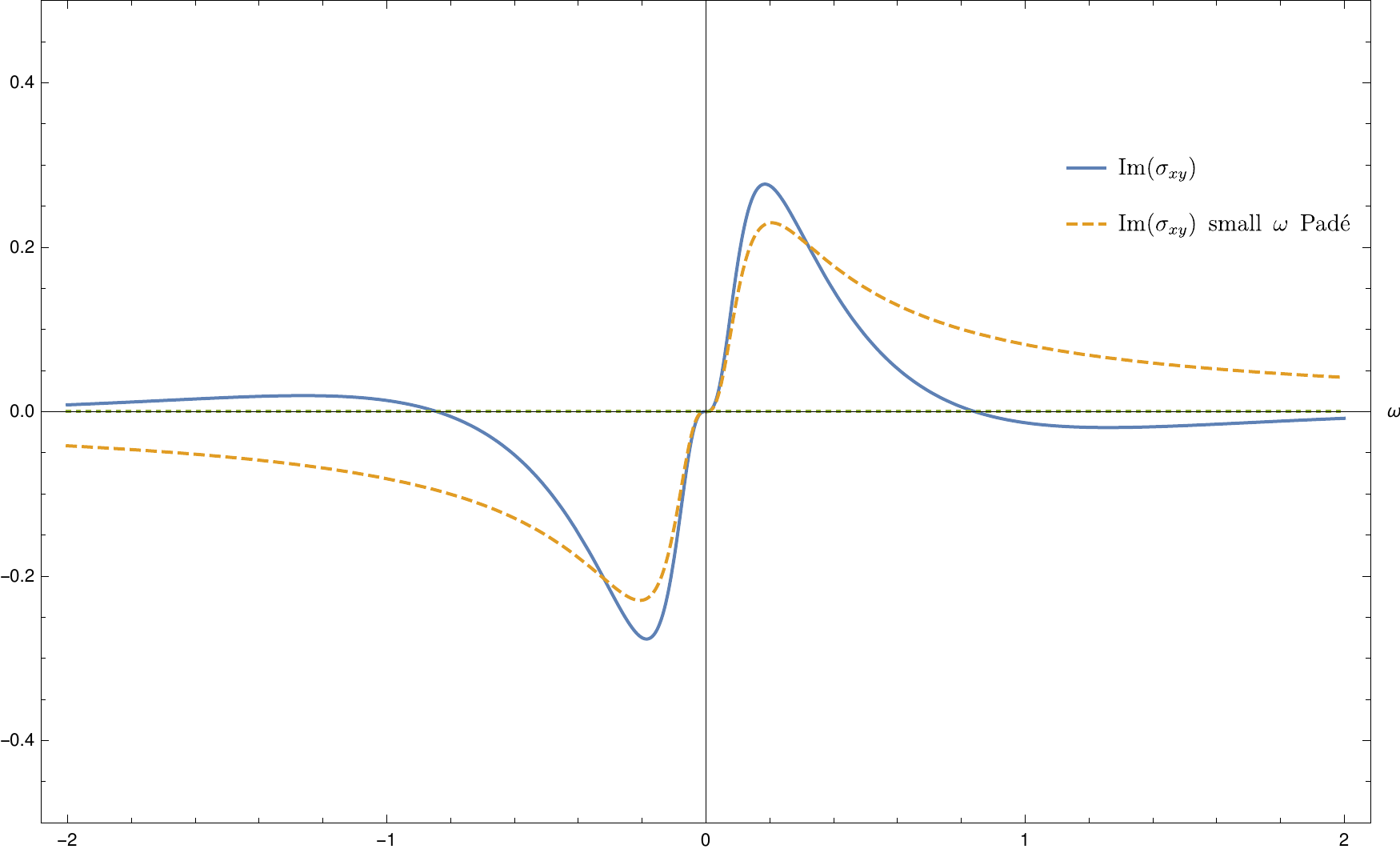}
\caption{Plots of the real and imaginary parts of the conductivities $\s_{xx}$ and $\s_{xy}$ as a function of a real frequency $\o$ for $H=0.28$, $Q=0.04$, $T=0.2324$ and $u_h=L=1$, together with the small frequency expansion (\ref{response-small-w}), the second Pad\'e approximant in (\ref{pades}), as well as the large $\o$ asymptotic behavior (\ref{lw-sigma}). Again, the location of the peaks is very well approximated by the Pad\'e approximant and is given by (\ref{poles-2}). Note that the poles have moved away from the real axis, which is why only peaks appear in these plots.}
\label{fig7} 
\end{figure}

\begin{figure}
\centering
\includegraphics[scale=0.30]{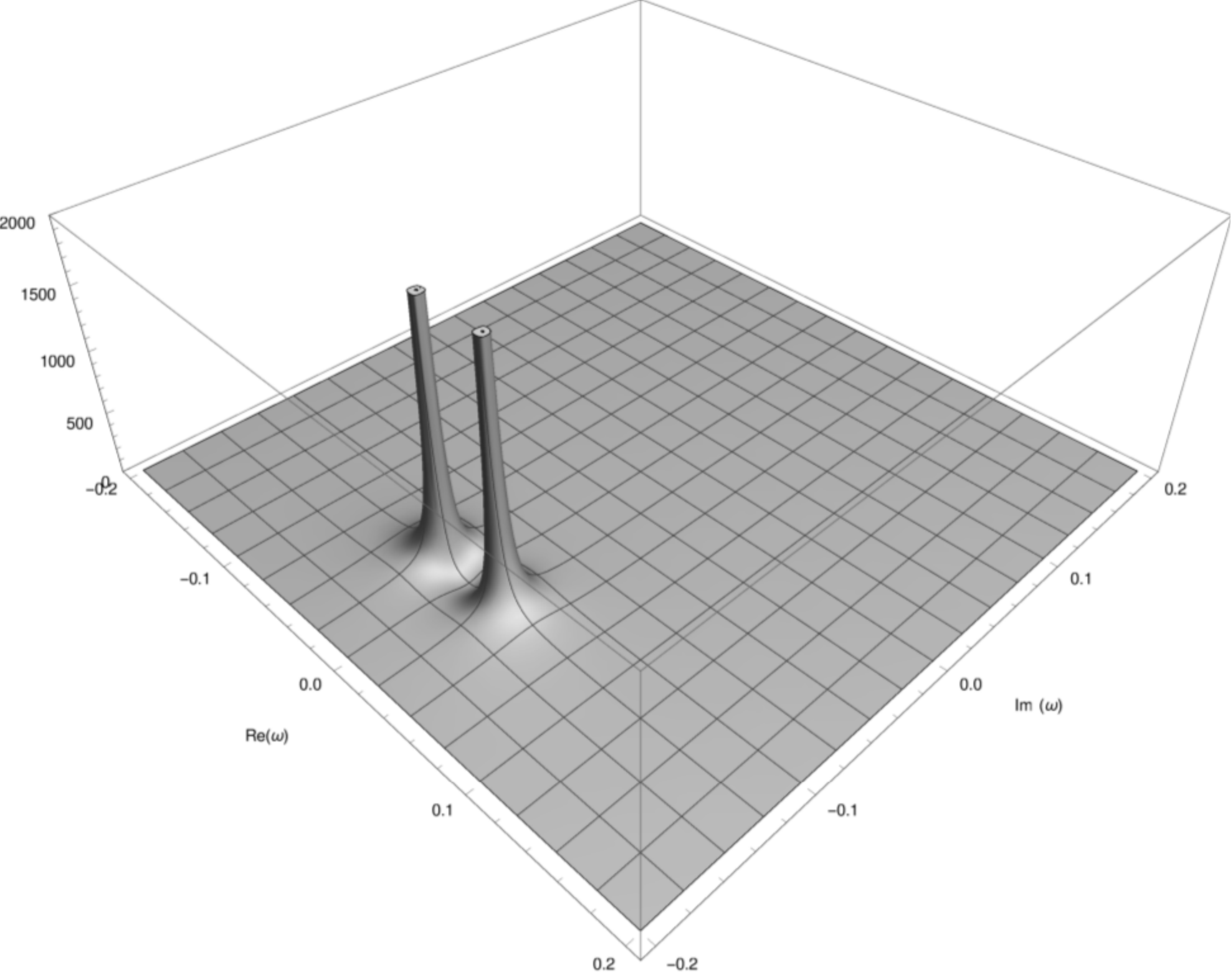}
\includegraphics[scale=0.30]{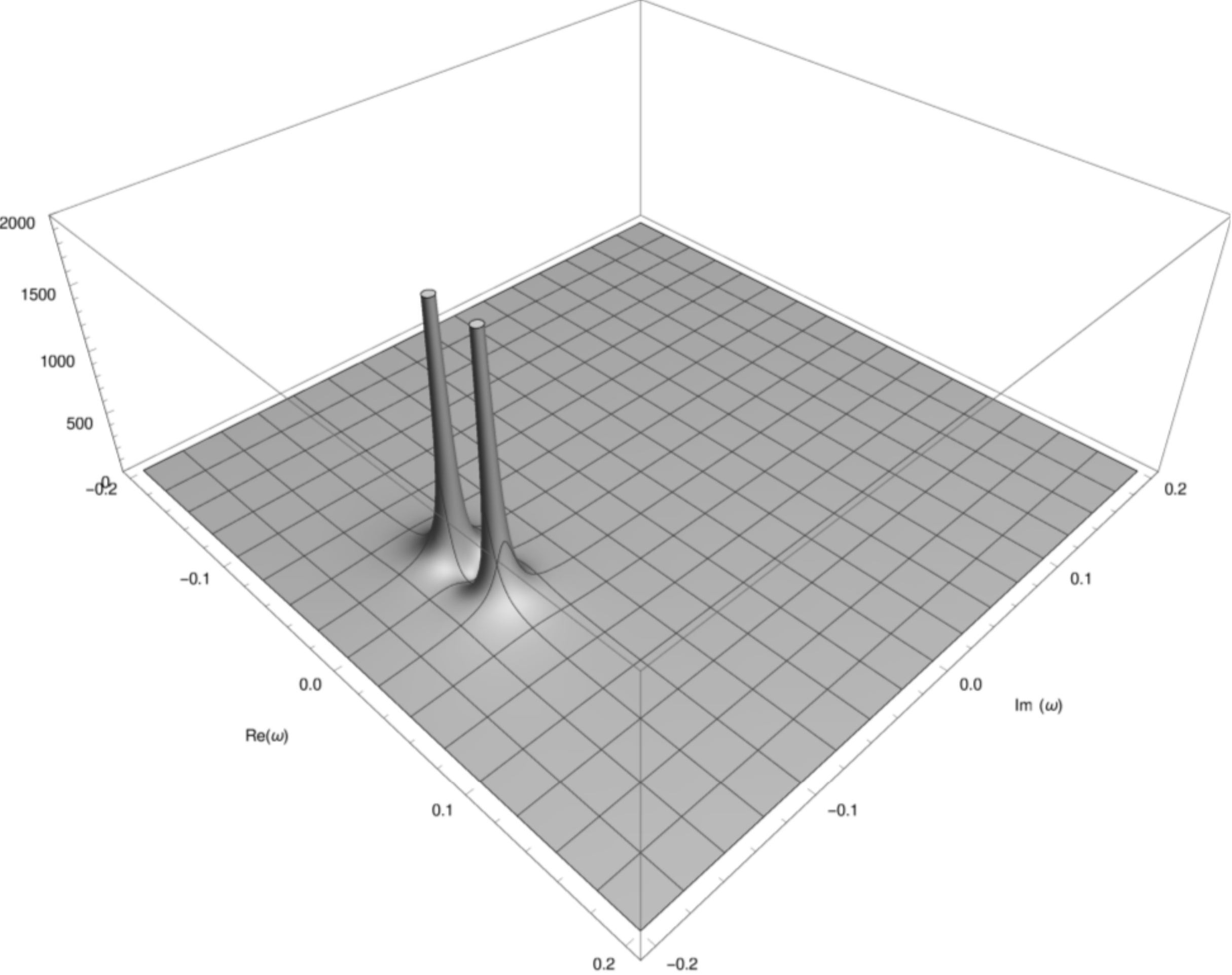}
\includegraphics[scale=0.30]{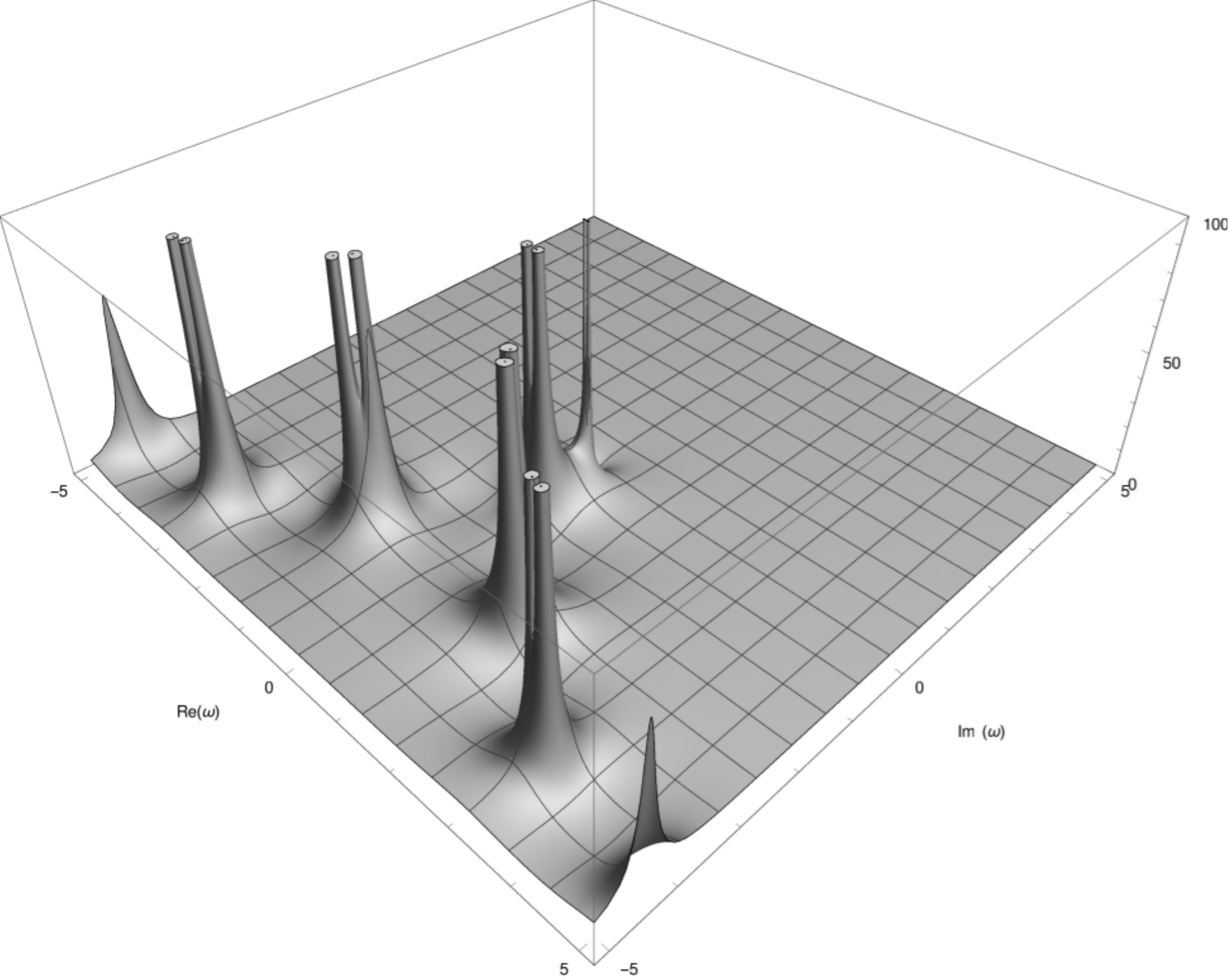}
\includegraphics[scale=0.30]{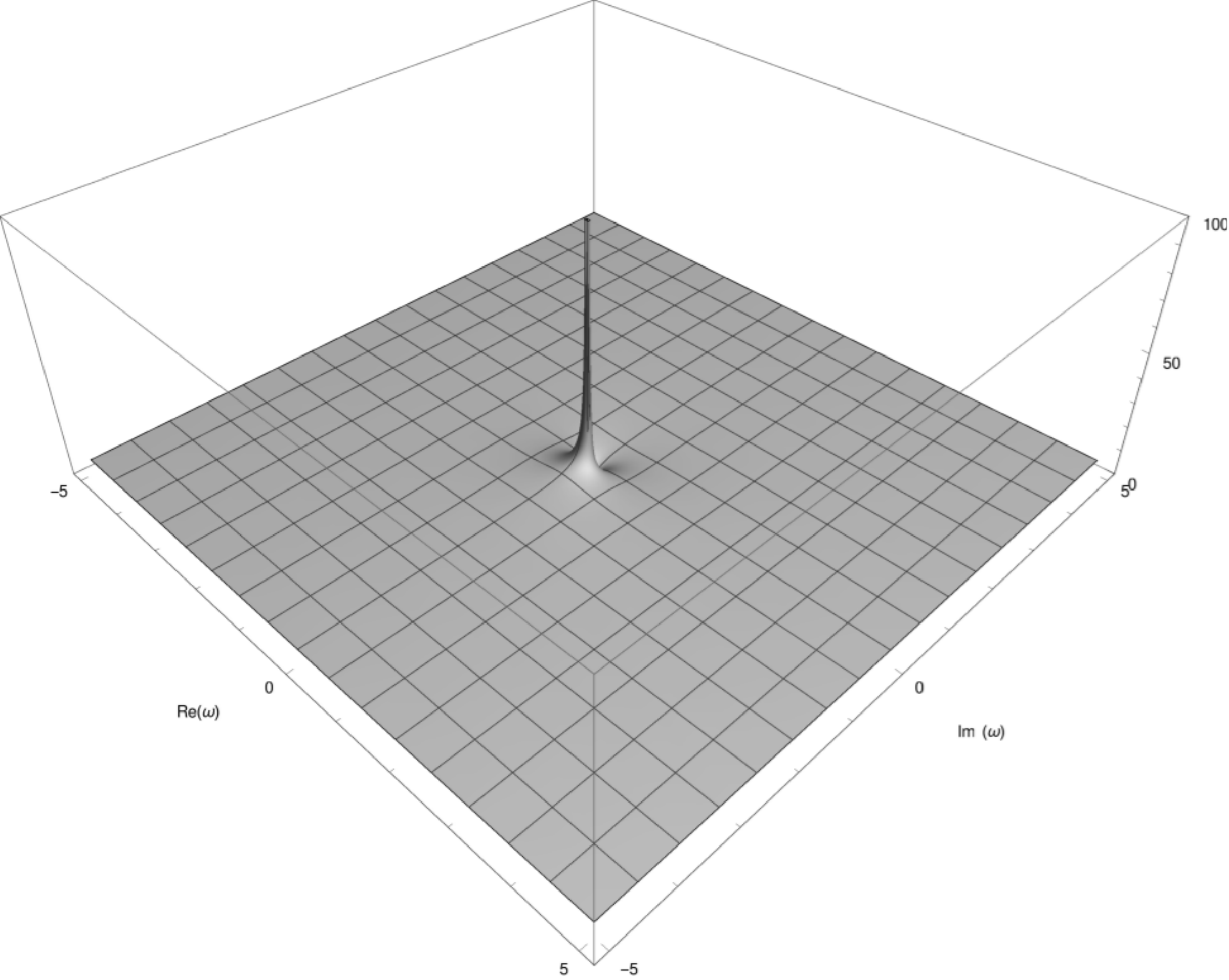}
\caption{Plot of $|\s_{xx}|^2$ as a function of a complex $\o$ for  $H=0.28$, $Q=0.04$, $T=0.2324$ (left), together with the second Pad\'e approximant in (\ref{pades}) (right). The plots on the top are zooming in on the poles closest to zero visible in the plots at the bottom. Again, the Pad\'e approximant captures the poles closest to the origin extremely accurately.}
\label{fig8} 
\end{figure}

\begin{figure}
\centering
\includegraphics[scale=0.3]{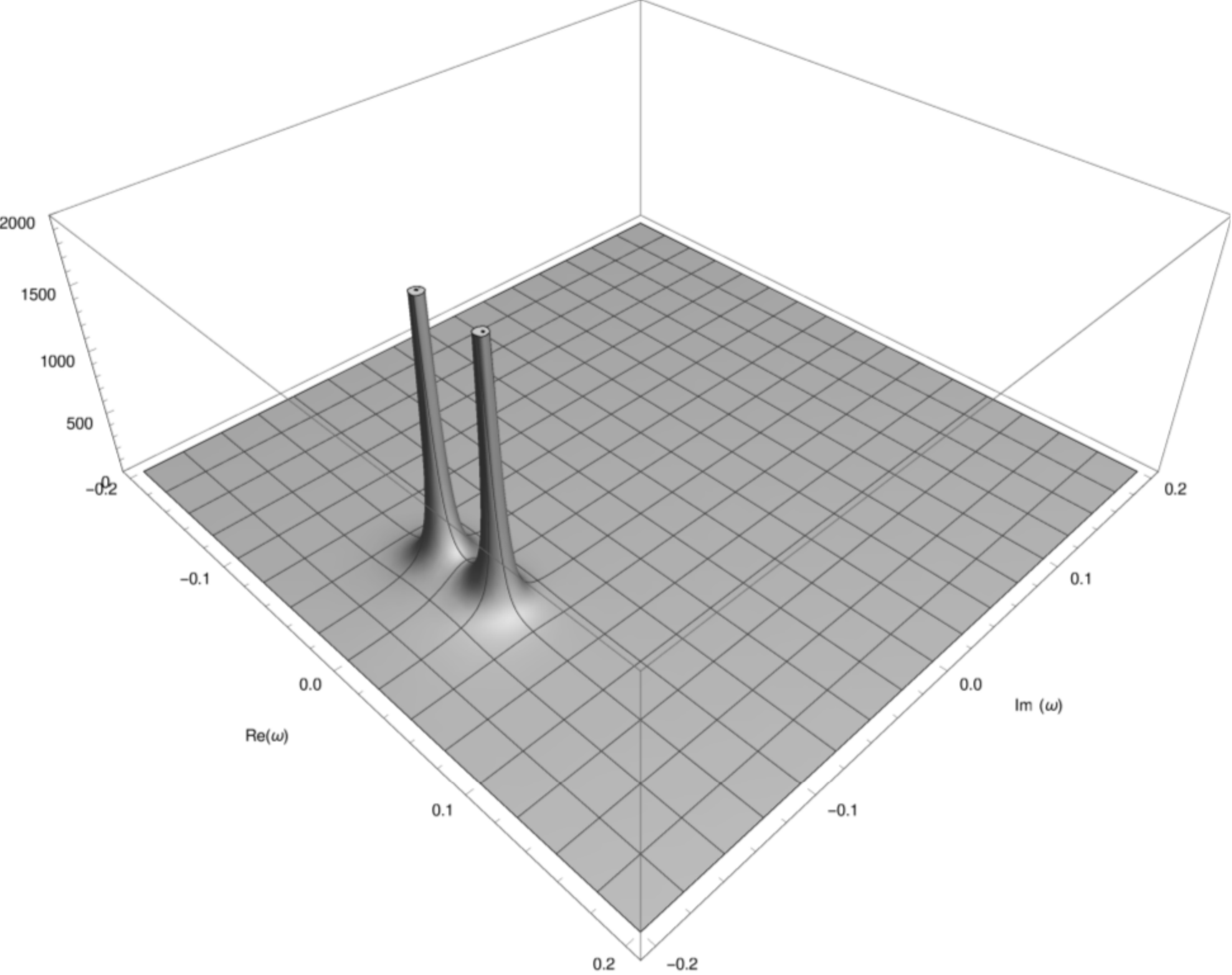}
\includegraphics[scale=0.3]{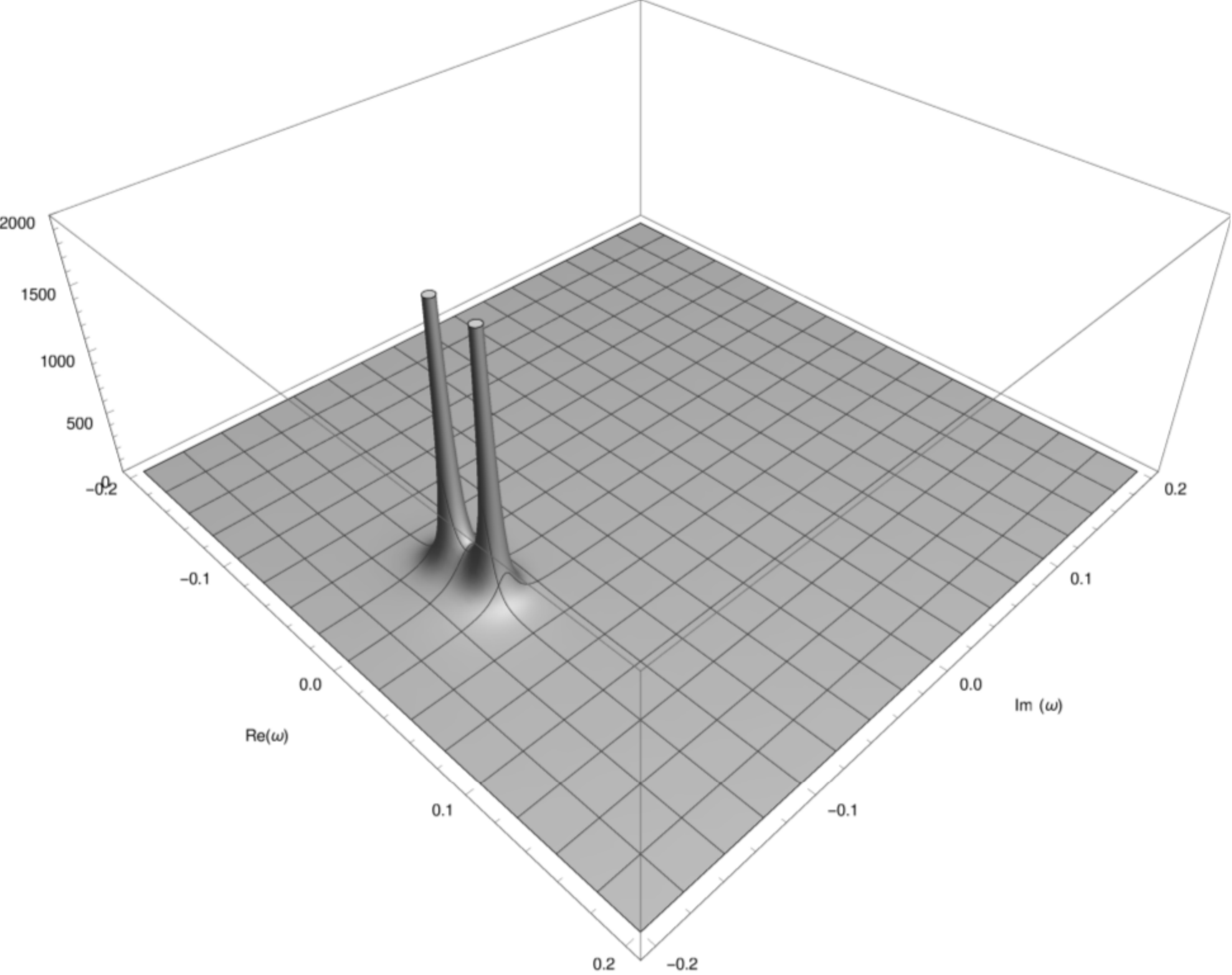}
\includegraphics[scale=0.3]{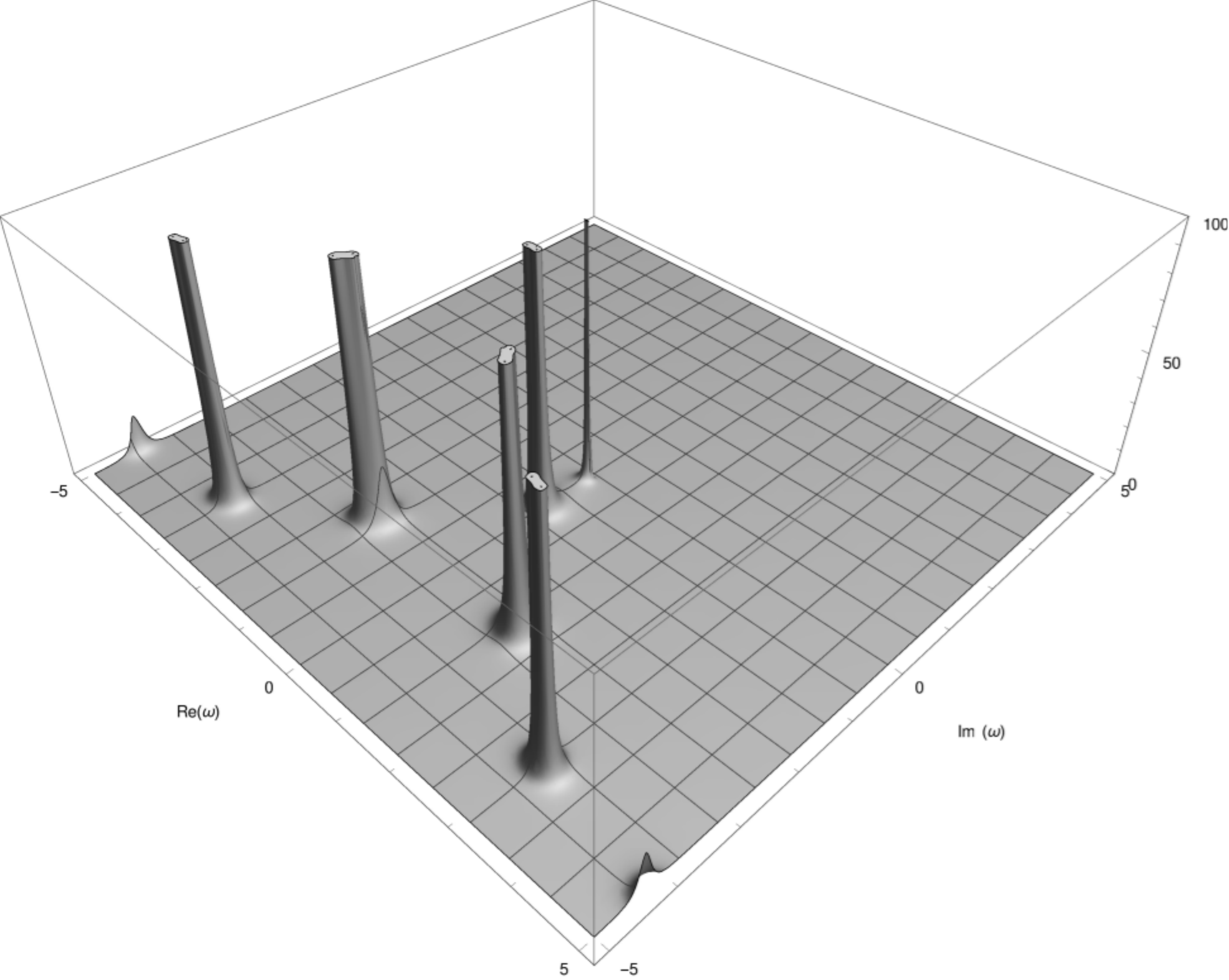}
\includegraphics[scale=0.3]{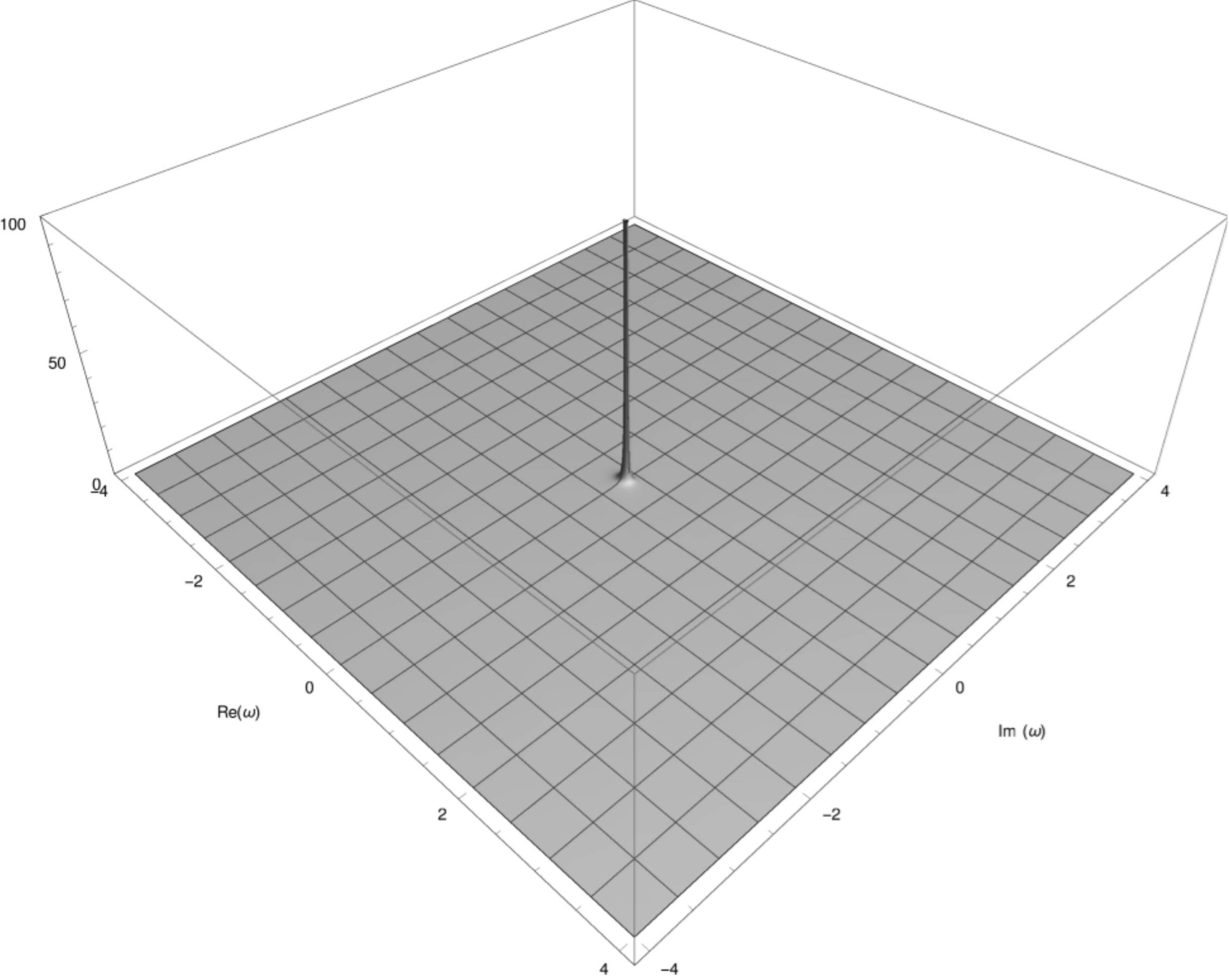}
\caption{Plot of $|\s_{xy}|^2$ as a function of a complex $\o$ for  $H=0.28$, $Q=0.04$, $T=0.2324$ (left), together with the second Pad\'e approximant in (\ref{pades}) (right). The plots on the top are zooming in on the poles closest to zero visible in the plots at the bottom. Again, the Pad\'e approximant captures the poles closest to the origin extremely accurately.}
\label{fig9} 
\end{figure}

In Figures \ref{fig1} and \ref{fig6} we plot the real and imaginary parts of the response function $\car_{-(1)}(\o)$ as a function of a real frequency for two different choices of the electric and magnetic fields, corresponding to different temperatures. The values of the electric and magnetic fields in Fig. \ref{fig1} correspond to a nearly extremal black hole, where the first Pad\'e approximant in (\ref{pades}) provides a good approximation of the response functions for small frequencies. In particular, the Pad\'e approximant very accurately reproduces the pole closest to the origin of the complex frequency plane as can be seen explicitly in the plot. The conductivities for these values of the electric and magnetic fields are plotted in Figures \ref{fig2}, \ref{fig3} and \ref{fig4}. The fact that there is no Drude peak and the poles are not located at zero frequency is a consequence of the broken translational symmetry due to the magnetic field \cite{Hartnoll:2007ai,Hartnoll:2007ip}. How the Drude peak is recovered in the limit of vanishing magnetic field is illustrated in Fig. \ref{R0vsH}. \hskip-0.8cm
The values of the electric and magnetic fields in Fig. \ref{fig6} are instead in the hydrodynamic regime where the second Pad\'e approximant in (\ref{pades}) provides a good approximation of the response functions for small frequencies, including the location of the poles, which have now moved away from the real axis according to (\ref{poles-2}). The corresponding conductivities are plotted in Figures \ref{fig7}, \ref{fig8} and \ref{fig9}.

\begin{figure}
\centering
\includegraphics[scale=0.7]{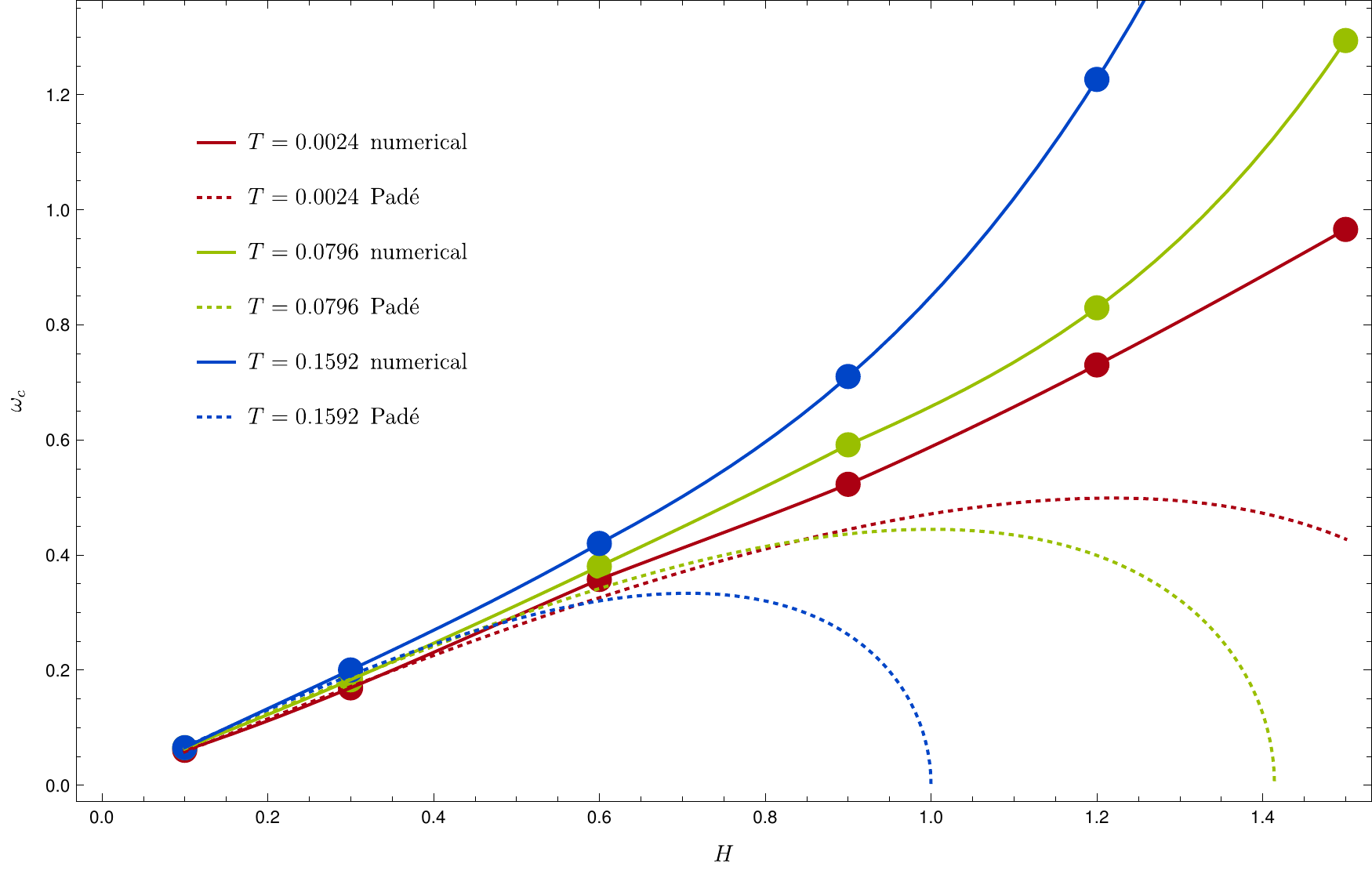}
\caption{Plot of the location of the pole of $\s_{xx}$ and $\s_{xy}$ closest to zero on the positive real $\o$ axis as a function of $H$ (solid lines), compared with the pole (\ref{poles-1}) predicted by the first Pad\'e approximant in (\ref{pades}). As expected, the agreement is best at small magnetic field and small temperatures.}
\label{fig:poles} 
\end{figure}

In Fig. \ref{fig:poles} we compare the location of the pole of the conductivities $\s_{xx}$ and $\s_{xy}$ nearest to zero on the real frequency axis as a function of the magnetic field $H$ at a number of different temperatures with the result (\ref{poles-1}) predicted by the first Pad\'e approximant in (\ref{pades}). As expected, the Pad\'e correctly gives the location of the pole at very small magnetic field, independently of the temperature. However, the lower the temperature the agreement extends to a higher value of the magnetic field. An analogous plot for the poles (\ref{poles-2}) in the hydrodynamic regime can be found in Fig. 3 of \cite{Hartnoll:2007ip}.  

\begin{figure}
\centering
\includegraphics[scale=0.76]{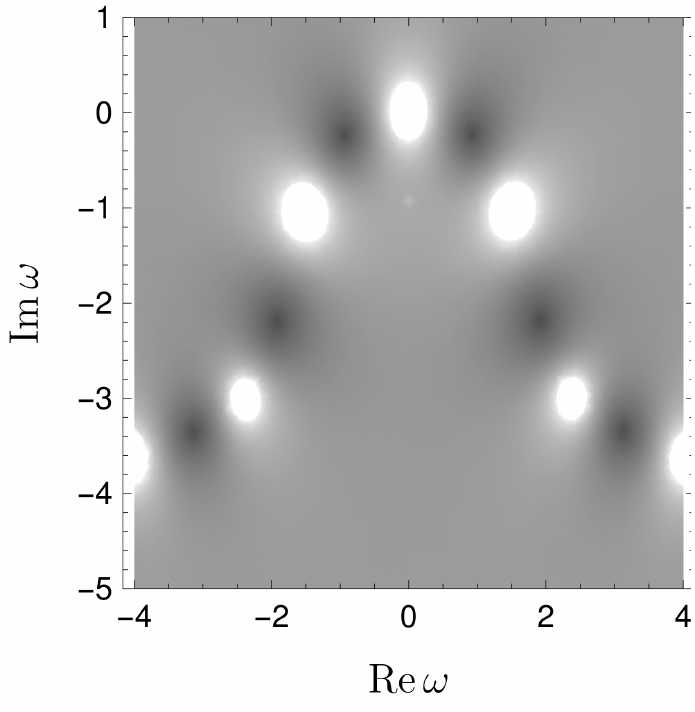}
\includegraphics[scale=0.76]{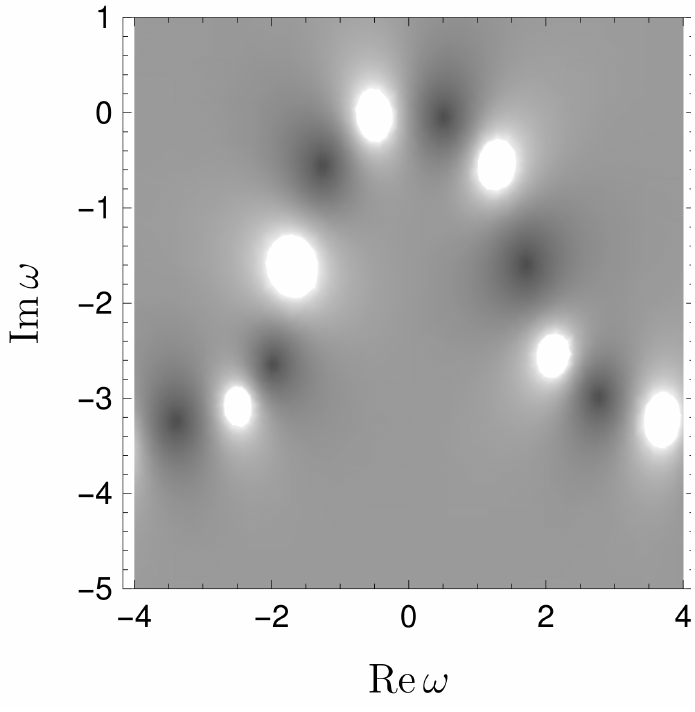}
\includegraphics[scale=0.76]{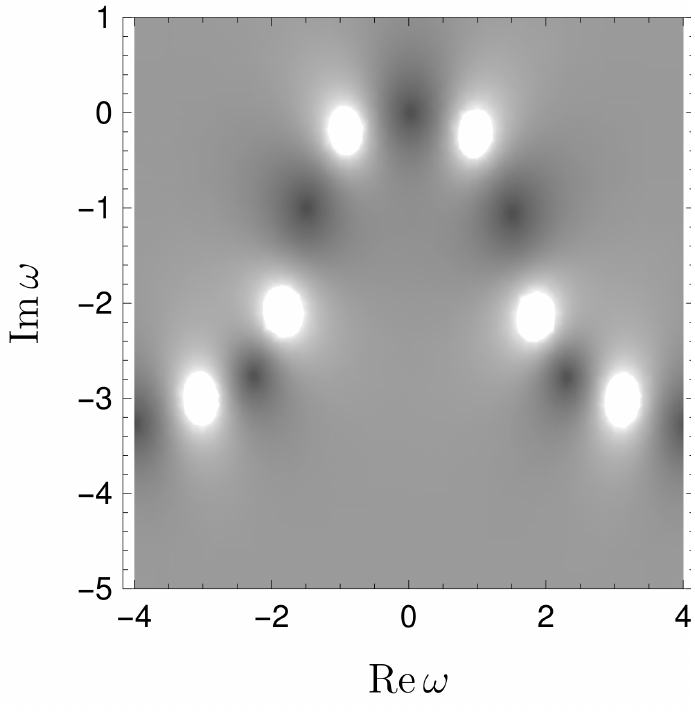}
\caption{Density plots of $|\s_+|$ as a function of complex frequency for $H=0$ and $Q=1$ (left), $H=Q=1/\sqrt{2}$ (center), and $H=1$ and $Q=0$ (right) at $T=1/2\p$ (cf.  Fig. 1 in \cite{Hartnoll:2007ip} and Fig. 9 in \cite{Kim:2015wba}). The white regions correspond to poles, while the blue regions to zeros. The location of the poles closest to zero forms a semicircle, along which the poles move as the values of $H$ and $Q$ are shifted, keeping $T$ fixed \cite{Hartnoll:2007ip}. A 3-dimensional version of the plot in the center is shown in Fig. \ref{fig9}.}
\label{fig11} 
\end{figure}

Finally, in Fig. \ref{fig11} we plot $|\s_+|$ as a function of the complexified frequency for a number of values of the electric and magnetic fields keeping $H^2+Q^2=1$ fixed, reproducing Fig. 1 in \cite{Hartnoll:2007ip} (see also Fig. 9 in \cite{Kim:2015wba}). To capture the non-liner regime where the cyclotron poles deviate from the semi-circle configuration it was necessary to impose the IR boundary condition using a Pad\'e approximant in the near horizon expansion, as discussed in Section \ref{RSolve}. This transition regime is not visible in the plots of \cite{Hartnoll:2007ip} and \cite{Kim:2015wba}.
\begin{figure}
\centering
\includegraphics[scale=0.3]{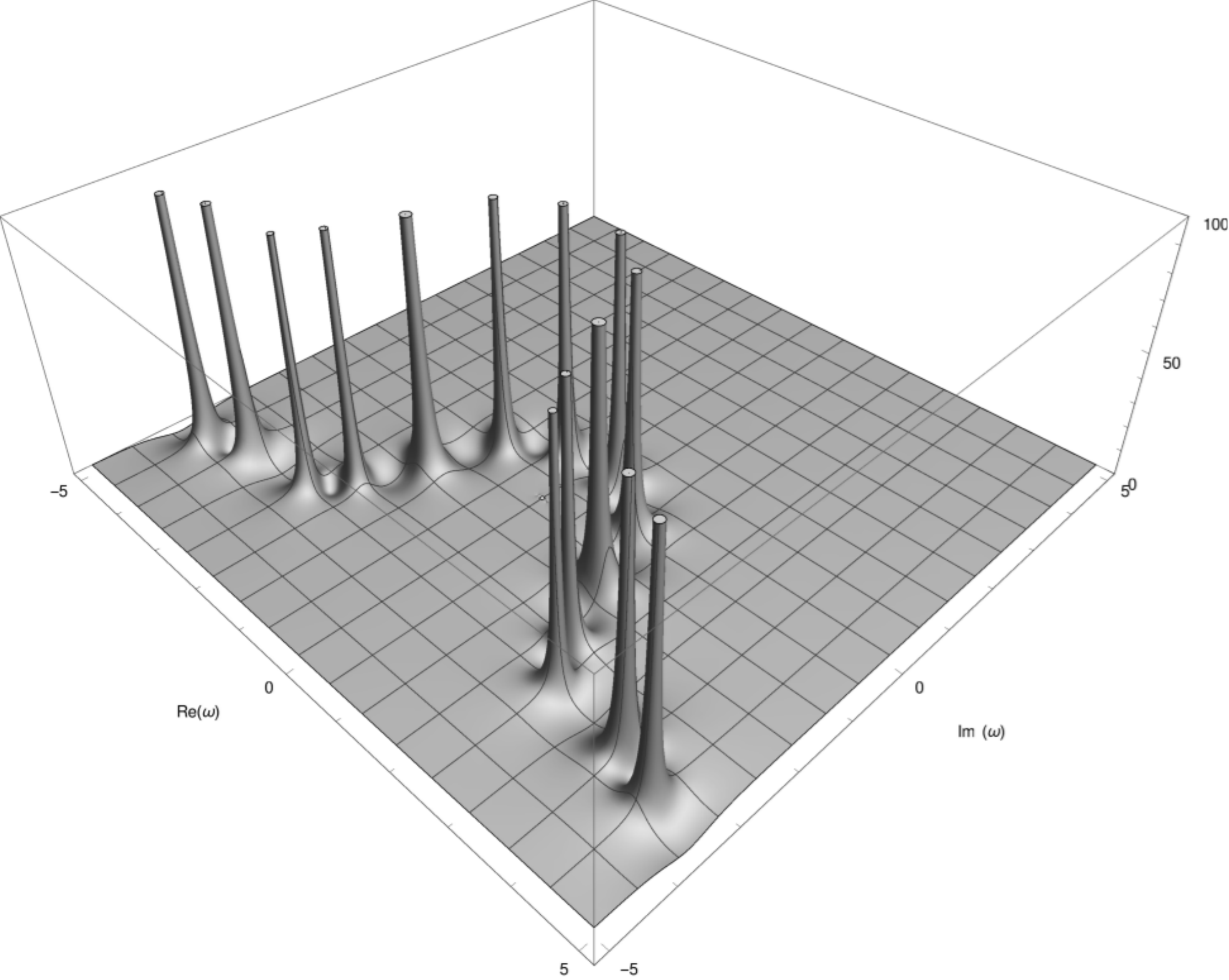}
\includegraphics[scale=0.3]{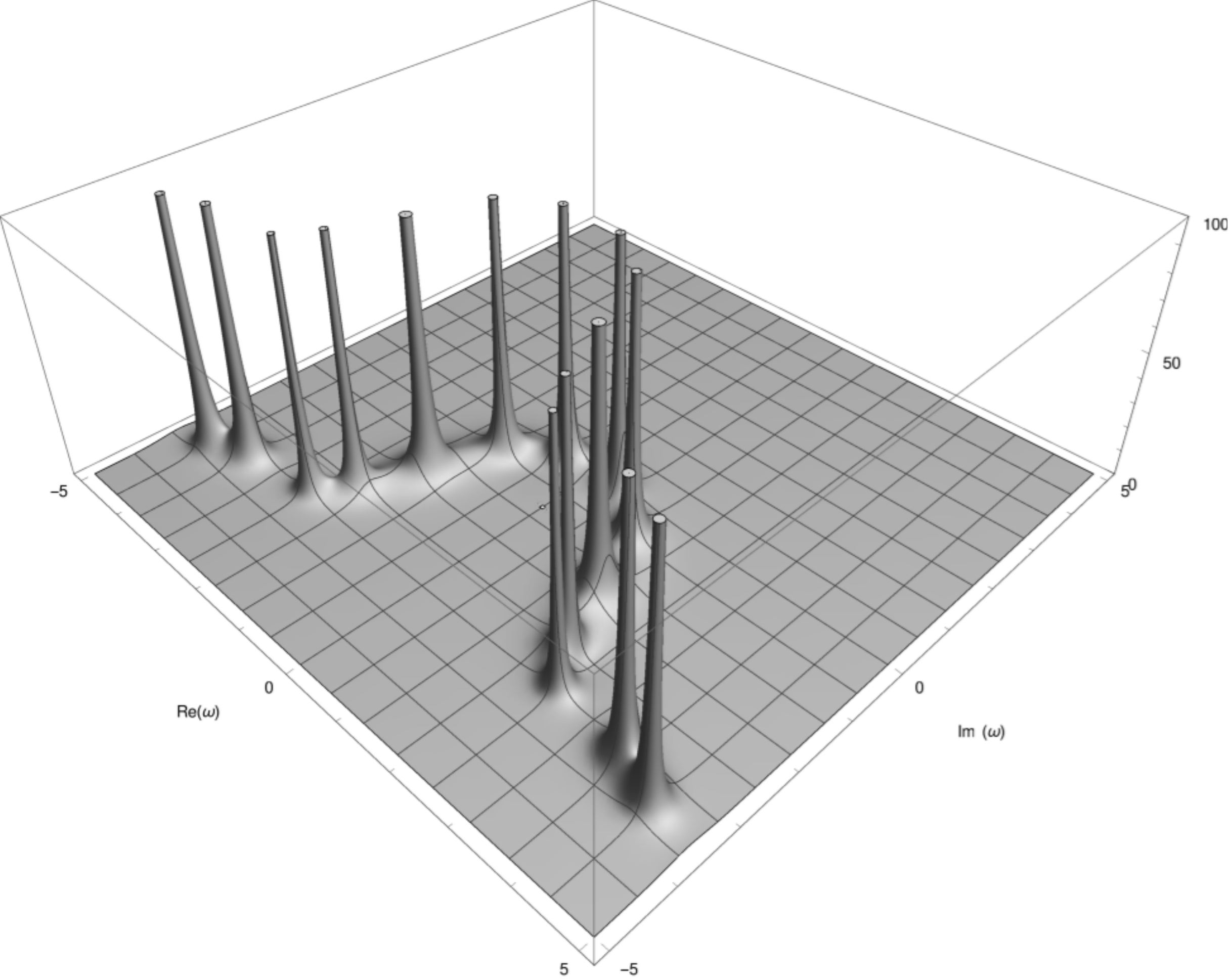}
\caption{Plots of $|\s_{xx}|^2$ (left) and $|\s_{xy}|^2$ as functions of complexified frequency for $H=Q=1/\sqrt{2}$, $T=1/2\p$. Note that the poles are arranged in a configuration between the near extremal case in Figures \ref{fig3} and \ref{fig4} and the hydrodynamic limit in Figures \ref{fig8} and \ref{fig9}.}
\label{fig10} 
\end{figure}

\newpage
	
\section{Concluding remarks} 

In this paper we presented a general framework for the holographic analysis of asymptotically AdS backgrounds with finite charge density and a constant magnetic field, including the systematic computation of the renormalized 1- and 2-point functions and the corresponding transport coefficients. The importance of such dyonic backgrounds in the holographic study of strongly coupled systems in both condensed matter and high energy physics was our main motivation for carrying out the analysis in a general and systematic fashion, in the hope that our results can be directly used in numerous applications. 

There are three important aspects in our general holographic prescription that we have tried to emphasize, all of which rely on a radial Hamiltonian formulation of the bulk dynamics. The first is a general recursive prescription for holographically renormalizing the theory. This is particularly important in the presence of running scalars, which can contribute to the UV divergences. Moreover, correctly renormalizing the theory is crucial to ensure that ultralocal and quasilocal terms in correlation functions (and hence transport coefficients) are compatible with the Ward identities. 

The second aspect we wanted to highlight is the fact that correlation functions are much more efficiently extracted directly from the solutions of the equations of motion instead of first evaluating the on-shell action and then taking derivatives. The (renormalized) radial canonical momenta are holographically identified with the 1-point functions of the dual operators in the presence of sources, which can be further differentiated with respect to the sources to give any desired $n$-point function. The only instance one actually needs to compute the on-shell action is in the computation of the free energy of the background solutions. 

The third aspect of our analysis that we consider important concerns the fluctuation equations for determining the 2-point (and higher) functions. These are in general a system of coupled linear second order equations and require boundary conditions in the UV and regularity conditions in the IR. Unless one can decouple these equations it is a priori tricky to identify which modes are the independent sources and which the responses, although in principle this can be addressed systematically using the symplectic form of the bulk theory \cite{Papadimitriou:2010as}. However, there is a straightforward alternative which utilizes the manifest symplectic structure of the Hamiltonian formalism. Namely, by trading the linear second order fluctuation equations for the corresponding first order Riccati equations one automatically eliminates the sources out of the problem and computes directly the correct response functions by imposing only regularity conditions in the IR. Besides automatically taking care of the identification of the sources and response functions, the Riccati equations can be used to directly holographically renormalize the 2-point functions (only computing the terms that are contributing to the particular 2-point function \cite{Papadimitriou:2004rz}), and they render the numerical solution of the fluctuation equations considerably simpler by eliminating the arbitrary sources from the start \cite{Papadimitriou:2013jca}.           

Finally, the radial Hamiltonian formulation of the bulk dynamics often leads to powerful solution generating techniques for background solutions, as we have demonstrated in Section \ref{exact} and Appendix \ref{W-solutions}. In particular, we found infinite families of exact RG flows interpolating between AdS in the UV and hyperscaling violating geometries in the IR, some of which we expect exhibit a gapped and discrete spectrum of fluctuations. Interestingly, even the purely electric version of these solutions is not strictly included in the classification of hyperscaling violating geometries discussed in \cite{Charmousis:2010zz,Gouteraux:2012yr,Gath:2012pg} since those assume a single exponential behavior for the scalar potential. However, the exact solutions we present here involve subleading terms in the scalar potential in an essential way. 

Computing the conductivities in these backgrounds is one of our immediate priorities \cite{rgflows}. Another interesting question is whether some of these solutions can be embedded in gauge supergravity. The potential $V(\f)$ can be easily embedded in supergravity by choosing $W_o(\f)$ to be (for example) the supersymmetric superpotential. However, the gauge kinetic function $\S(\f)$ is then determined as well, which renders the embedding not a completely trivial question. Finding exact families of dyonic black hole solutions with scalar hair and adding momentum relaxation are two other interesting directions we plan to explore \cite{caldarelli}.

\section*{Aknowledgments}
	
We would like to thank Blaise Gut\'eraux, Chris Herzog, Elias Kiritsis, Matthew Lippert, Ren\'e Meyer and Mikhail Stephanov for useful comments and discussions.  	
IP would like to thank the Galileo Galilei Institute for Theoretical Physics and the Vrije Universiteit Brussel for the hospitality, as well as 
the INFN for partial support during the completion of this work. The work of EJL, AT and JV was supported in part by the Belgian Federal Science Policy Office through the Interuniversity Attraction Pole P7/37, by FWO-Vlaanderen through project G020714N, and by the Vrije Universiteit Brussel through the Strategic Research Program ``High-Energy Physics". 
The work of EJL was partially supported by the ERC Advanced Grant ``SyDuGraM", by IISN-Belgium (convention 4.4514.08) and
by the ``Communaute Francaise de Belgique" through the ARC program. The work of AT is partially funded by the VUB Research Council. JV is Aspirant FWO.


\appendix

\renewcommand{\theequation}{\Alph{section}.\arabic{equation}}

\setcounter{section}{0}

\section*{Appendices}
\setcounter{equation}{0}

\section{Exact superpotentials and RG flows in various dimensions}
\label{W-solutions}

In this appendix we provide a few more examples of solutions to the superpotential equation (\ref{superpotential}), leading to exact RG flows in various dimensions. 
\begin{itemize}

\item[(i)] $d=3$:

The solution (\ref{exactRG}) can be generalized to include a non-trivial axion, namely 
\be\label{exactRG-axion}
W(A,\f,\c)=W_o(\f,\c)\sqrt{1+q^2e^{-4A}},
\ee
with the potentials given by 
\bal\label{cond}
&V(\f,\c)=W_{o\f}^2+Z^{-1}(\f)W_{o\c}^2-\frac32W_o^2,\NO\\
&H^2\S(\f)+(\wt Q-2\P(\c) H)^2\S^{-1}(\f)=\frac{q^2}{2}\left(W_{o\f}^2+Z^{-1}(\f)W_{o\c}^2+\frac12W_o^2\right).
\eal
However, contrary to the purely dilatonic case discussed in Section \ref{exact}, the function $W_o(\f,\c)$ cannot be chosen arbitrarily here since it is constrained by the second equation in (\ref{cond}).

\item[(ii)] $d$ arbitrary, $H=0$, $\wt Q\neq 0$:

For zero magnetic field the superpotential equation (\ref{superpotential}) admits a solution of the form 
\be\label{}
W(A,\f)=W_o(\f)\sqrt{1+q^2e^{-2(d-1)A}},
\ee
in any dimension, with $\wt Q^2\S^{-1}_0=q^2L^{-2}$ and  
\be
V(\f)=W_{o}'^2-\frac{d}{d-1}W_o^2,\quad \S(\f)=\frac{2\S_0}{L^2(W_{o}'^2+\frac{d-2}{d-1}W_o^2)}.
\ee
As in the example in Section \ref{exact} the function $W_o(\f)$ can be prescribed at will. The corresponding metrics again flow to a hyperscaling violating geometry in the IR with exponents 
\be
1< z < \frac{2d-3}{d-2}, \quad \th=d+z-2,
\ee
and it is important to keep in mind that $d$ here includes time.  

\item[(iii)] $d=5$, $\wt Q-2\P_0 H\neq 0$:

In this case a possible solution of (\ref{superpotential}) is
\be
W(A,\f)=W_0(\f)+(\wt Q-2\P_0 H)W_1(\f)e^{-4A},
\ee
where 
\bal
&V(\f)=W_{0}'^2-\frac54W_0^2,\quad \S(\f)=\frac{2}{W_1'^2+\frac34W_1^2},\NO\\
&\left(W_1'^2+\frac34W_1^2\right)\left(W_0'W_1'+\frac14W_0W_1\right)=\frac{2H^2}{\wt Q-2\P_0 H},
\eal
The last equation should be viewed as a differential equation for $W_1(\f)$, and the ratio $H^2/(\wt Q-2\P_0 H)$ must be treated as a parameter of the theory, unless it drops out of the combination $W_1'^2+\frac34W_1^2$. Again, $W_0(\f)$ can be any function. 

\end{itemize}

\section{Gauss-Codazzi equations}
\label{GC-appendix}

In this appendix we write for completeness the full set of Gauss-Codazzi equations for the model described by the action (\ref{S}) in the gauge
\be
	ds^2=dr^2+\g^{ij}dx^idx^j, \qquad A=A_idx^i,
\ee
which is used throughout our analysis. These equations are the starting point for deriving the general fluctuation equations, which we give in Appendix \ref{gen-fluct}. The Gauss-Codazzi equations following from the equations of motion (\ref{EOMS}) are 
\begin{subequations}\label{GC}
	\begin{align}
			&\mbox{\bf Einstein $rr$:}\NO\\
			& K^2-K_{ij}K^{ij}=R[\g]+\dot{\f}^2-\pa^i\f\pa_i\f+Z(\f)\left(\dot{\c}^2-\pa^i\c\pa_i\c\right) \notag\\
			&\rule{4.5cm}{0cm}+\S(\f)\left(2\g^{ij}\dot{A}_i\dot{A}_j-F_{ij}F^{ij}\right)-V(\f,\c), \\
			&\mbox{\bf Einstein $ri$:}\NO\\
			& D_jK^j_i-D_iK=\dot{\f}\pa_i\f+Z(\f)\dot{\c}\pa_i\c+2\S(\f)F_{ij}\g^{jk}\dot A_k, \\
			&\mbox{\bf Einstein $ij$:}\NO\\
			&\dot K^i_j+KK^i_j=R^i_j[\g]-\pa^i\f\pa_j\f-Z(\f)\pa^i\c\pa_j\c-2\S(\f)\left(\g ^{ik}\dot A_k \dot A_j+F^i{}_{k}F_j{}^k\right)\NO\\
			&\rule{4.0cm}{0cm}+\frac{1}{(d-1)}\d^i_j \left(\S(\f)\left(2\g^{k l} \dot A_k \dot A_l+F_{kl}F^{kl}\right)-V(\f,\c)\right),
	\end{align}
\end{subequations}
\begin{subequations*}
	\begin{align}
			&\mbox{\bf Scalar $\f$:}\NO\\
			&\ddot{\f}+K\dot{\f}+\square\f-\frac{1}{2}\left(V_\f+Z_\f\left(\dot{\c}^2+\pa^i\c\pa_i\c\right)+ \S_\f\left(2\g^{ij}\dot A_i \dot A_j+F_{ij}F^{ij}\right)\right)=0, \\
			&\mbox{\bf Scalar $\c$:}\NO\\
			& Z \left(\ddot{\c}+K\dot{\c}+\square\c\right)+Z_\f\left(\dot{\c}\dot{\f}+\pa^i\f\pa_i\c\right)-\frac{1}{2}\left( V_\c+4\P_\c\e^{ijk}\dot A_iF_{jk} \right)=0, 
	\end{align}
\end{subequations*}
\begin{subequations*}
	\begin{align}
			&\mbox{\bf Maxwell $i$:}\NO\\
			&\frac{1}{\sqrt{-\g}}\pa_r\left(\sqrt{-\g}\left(\S(\f)\g^{ij}\dot A_j+\P(\c)\e^{ijk}F_{jk}\right)\right)= D_j\left(\S(\f)F^{ij}+2\P(\c)\e^{ijk}\dot A_k\right), \\
			&\mbox{\bf Maxwell $r$:}\NO\\
			& D_i\left(\S(\f)\g^{ij}\dot A_j+\P(\c)\e^{ijk}F_{jk}\right)=0.
	\end{align}
\end{subequations*}

\section{General fluctuation equations for $d=3$}
\label{gen-fluct}

Although in the main body of the paper we consider only certain time-dependent fluctuations around the backgrounds (\ref{Bans}) with no spatial dependence, in this appendix we provide the complete set of fluctuation equations for $d=3$  following from the Gauss-Codazzi equations in Appendix \ref{GC-appendix}, with generic fluctuations around the backgrounds (\ref{Bans}) of the form
\be
\g_{ij}=\g^{B}_{ij}(r)+h_{ij}(r,\mathbf{x}), \quad	
A_i= A^{B}_i(r,\mathbf{x})+a_i(r,\mathbf{x}), \quad
\f=\f_B(r)+\vf(r,\mathbf{x}), \quad
\c=\c_B(r)+\t(r,\mathbf{x}).
\ee
Introducing the quantities $S_i^j\equiv\g_B^{jk}h_{ki}$, $S\equiv S_i^i$ and $S_{\perp}\equiv S_i^i-S_0^0$, we have the following useful identities:
\begin{subequations}\label{fluct-exp}
	\begin{align}
	& R_{ij}[\g_B]=0, \quad D^B_i=\pa_i, \\
	&\overset{(0)}{K_i^j}=\dot{A}\d_i^j+\frac{\dot{f}}{2f}\d_0^j\d_i^0, \quad
		\overset{(0)}{K}=d\dot{A}+\frac{\dot{f}}{2f}, \quad
		\overset{(0)}{K_{ij}}=\frac{1}{2}\overset {(0)}{\dot{\g}_{ij}}=e^{2A}\left(\dot{A}\d_{ij}-\frac{1}{2}\left(\dot{f}+2\dot{A}(1+f)\right)\d_{i0}\d_{j0}\right),  \\
	&\overset
		{(0)}{\left(\ddot{\g}_{ki}\g^{kj}\right)}=(4\dot{A}^2+2\ddot{A})\d_i^j+\frac{4\dot{A}\dot{f}+\ddot{f}}{f}\d_{i0} \d^{0j},\\
	&\overset {(1)}{K_i^j}=\frac{1}{2}\dot{S}_i^j+\frac{\dot{f}}{2f}(S_i^0\d^j_0-S^j_0\d_i^0), \quad
		\overset{(1)}{K}=\frac{1}{2}\dot{S}, \quad
		\overset{(1)}{R}=\pa^i\pa_jS_i^j-\square_BS, \\
	&\overset{(1)}{\G^i_{jk}}=\frac{1}{2}\left(\pa_jS_{k}^i+\pa_{k}S_j^i-\pa^ih_{jk}\right), \quad
		\overset{(1)}{R_i^j}=\frac{1}{2}\left(\pa^k\pa_iS_{k}^j+\pa_{k}\pa^jS_i^k-\square_BS_i^j-\pa_i\pa^jS\right).
	\end{align}
\end{subequations}
Expanding the Gauss-Codazzi equations in Appendix \ref{GC-appendix} and using these identities leads to the following set of fluctuation equations for $d=3$:
\begin{subequations}
	\begin{align}
	&\mbox{\bf Einstein $ij$:}\NO\\
	&\left(\pa_r^2+\left(d\dot{A}+\frac12f^{-1}\dot f\right)\pa_r+\square_B\right)S_i^j-f^{-1}\dot f\left(\d_i^0\dot S_0^j-\d^j_0\dot S^0_i\right)+\frac{\dot f}{2f}\dot{S}\d_i^0\d^j_0 -\left(\pa^j\pa_{k}S_i^k+\pa_i\pa^kS^j_{k}-\pa^j\pa_iS\right)\NO \\
	&+4\S e^{-4A}\left(e^{2A}f^{-1}\dot\a^2+H^2\right)\left(S^0_i\d^j_0-S^j_0\d^0_i\right)+\d_i^j\left(\dot{A}\dot{S}+\frac{2}{d-1}(V_{\f}\vf+V_{\c}\t)\right) \NO \\
	&+4\S_{\f}\vf\left(-f^{-1}e^{-2A}\d_i^0\d_0^j\dot{\a}^2+e^{-4A}(\d_i^x\d_x^j+\d_i^y\d_y^j)H^2\right)+4\S\left(-f^{-1}e^{-2A}\dot{\a}\left(\d_0^j\dot{a}_i-\dot\a\d^0_iS^j_0\right)+\g^{jk}_B\d^0_i\dot{\a}\dot{a}_k\right) \NO \\
	&-4H\S e^{-2A}\left(He^{-2A}\left((\d^j_x\d^x_i+\d^j_y\d^y_i)(S^x_x+S^y_y)+\d^j_0(S^0_x\d^x_i+S^0_y\d^y_i)\right)\right.\NO\\
	&\left.-\g^{jk}_B\left((\d_{k}^x\d_y^l-\d_{k}^y\d_x^l)\d^p_i+(\d_i^x\d_y^l-\d_i^y\d_x^l)\d^p_k\right)(\pa_pa_l-\pa_l a_p)\right) \NO \\
	&-\frac{4}{d-1}\d_i^j\S e^{-2A}\left(-2f^{-1}\dot{a}_t\dot{\a}+f^{-1}S_0^0\dot{\a}^2+2He^{-2A}(\pa_xa_y-\pa_ya_x)-e^{-2A}H^2(S_x^x+S_y^y)\right) \NO \\
	&-\frac{4}{d-1}e^{-2A}\d_i^j\S_{\f}\vf\left(-f^{-1}\dot{\a}^2+H^2e^{-2A}\right)=0,
	\end{align}
\end{subequations}	
\begin{subequations*}
	\begin{align}
	&\hskip-0.7in \mbox{\bf Einstein $ri$:}\NO\\
	&\hskip-0.7in \pa_j\left(\dot{S}_i^j+f^{-1}\dot f(S_i^0\d^j_0-S^j_0\d_i^0)+\frac12 f^{-1}\dot f\d_i^0\d_0^jS\right)-\pa_i\left(\dot{S}+\frac12 f^{-1}\dot fS_0^0+2\dot{\f}_B\vf+2Z\dot{\c}_B\t\right)=\NO \\
	&\hskip-0.7in -4\S e^{-2A}f^{-1}\dot{\a}(\pa_ia_t-\pa_0a_i)+4\S He^{-2A}(\d_i^x\dot{a}_y-\d_i^y\dot{a}_x)-4\S He^{-2A}(\d_i^xS_y^0-\d_i^yS_x^0)\dot{\a}, \\\NO\\
	&\hskip-0.7in \mbox{\bf Einstein $rr$:}\NO\\
	&\hskip-0.7in (d-1)\dot{A}\dot{S}+\frac12f^{-1}\dot f\dot{S}_{\perp}=\left(\pa^i\pa_j-\d_j^i\square_B\right)S_i^j+2\dot{\f}_B\dot{\vf}+2Z\dot{\c}_B\dot{\t}+\left(Z_{\f}\dot{\c}_B^2-V_{\f}\right)\vf-V_{\c}\t \NO \\
	&\hskip-0.7in -2\S_{\f}\vf e^{-2A}\left(f^{-1}\dot{\a}^2+H^2e^{-2A}\right) \NO \\
	&\hskip-0.7in +2e^{-2A}\S\left(f^{-1}\dot{\a}^2S_0^0-2f^{-1}\dot{\a}\dot{a}_t-2He^{-2A}(\pa_xa_y-\pa_ya_x)+H^2e^{-2A}(S_x^x+S_y^y)\right), 
	\end{align}
\end{subequations*}
\begin{subequations*}
	\begin{align}
	&\hskip-0.9in \mbox{\bf Scalar $\vf$:}\NO\\
	&\hskip-0.9in \ddot{\vf}+2(d\dot{A}+\frac12f^{-1}\dot f)\dot{\vf}+\square_B\vf-Z_{\f}\dot{\c}_B\dot{\t} +\frac12\dot{\f}_B\dot{S}-\frac12\left(V_{\f\f}+Z_{\f\f}\dot{\c}_B^2\right)\vf-\frac12V_{\c\f}\t \NO \\
	&\hskip-0.9in -\S_{\f\f}\vf e^{-2A}\left(H^2e^{-2A}-f^{-1}\dot{\a}^2\right)\NO \\
	&\hskip-0.9in -\S_{\f}e^{-2A}\left(-2f^{-1}\dot{\a}\dot{a}_t+f^{-1}S_0^0\dot{\a}^2-H^2e^{-2A}(S_x^x+S_y^y)+2He^{-2A}(\pa_xa_y-\pa_ya_x)\right)=0,\\\NO\\
	&\hskip-0.9in \mbox{\bf Scalar $\t$:}\NO\\
	&\hskip-0.9in Z\left(\ddot{\t}+(d\dot{A}+\frac12f^{-1}\dot f+Z^{-1}Z_{\f}\dot{\f}_B)\dot{\t}+\square_B\t\right)+Z_{\f}\dot{\c}_B\dot{\vf}+\frac12Z\dot{\c}_B\dot{S}-\frac12V_{\c\c}\t\NO \\
	&\hskip-0.9in +\left(Z_{\f}\left(\ddot{\c}_B+(d\dot{A}+\frac12f^{-1}\dot f)\dot{\c}_B\right)+Z_{\f\f}\dot{\f}_B\dot{\c}_B -\frac12V_{\c\f}\right)\vf  \NO\\
	&\hskip-0.9in +2f^{-1/2}e^{-dA}\left(-2\P_{\c\c}\t\dot{\a}H+\P_{\c}S\dot{\a}H-2\P_{\c}H\dot{a}_t-2\P_{\c}\dot{\a}(\pa_xa_y-\pa_ya_x)\right)=0,
\end{align}
\end{subequations*}
\begin{subequations*}
	\begin{align}
	&\mbox{\bf Maxwell $r$:}\NO\\
	&\S\dot\a\left( \pa_iS^i_0-\frac12\pa_0S\right)-\S_\f \dot\a\pa_0\vf+f\S\left(-f^{-1}\pa_0\dot a_t+\pa_x\dot a_x+\pa_y\dot a_y\right)+2H\P_\c f^{1/2}e^{-(d-2)A}\pa_0\t=0,
	\\\NO\\
	&\mbox{\bf Maxwell $i$:}\NO\\
	&\pa_r\left(f^{1/2}e^{dA}\left(f^{-1}e^{-2A}\dot\a\left(-\d^i_0\S_\f\vf+\S S^i_0-\frac12S\d^i_0\right)+\S\g_B^{ij}\dot a_j\right)+2H\P_\c\d^i_0\t\right)+2\P_\c\dot\c_B\bar\e^{ijk}\pa_ja_k=	\NO\\
	& 2\P_\c \dot\a(\d^i_x\d^j_y-\d^i_y\d^j_x)\pa_j\t +f^{1/2}e^{dA}\S\pa_j(\pa^ia^j-\pa^ja^i)\NO\\
	&+Hf^{1/2}e^{(d-4)A}\pa_j\left(\left(\S_\f\vf+\frac12\S S\right)(\d^i_x\d^j_y-\d^i_y\d^j_x)-\S S^i_k(\d^k_x\d^j_y-\d^k_y\d^j_x)-\S S^j_k(\d^i_x\d^k_y-\d^i_y\d^k_x)\right).	
	\end{align}
\end{subequations*}
Note that $\bar\e^{ijk}$ denotes the totally antisymmetric symbol in flat space.



\bibliographystyle{JHEP}
\bibliography{2ptfns}
	
\end{document}